\documentclass[final,5p]{elsarticle}
\graphicspath{{figures/}{plots/}{plots/tracking_performance/}{plots/trigger_performance/}}
\usepackage{amssymb,amsmath}
\usepackage{amsfonts}
\usepackage{multirow, makecell}
\usepackage{booktabs}
\usepackage{float}
\usepackage[tableposition=top]{caption}
\usepackage{color}
\usepackage[unicode=true,colorlinks=true,citecolor=red,linkcolor=black]{hyperref}
\usepackage{url}
\usepackage{doi}
\usepackage{siunitx}
\usepackage[caption=false]{subfig}
\usepackage{microtype}
\microtypesetup{expansion=true}
\microtypesetup{protrusion=true}

\newcommand*{\pT}{\mathop{}\!{p_\mathrm{T}}}

\newcommand*{\pTtru}{\mathop{}\!{p_\mathrm{T}^\mathrm{tru}}}

\newcommand{\asin}{\textrm{arcsin}}
\newcommand{\acot}{\textrm{arccot}}
\usepackage{lineno}
\usepackage{enumitem, kantlipsum}
\DeclareSIUnit\barn{b}
\usepackage{xcolor}

\usepackage{apptools}
\AtAppendix{\renewcommand\appendixname{\relax}}

\title{A Triplet Track Trigger for the FCC-hh to improve the measurement of Di-Higgs production and the Higgs self-coupling}

\author[unihd]{T.~Kar}
\ead{kar@physi.uni-heidelberg.de}

\author[unihd]{A.~Sch\"oning}
\ead{otheremail@dress.com}

\address[unihd]{Physics Institute, Heidelberg University, Im Neuenheimer Feld 226, 69120 Heidelberg, Germany}

\journal{Nuclear Instruments and Methods A}

\begin{document}

\begin{abstract}

A new concept, the Triplet Track Trigger (TTT), is proposed for stand-alone tracking at the first trigger level of the FCC-hh detector.
The concept is based on a highly scalable monolithic pixel sensor technology and uses a very simple and fast track reconstruction algorithm that can be easily implemented in hardware processors.
The goal is to suppress the enormous pileup of $\sim$ 1000 minimum bias collisions expected at the FCC-hh experiment and to identify the hard-interaction vertex and the corresponding tracks as a basis for a trigger decision.
In the barrel region, the TTT consists of three closely stacked, highly granular pixel detector layers at radii of $\sim \SI{1}{m}$.
An extension of the TTT to the endcap region increases the geometrical acceptance.

We present full Geant4 simulations and reconstruction performance of a modified FCC-hh reference tracker that includes TTT barrel and endcap detector layers. 
The stacking of TTT layers results in excellent track purity, and the large lever arm ensures very good momentum resolution.
Additionally, sub-mm $z$-vertex resolution is achieved, which allows for very efficient pileup suppression.
By reconstructing pileup suppressed track-jets, the primary vertex of the hard interaction is successfully identified, even at a pileup rate of $\langle \mu \rangle = 1000$ and at trigger level.

The multi-jet signature, ${pp} \rightarrow {HH} \rightarrow b\overline{b} b \overline{b}$  
is used as a showcase to study the trigger performance of the TTT 
and compare it to an emulated calorimeter trigger (calo-trigger).
The TTT allows for significantly lower trigger thresholds and higher trigger efficiencies compared to a calo-trigger. Furthermore, the TTT is very robust against fluctuations in the pileup rate in contrast to the calo-trigger.
As a result, a significant increase in the statistics of di-Higgs events is expected, in particular at low transverse momentum, where the sensitivity to the trilinear Higgs self-coupling ($\lambda$) is the highest.

\end{abstract}
\begin{keyword}
FCC-hh \sep MAPS \sep track trigger \sep calorimeter trigger \sep fast track reconstruction \sep track-jets \sep di-Higgs
\end{keyword}
\maketitle
\section{Introduction}
\label{sec:intro}

Real-time reconstruction of tracks in high rate particle physics experiments is generally extremely demanding, especially in terms of readout bandwidth, the hit combinatorial problem, and computational power.
This is particularly true for the proposed hadron-hadron Future Circular Collider (FCC-hh) experiment.
The baseline is a proton-proton ($pp$) center of mass energy of \SI{100}{\TeV} and an unprecedented luminosity corresponding to about 1000 $pp$-collisions at a bunch crossing rate of \SI{40}{MHz}~\cite{CDR_FCC}.
About $1-\SI{2}{\peta \byte \per \second}$ of raw data is expected from the FCC-hh tracking system alone.
Whether such a huge amount of data can be fully read out with a trigger-less system even $20 - 30$ years from now remains a big question.

The large number of $pp$-collisions (pileup) also has a sizeable impact on almost all the reconstructed objects of an experiment, and together with computational, electrical and storage limitations, it compels trigger systems to increase the trigger thresholds in general.
High trigger thresholds, however, restrict physics analyses of signatures based on ``low-momentum" objects; this makes measurements and systematic studies already at the electroweak scale difficult or even impossible.

An example that demands coverage of the ``low-\-mo\-men\-tum" phase space is di-Higgs production and the related measurement of the trilinear Higgs self-coupling ($\lambda$).
Precise measurement of $\lambda$ and hence understanding the nature of the Higgs potential is one of the crucial next steps for future particle collider experiments.
Unfortunately, di-Higgs production is an extremely rare process, even at the FCC-hh, and therefore a big experimental challenge.

Furthermore, depending on the Higgs decay topology, one or several final state objects can involve hadronic jets with low transverse momenta $\pT_i \lesssim \SI{50}{\GeV/c}$, which will be impossible to trigger relying solely on calorimetry.
To fully exploit the potential of this process in measuring $\lambda$, di-Higgs production must be studied in events in which the Higgs bosons have low transverse momentum ($\pT$)\,\cite{ggFKfactor}.

In order not to lose these low-$\pT$ di-Higgs events, low-momentum objects must be reconstructed at the earliest possible trigger level.
A track trigger, which includes the reconstruction of all tracks and low-energy objects at bunch crossing frequency, offers an attractive but also very challenging solution here.
Tracks not only provide the best handle to identify and suppress event pileup but also allow to trigger physics processes which are difficult to trigger otherwise. 
Prominent examples are tau-leptons or other exotic charged particles~\cite{Alimena:2021mdu}. 
Another example is the ${pp} \rightarrow {HH} \rightarrow b\overline{b} b \overline{b}$ process, which has a multi-jet final state signature very similar to ordinary QCD background and is of high relevance for measuring the trilinear Higgs self-coupling.

The three big challenges in real-time track reconstruction are the limited bandwidth for hit readout, solving the combinatorial hit problem and computational power.
All three challenges are addressed by the Triplet Track Trigger (TTT) proposal \cite{Aschoening1}.
Initial studies of the TTT concept have been performed for the ATLAS and an FCC-hh detector \cite{Tkar_thesis, Fawcett} and are significantly extended here for the latter.
The TTT consists of three dedicated high-resolution tracking layers positioned relatively far from the $pp$-interaction region, where hit rates and occupancies are moderate.
The hit combinatorial problem is largely reduced by closely stacking the tracking layers and instrumenting all TTT layers with high-resolution pixel detectors. 

The instrumentation of large areas of  $\mathcal{O}$(\SI{100}{\m^2}) with pixel detectors has been unthinkable until recently.
Commercialisation of the production process is imperative, and cost will play an important role in the choice of technology.
The sensor technology choice also depends on various other factors: the expected particle rates, the radiation environment, and the targeted momentum range, which defines the magnitude of the multiple Coulomb scattering (MS) 
and hence the maximum allowable amount of material.
The advent of high-voltage monolithic active pixel sensors (HV-MAPS)~\cite{I_Peric}, which also have been proven to be radiation tolerant~\cite{ATLASPix, Augustin:2017guc}, is hence a game changer in many respects since it enables the use of monolithic sensors in high rate applications. 
In contrast to standard monolithic active pixel sensors (MAPS) \cite{SNOEYS1993144, TURCHETTA2001677}, which collect charge by diffusion and are prone to radiation damage,  the substrate of HV-MAPS is depleted such that charge is collected by drift in a large electric field. 
It is therefore not surprising that several other types of depleted monolithic active pixel sensors (D-MAPS) have since been developed \cite{Cardella:2019ksc, Caicedo_2019}, demonstrating the big interest in this technology.  
The radiation hardness of D-MAPS has already been studied and confirmed in the context of the HL-LHC upgrade \cite{ATLAS:2017svb}.
In the context of the planned ITS-3 upgrade of the ALICE experiment~\cite{ALICE_ITS3TDR}, a modified \SI{65}{nm} CMOS has been studied~\cite{Buckland_2024}, also with the goal to increase the radiation tolerance. 
Therefore, we consider D-MAPS sufficiently radiation tolerant and to be suitable for FCC-hh reference detector design, for which a \SI{1}{\MeV}\, neutron equivalent fluence of $\lesssim \SI{1e16}{\!\per \cm^2}$ is projected for radii greater than \SI{30}{\cm} and a fluence of $\approx \SI{1e15}{\!\per\cm^2}$ at a radius of  \SI{85}{\cm} (TTT barrel layers) \cite{CDR_FCC}.

The monolithic concept offers a variety of advantages over hybrid pixel sensors:
Monolithic pixel sensors are based on commercially available standard CMOS processes, which guarantee high availability. 
The compact design, with the implementation of the readout circuitry and charge-collecting diode in the same die, simplifies the production of detector modules and boosts the scalability of the design.
They offer a higher integration level and can, therefore, be produced more cost-effectively than hybrid solutions.
Recent examples include the ALICE ITS tracker~\cite{ALICE_ITS_TDR}, the Mu3e pixel 
tracker~\cite{Mu3e:2020gyw}
and the LHCb mighty tracker project~\cite{Hammerich_2022}.
Because of these advantages, it is reasonable to assume that tracking detectors in future particle physics experiments, like the proposed FCC-hh tracker, will exploit MAPS technologies, either fully or at least to a large extent.

In the following, we assume that the TTT layers can be realised with pixel detectors providing a granularity between $\SI{50}{\micro \meter} \times \SI{40}{\micro \meter}$ and $\SI{20}{\micro \meter} \times \SI{40}{\micro \meter}$, which is state-of-the-art as similar pixel sizes were already implemented in D-MAPS \cite{Cardella:2019ksc}.
A small pixel size is important for the TTT concept as this ensures very good track parameter resolutions. 
In particular, the $z_0$ track parameter resolution is crucial for reconstructing the $z$-position of the primary $pp$ interaction vertex (PV)\footnote{The primary vertex is defined as the point where the hardest $pp$ interaction occurs.} and the aimed pileup suppression at the first trigger level.
\vspace{0.3cm}

In this work, we show that the TTT facilitates the identification of the PV with high performance, even for multi-jet final states, which do not exhibit prominent features like high-transverse momentum tracks (e.g.\ leptons). 
We also show that the process 
${pp} \rightarrow {HH} \rightarrow b\overline{b} b \overline{b}$, which exhibits a multi-jet final state signature,  
can be very efficiently triggered by the TTT at a trigger rate of $\SI{1}{} - \SI{4}{MHz}$ without any b-tagging information. 
This corresponds to a reduction of $10000$ -- $40000$ in proton-proton collision rate and $10$ -- $40$ in bunch crossing rate.
Furthermore, we show that the TTT outperforms a calo-trigger, which has no information about the origin of the particles (vertex) and has only very limited means to distinguish signal and pileup.

The paper is organised as follows: In \autoref{sec:TTTC}
the main concept of the TTT is introduced.
The implementation of the TTT concept in the FCC-hh reference detector and its simulation is described in \autoref{sec:TTTforFCC}.
The track reconstruction algorithm is detailed in \autoref{sec:TTTR}, followed by its implementation in hardware and software in \autoref{sec:TTTAlg_Implementation} and a discussion of the optimal TTT geometry and the track selection cuts in \autoref{sec:RecOpt}.
The presentation of the results is divided into the performance of the track reconstruction in \autoref{sec:trk_perf} and the trigger performance with the ${pp} \rightarrow {HH} \rightarrow b\overline{b} b \overline{b}$ showcase study in \autoref{sec:trig_perf}. 
The summary is given in \autoref{sec:summary}.

\section{The TTT concept}
\label{sec:TTTC}

\begin{figure}[!htb]
  \centering
  \includegraphics[width=0.85\linewidth]{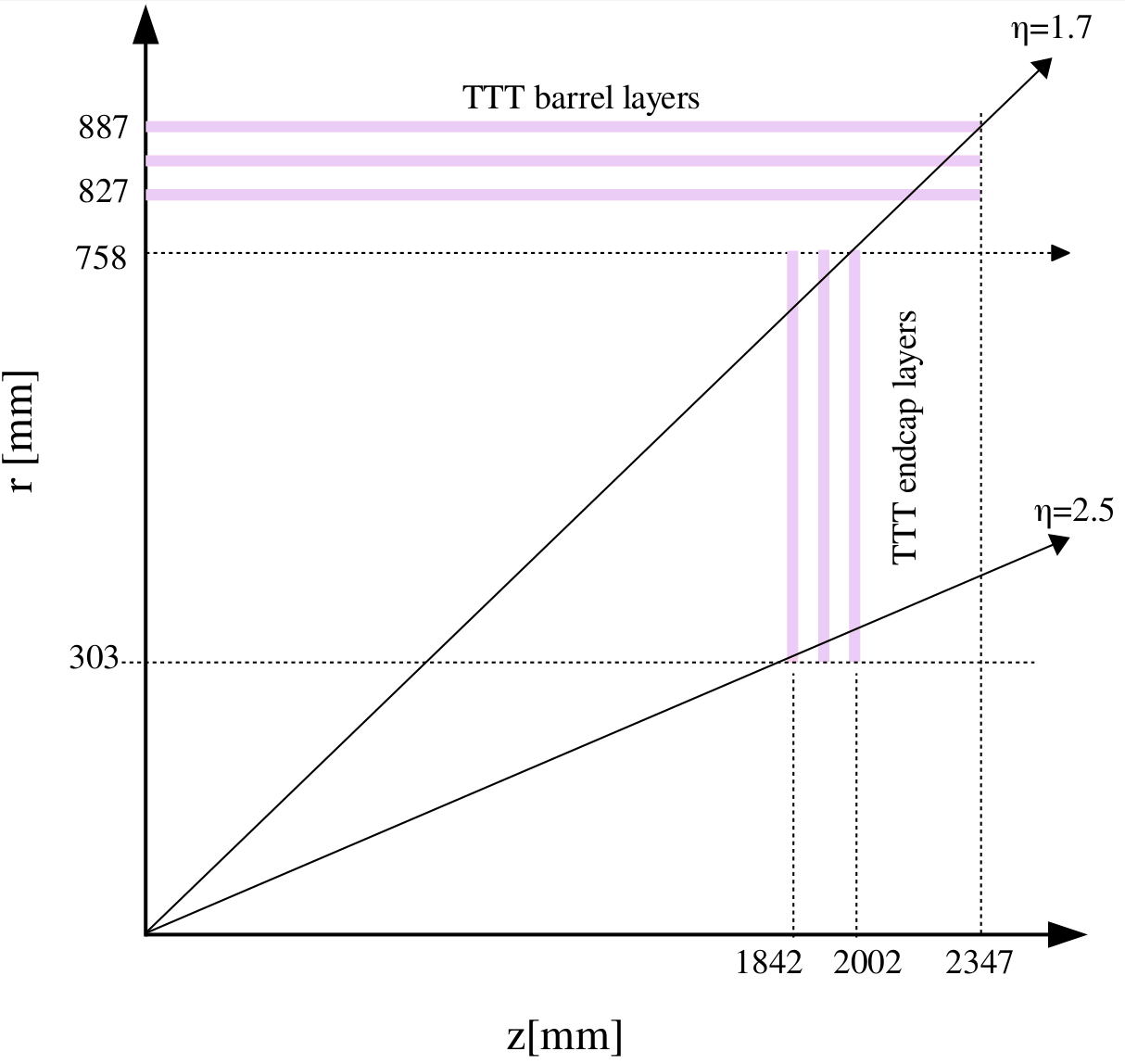}
  \caption{Sketch of the TTT barrel and endcap layers in the $r-z$ plane. The TTT barrel layers cover a pseudorapidity of $|\eta| \leq 1.7$, and the TTT endcap discs extend the coverage to $1.7 < |\eta| \le 2.5$.
  In the simulation, the TTT barrel layers have a pixel size of $40\times \SI{40}{\micro \meter^2}$.
  The endcap layers have a pixel size ranging from $50\times\SI{40}{\micro\meter^2}$ at the outer radius to $20\times\SI{40}{\micro \meter^2}$ at the inner radius. 
}
  \label{fig:BrECsketch}
\end{figure}

The TTT concept is based on the reconstruction of hit triplets in three closely spaced, highly granular pixel detector layers placed in a uniform magnetic field ($B$) at large radii (TTT barrel), see \autoref{fig:BrECsketch}.
There are two main reasons for placing the pixel detectors at a radial distance of
$\cal O$(\SI{1}{m})
from the beamline.
Firstly, at large radii, the hit rate per area is low enough to read out and process all hits. 
Secondly, the tilt angle of a track increases with the radius in the magnetic field; thus, more accurate transverse momentum information is acquired by measuring the tilt angle at large radii.
The TTT concept can also be applied in the endcap regions where triplets of highly granular pixel discs are placed at a large longitudinal distance 
$\approx \SI{2}{m} $ from the collision region and at radii $> \SI{0.3}{m}$
(TTT endcap), see \autoref{fig:BrECsketch}.

\begin{figure}[!htb]
  \centering
    \subfloat[transverse plane]{\includegraphics[width=0.95\linewidth]{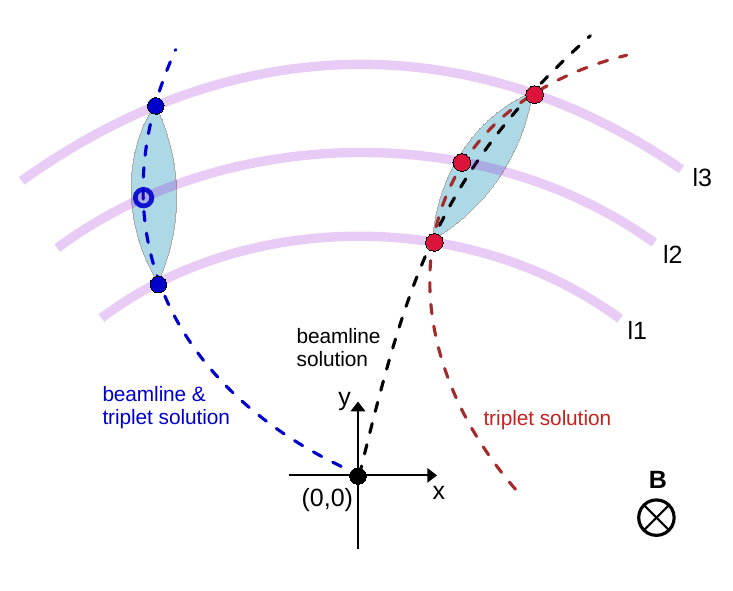}}\\
    \subfloat[longitudinal plane]{\includegraphics[width=0.95\linewidth]{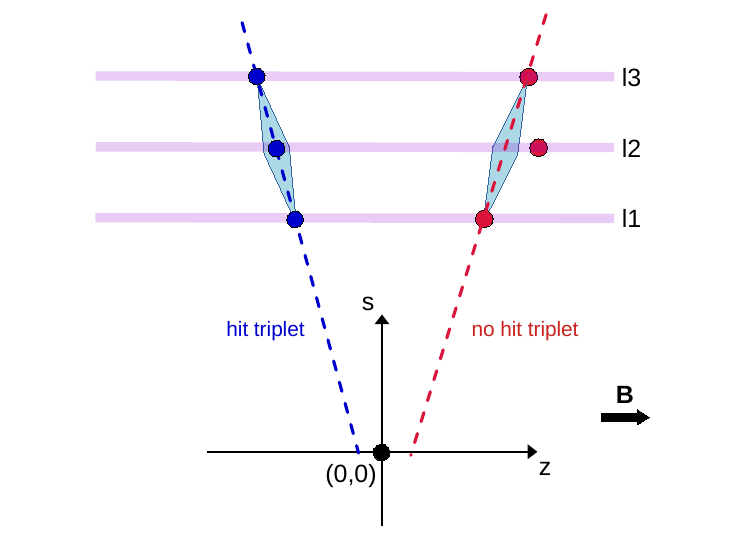}}
  \caption{Sketch of the TTT track reconstruction concept in (a) the transverse (bending) plane  and (b) the longitudinal (non-bending) plane. 
  Triplets are reconstructed in search regions defined by the inner and outer layer hits (\emph{blue} regions).
  A track candidate is successfully reconstructed in the transverse plane if the triplet and beamline solutions match. A track candidate is successfully reconstructed in the longitudinal plane if all three hit positions match a straight line.
  }
  \label{fig:TTTConcept}
\end{figure}

Why three tracking layers? -- At least three space points (a \emph{triplet}) are required to unambiguously define the trajectory of a charged particle in a magnetic field if the particle momentum is not known. 
The trajectory, which is described by a helix in a homogeneous magnetic field, is over-constrained by one degree of freedom.
This constraint comes from the non-bending plane in which the three hits essentially line up as  a straight line, see \autoref{fig:TTTConcept}\,(b).\footnote{This is the reason why the TTT tracking layers need a high segmentation in $z$-direction, and a pixel technology is required.} 
Furthermore, it is possible to apply an additional \emph{beamline constraint} in the bending plane by requiring that the particle originates from the beamline ($x=0$, $y=0$), see \autoref{fig:TTTConcept}\,(a).
A consistency cut ensures that the track curvature determined from the triplet alone (circle fit) is consistent with the one determined from only two of the three hits and the beamline position.
Technically, this is achieved by calculating the track curvature
using two complementary yet simple methods and checking their consistency (\emph{curvature consistency cut}, see \autoref{sec:TTTR}).

Both constraints provide a very efficient handle to suppress combinatorial background and secondary tracks which do not originate from the proton-proton collision.
Compared to other track trigger proposals based on doublet tracking layers only~\cite{Garcia-Sciveres:2010xbz,CMS:2017lum}, the over-constrained kinematics is the biggest advantage of the TTT proposal.
It is quite obvious that an additional fourth tracking layer would add even more redundancy and further improve the performance.
However, in this study, we demonstrate that three 
pixel tracking layers
are sufficient to (a) reconstruct all tracks and (b) provide a sufficiently high track purity for a track trigger.
Note that a high track purity is crucial for identifying the PV of the hard interaction at the FCC-hh conditions with a pileup rate of $\langle \mu \rangle = 1000$.

By construction, the TTT concept is quite agnostic concerning secondary particles from decays of long-lived particles. 
Due to the large track extrapolation from the TTT layers to the collision region, the impact parameter resolution is very moderate.
Because of the beamline constraint in the bending plane, secondary particles can only be identified in the non-bending plane.

\vspace{0.3cm}

\begin{figure}[!htb]
  \centering
{\includegraphics[width=0.99\linewidth]{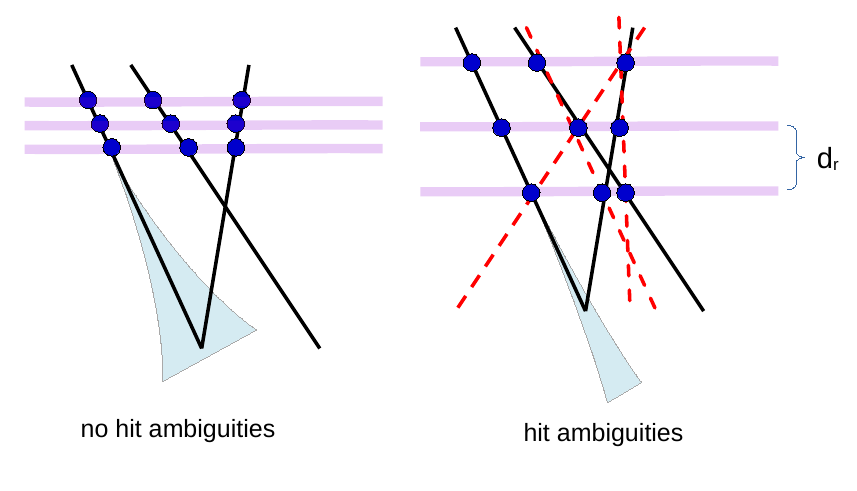} }
\caption{Sketch of closely (\emph{left}) and widely (\emph{right}) stacked tracking layers with the gap distance defined by $d_r$. Black solid (red dashed) lines represent tracks reconstructed from the correct (wrong) hit combinations. The extrapolation uncertainty is visualised by bright-blue bands.} 
  \label{fig:TTTPS}
\end{figure}

\autoref{fig:TTTPS} elucidates one of the main advantages of closely stacked tracking layers for fast track reconstruction.
The close stacking 
allows to run many \emph{Local Track Finder} engines in parallel and reduces hit combinatorics in comparison to more widely stacked tracking layers. 
This is because the phase space for the combinatorial background increases with the distance between the tracking layers.
A disadvantage of the concept is the small
lever arm of the closely stacked layers that compromises the $z_0$ track parameter resolution and, in particular, the momentum resolution.
In the bending plane (\autoref{fig:TTTConcept}\,(a)), 
this is mitigated by applying the beamline constraint, which is equivalent to adding an additional virtual space point at $(x,y)=(0,0)$.
Under the assumption that the particle originates from the beamline, the momentum is then precisely reconstructed. 
This implies that the momentum reconstruction with beamline constraint gives only correct results if the particle originates from or close to the beamline (e.g.\ from particle decays with small impact parameters), as otherwise, the momentum is wrongly reconstructed.

Track finding and reconstruction in the TTT layers are fairly simple and modular, thanks to the stacking of the TTT layers. 
This allows the implementation of the tracking algorithm in hardware, e.g.\ in field programmable gate arrays (FPGAs) or small application-specific integrated circuits (ASICs), as discussed in \autoref{sec:TTTAlg_Implementation}.
The high parallelizability and small latency of these devices allow real-time track reconstruction of all tracks for each hadron-hadron collision and can thus serve as input for a first-level trigger decision.

Concerning the geometrical design of the TTT, a compromise between high track purity and good track parameter resolution has to be found by choosing an optimum gap size for both the barrel and endcap layers.
This optimisation must also include the $z_0$ resolution of tracks, which is crucial for identifying and suppressing pileup.


\section{TTT design for the FCC-hh and Simulation}
\label{sec:TTTforFCC}

\begin{figure*}[!htbp]
  \centering
 \subfloat[The FCC-hh reference tracker]{\includegraphics[width=0.44\linewidth, trim={0 0 -2cm 0cm}, clip]{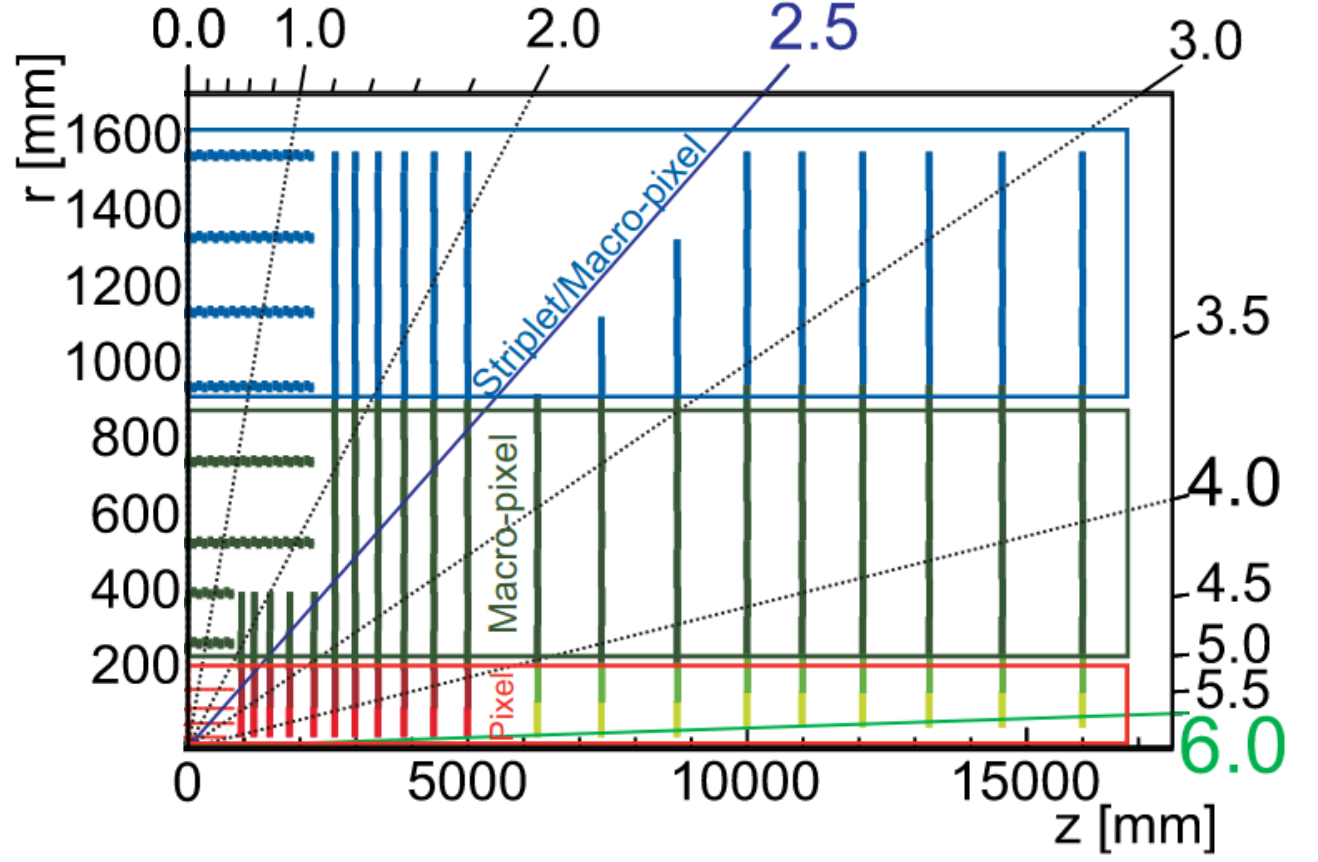} } \\  
 \subfloat[TTT \emph{endcap design}]{\includegraphics[width=0.39\linewidth, trim={0 0 -6cm 0cm}, clip]{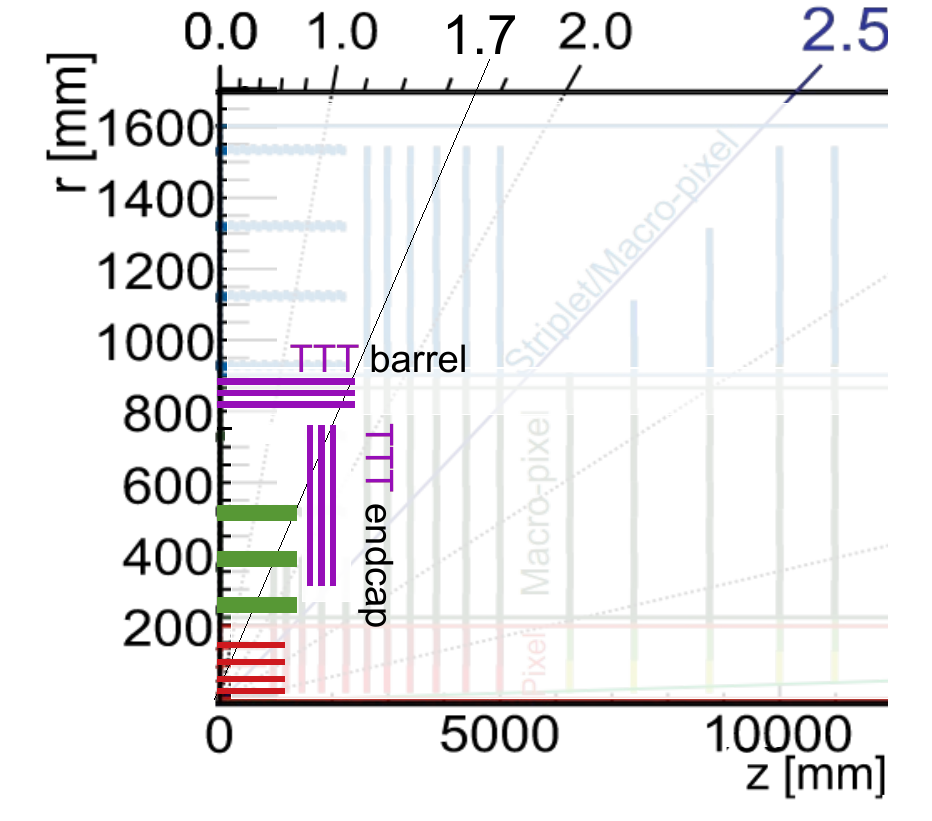}} \qquad \qquad
 \subfloat[TTT \emph{extended design}]{\includegraphics[width=0.39\linewidth, trim={0 0 -6cm 0cm}, clip]{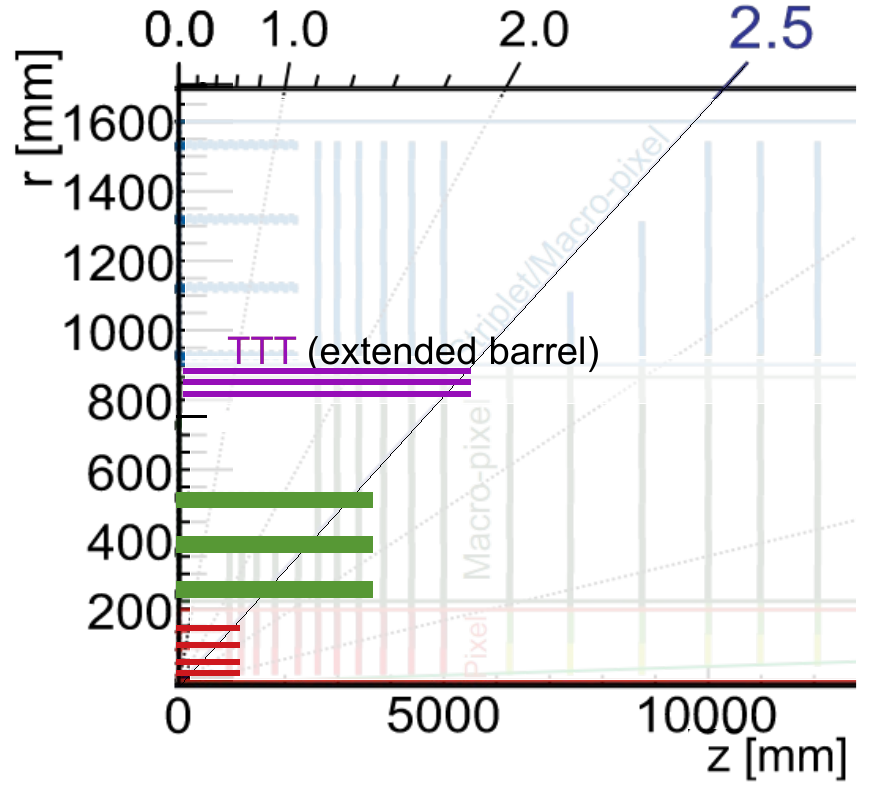}}
  \caption{Tracker layouts ($r-z$ view) for the FCC-hh detector: (a) reference design taken from the FCC-hh CDR\,\cite{CDR_FCC} with a pseudorapidity coverage of $|\eta|\leq6.0$, (b) TTT \emph{endcap design} with TTT barrel layers covering $|\eta|\leq1.7$ and TTT endcap discs in the range $1.7<|\eta|\leq2.5$, and (c) TTT \emph{extended design} with very long TTT barrel layers covering $|\eta|\leq2.5$.
  The pixel, macro-pixel, and strip layers are shown in red, green, and blue, respectively.
  In (b) and (c), the TTT layers are shown in magenta.
  }
  \label{fig:FCC_layout}
\end{figure*}

By modifying the all-silicon FCC-hh reference tracker shown in \autoref{fig:FCC_layout}\,(a),
two different TTT designs for the FCC-hh are studied here,
namely, the TTT \emph{endcap design} \autoref{fig:FCC_layout}\,(b) and the TTT \emph{extended design} \autoref{fig:FCC_layout}\,(c).
Both designs cover a pseudorapidity range of $|\eta|\leq2.5$.
The \emph{extended design} comprises long barrel layers extending up to $|\eta|\leq2.5$, whereas the endcap design uses endcap discs for $1.7<|\eta|\leq2.5$.
Note that the \emph{extended design} might be challenging in terms of installation and routing of services.

\subsection{Geometry}
\label{sec:simulation}

\begin{figure*}[!htbp]
    \centering
    \includegraphics[width=0.8\linewidth, trim={2cm 0.3cm 0.15cm 0.2cm}, clip]{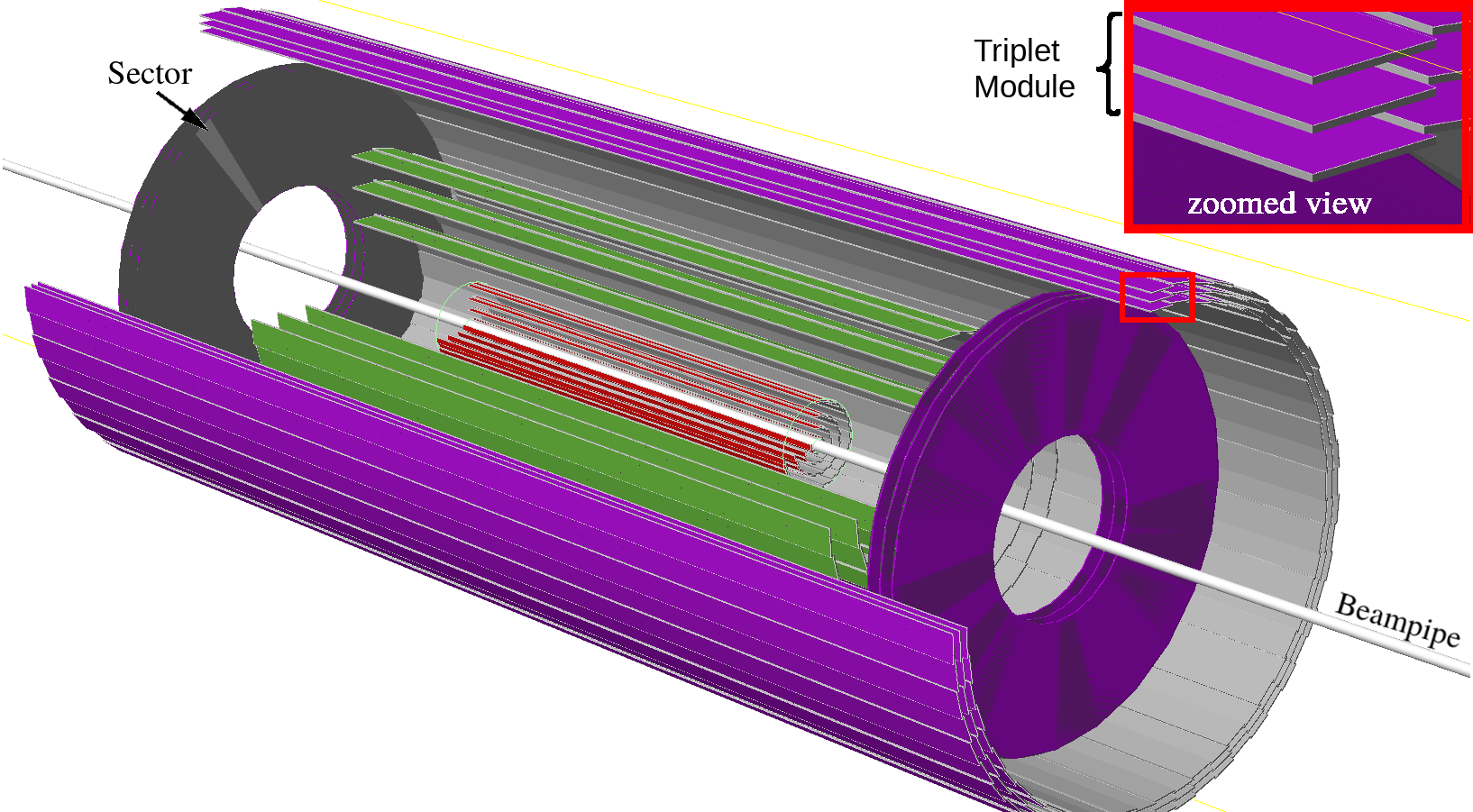}
    \caption{
    3D view of the solid detector geometry implementation of the TTT \emph{endcap design} in Geant4.
    The detector is cut open to present a detailed view of all the layers implemented in the simulation.
    Different colours are used to differentiate between the TTT (magenta), macro-pixel (green), pixel layers (red) and the beampipe (grey).
    A zoomed view of the TTT barrel layers shows the castellation of the triplet modules.
    One of the 58 sectors in the rear endcap disc is highlighted.
    }
    \label{fig:G4BrEC_3D}
\end{figure*}
For both, the \emph{endcap} and \emph{extended design},
the experimental volume consists from the inside out of a beryllium beampipe (radius of \SI{20}{\mm}), four pixel and three macro-pixel silicon layers, followed by the TTT layers and a cylindrical solenoid magnet with uniform field strength $B = \SI{4}{\tesla}$ along the $z$-direction.
The inner pixel and macro-pixel layers of the FCC-hh reference tracker have relative radiation lengths $(X/X_0)$ as specified in \autoref{table:FCC_tracker}, with a total radiation length of at least \SI{11}{\percent}.
In this study, these inner layers act as dead material in front of the TTT. The impact of the TTT on the offline tracking performance is not studied here.

In both designs, the middle TTT barrel layer is positioned at a radius of \SI{857}{\mm} with an equidistant gap of $d_r=\SI{30}{\mm}$ to the inner and outer barrel layers.
The middle layer of the TTT endcap is positioned at $z = \SI{1922}{\mm}$ with an equidistant gap of $d_z = \SI{80}{\mm}$ to the first and last disc layers.
The chosen gap sizes, $d_r$ and $d_z$, are the result of an optimisation study presented in \autoref{sec:geo_opt}.
In the case of the \emph{endcap design},
the transition between the barrel and the endcap discs is chosen to be at $|\eta|=1.7$.
The outer and inner radii of the endcaps are \SI{758}{\mm} and \SI{303}{\mm}, respectively.

\subsubsection{TTT barrel detector}
\label{sec:Barrel_geo}
\begin{table*}[!htb]
\centering
\caption{Geometry specifications of the FCC-hh modified tracker with a pseudorapidity coverage of $|\eta|<2.5$.
The inner pixel ($P_1,\, P_2,\, P_3,\, P_4$) and macro-pixel ($M_1,\, M_2,\, M_3$) layers are simulated as dead material.
The relative radiation lengths include all materials, including sensors, support structure, readout, cooling and other services. 
Details of the material composition can be found in~\cite{Tkar_thesis}.
}
\begin{tabular}{@{}lccccccccc@{}}
\toprule
              & \multicolumn{8}{c}{Barrel Layers}  &     Endcap Discs                                                                     \\ \cmidrule{2-10}
Properties       & $P_1$       & $P_2$      & $P_3$       & $P_4$      & $M_1$      & $M_2$      & $M_3$      & TTT($\times 3$) & TTT($\times 3$)      \\  \midrule
Radius {[}mm{]}          & 25          & 60         & 100         & 150        & 270        & 400        & 520        & 827 -- 887  & 303 -- 758    \\ 
$X/X_{0}\,\text{per layer}~[\si{\percent}]$    & 1.0 & 1.0 & 1.5 & 1.5 & 2.0 & 2.0 & 2.0 & 1.5 & 1.5\\
Stave thickness {[}mm{]}    & 3.3 & 3.3 & 5.0 & 5.0 & 6.6 & 6.6 & 6.6 & 5.0 & 5.0
\\ 
 max.\ pixel granularity [$\si{\micro \meter}^2$] & \multicolumn{7}{c} {inactive}         &$40 \times 40$                & $20\times40$      \\ \bottomrule
\end{tabular}
\label{table:FCC_tracker}
\end{table*}

The basic element of all the barrel layers is a \SI{100}{\micro \meter} thick, fully efficient $2\times\SI{2}{\cm^2}$ silicon sensor with a pixel size of $40\times \SI{40}{\micro \meter^2}$, of which \SI{50}{\micro \meter} are considered to be active and collect charge.
Twenty-five of such sensors are combined to form a $10 \times  \SI{10}{\cm^2}$ \emph{sensor module}.\footnote{ 
For the simulation a sensor efficiency of \SI{100}{\percent} and no dead areas between sensors are assumed.}
Numerous sensor modules are placed along the barrel axis over a long and $3.3 -\SI{6.6}{\mm}$ thick stave to form a \emph{barrel module}.
A stack of three barrel modules, separated by the radial gap size $d_r$, define a \emph{triplet module}, shown in the zoomed view in \autoref{fig:G4BrEC_3D}.
Finally, a TTT  layer is constructed by castellating several triplet modules along the azimuth angle, $\phi$, at respective radii.
58 triplet modules form the TTT barrel layer for a radial gap size of $d_r = \SI{3}{\cm}$.
Each TTT layer has a total radiation length of $1.5\%$ corresponding to a stave thickness of \SI{5}{\mm}.
In the \emph{endcap design}, the TTT barrel layers are \SI{4.8}{\m} long, with an acceptance of $|\eta|\leq 1.7$;
in the \emph{extended design}, they have a length of \SI{10.4}{\meter} and an acceptance of $|\eta|\leq 2.5$.

\subsubsection{TTT endcap disc detector}
\label{sec:Endcap_geo}
Similar to the barrel layers, the endcap discs are constructed by combining 58 endcap modules or \emph{sectors}, which consist of several trapezoidal sensors,
with the pixel positions defined by the azimuthal angle $\phi$ and the radius $r$. 
This simplified design is chosen for the sake of simplicity.
For illustration, one of the 58 sectors of an endcap disc is 
highlighted in \autoref{fig:G4BrEC_3D}.
The material composition of the endcap discs is the same as for the TTT barrel layers.
Unlike the barrel layers, the endcap disc geometry is kept simple and involves no castellation of endcap sectors.
Furthermore, the sensors have a trapezoid-like shape consisting of pixels with dimensions ($r \Delta \phi \times \Delta r$) ranging from $50\times\SI{40}{\micro\meter^2}$ at the outer radius to $20\times\SI{40}{\micro \meter^2}$ at the inner radius.
The rapidity coverage of the endcap discs is $1.7 \leq |\eta| \leq 2.5$.
Around $\SI{7}{\cm}$ gap between the endcap and the innermost TTT barrel layer is not instrumented to provide space for services.
To ensure a smooth transition of the radial size of the hit triplets\footnote{The radial size of the hit triplet determines the momentum resolution.} in the barrel-endcap transition region, 
the $z$-positions of the endcap discs are placed at  
$|z| = 1842, 1922, \text{ and } \SI{2002}{\mm}$.

\subsection{Simulation and Monte Carlo Samples}
\label{sec:MC_sample}
The tracking detector geometry of the FCC-hh all-silicon detector, including the TTT, is simulated using Geant4\,\cite{Geant4}. 
\autoref{fig:G4BrEC_3D} shows a 3D view of the \emph{endcap design} implementation.

For the tracking and trigger performance studies, the following signal and background Monte Carlo (MC) samples are used: $HH \rightarrow 4b$ and $pp \rightarrow 4b$, respectively, generated at leading order employing the matrix element generator \texttt{MADGRAPH5\_aMC@NLO}\,\cite{MG5aMCNLO1}, and \texttt{PYTHIA8}\,\cite{pythia8} for parton showering.
The matrix elements are generated for $pp$ collisions at a center of mass energy of $\sqrt{s} = \SI{100}{\TeV}$. 
The luminous region has a Gaussian beam profile with $\sigma_{z_0} = \SI{57}{\mm}$ and $\sigma_{x_0 (y_0)} = \SI{6.8}{\micro\meter}$.
Minimum bias samples with the same beam conditions are generated and merged with the signal and the background samples to simulate the FCC-hh like conditions with an average pileup of $\langle \mu \rangle = 1000$.
For further details on event generation, we refer to reference\,\cite{Tkar_thesis}.

Geant4 simulations of the FCC-hh all-silicon tracker, including the TTT layers, are performed for all Monte Carlo samples.
Charged particles leave hits on the active detector volume, which are then digitised to global hit coordinates and passed through a simple clusterisation algorithm, which combines
neighbouring hit pixels into a cluster object before being fed into the TTT reconstruction algorithm.
Detector noise is assumed to be much smaller than the huge hit rate from charged particles and, therefore, neglected.

\subsection{Readout and Bandwidth Limitation}
\label{sec:ROandBW}

\begin{figure}[!bht]
\centering
\includegraphics[width=0.9\columnwidth, trim={6 5 8 0.5cm}, clip]{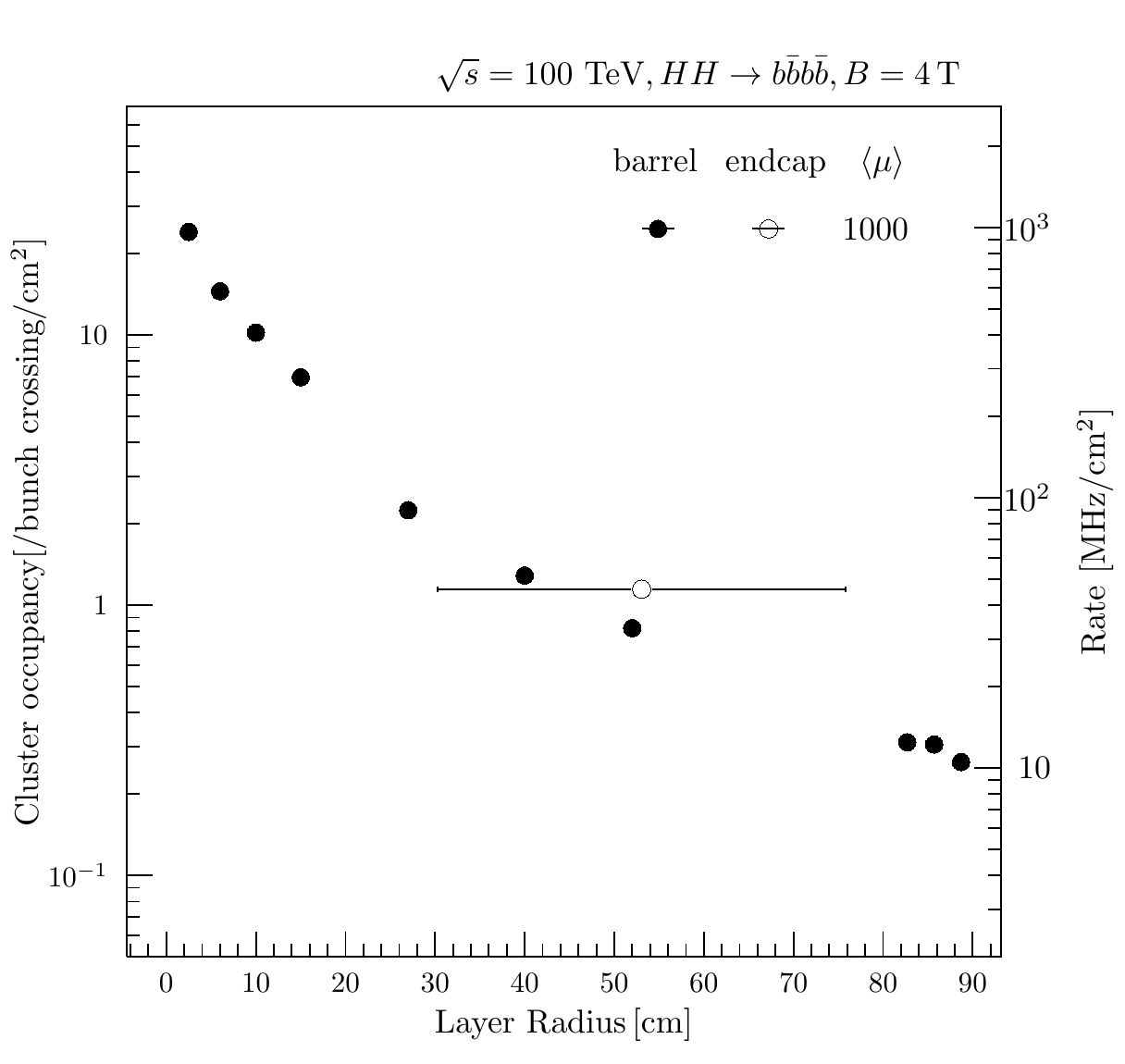}
\caption{Average cluster occupancy per bunch crossing (\SI{25}{\nano\second}) per square cm (left) and the readout rate (right) in \si{M\Hz} per square cm as a function of the layer radius for an average pileup of 1000 events.
The barrel and the endcap (with its average radius) layers are indicated with filled and open markers, respectively.}
\label{fig:HitOcc}
\end{figure}

For the following, we assume that hit clusters are combined on the sensor, yielding a data reduction of about $1.6$ for the TTT layers.
\autoref{fig:HitOcc} shows the average cluster occupancy per \SI{25}{\nano \second} bunch crossing (BX) per  \si{\cm^2}  and the readout rate per \si{\cm^2} of the barrel and endcap discs as a function of their radii for $HH \rightarrow 4b$ events with $\langle \mu \rangle = 1000$.
As expected, the cluster occupancy increases with increasing proximity of the layers to the beamline ($r \sim 0$).
For the TTT barrel layers, on average, about 0.3\,cluster/BX/\si{\cm^2} is expected for a pileup of 1000.
Furthermore, a mean maximum hit occupancy of $\sim 100$\,hits/BX/$(10 \times 10)\,\si{\cm^2}$ sensor module (before cluterisation) 
is obtained from simulation for the outermost TTT barrel layer.
This corresponds to roughly 1\,hit/BX/\si{\cm^2}.

With current readout technologies, we estimate that not more than 1\,hit/BX/\si{\cm^2} can be read out.
Even with anticipated improvements in the readout technology, we consider it impossible to have a complete readout of all hits in the tracker at \SI{40}{\MHz} for $\langle\mu\rangle=1000$.
Therefore, we restrict the \SI{40}{\MHz} readout to radii larger than \SI{30}{cm}.

Assuming a hit representation of \SI{32}{\bit\per hit}, the data rate of 1\,hit/BX/\si{\cm^2}
\sloppy corresponds to a data rate of \SI{1.28}{G\bit/\second} per \si{\cm^2}.
This implies that links with about \SI{5}{G\bit/\second} data rate will be needed to read out sensors of the size of $2\times\SI{2}{\cm^2}$ in the barrel.
An average occupancy of about $1$\,cluster/BX/\si{\cm^2} is expected for the endcap discs that extend from $\sim 30 - \SI{76}{\cm}$ in radius.

\section{TTT Reconstruction Algorithm}
\label{sec:TTTR}

\begin{figure*}[!htb]
  \centering
  \subfloat[]{\includegraphics[width=0.35\linewidth, trim={6.5cm 0.3cm 0.3cm 0.2cm}, clip]{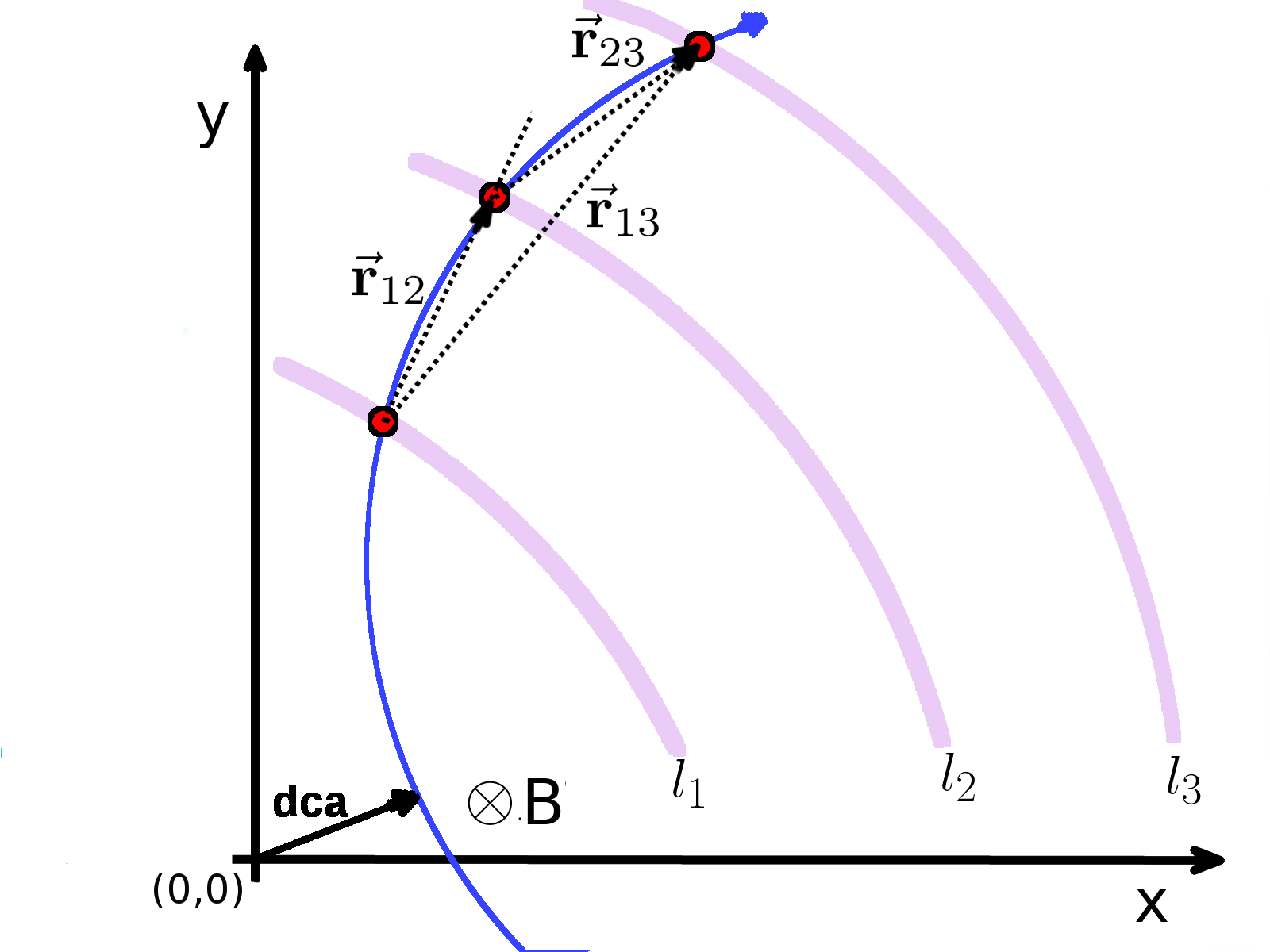}}\qquad \qquad
  \subfloat[]{\includegraphics[width=0.35\linewidth, trim={6.5cm 0.3cm 0.3cm 0.2cm}, clip]{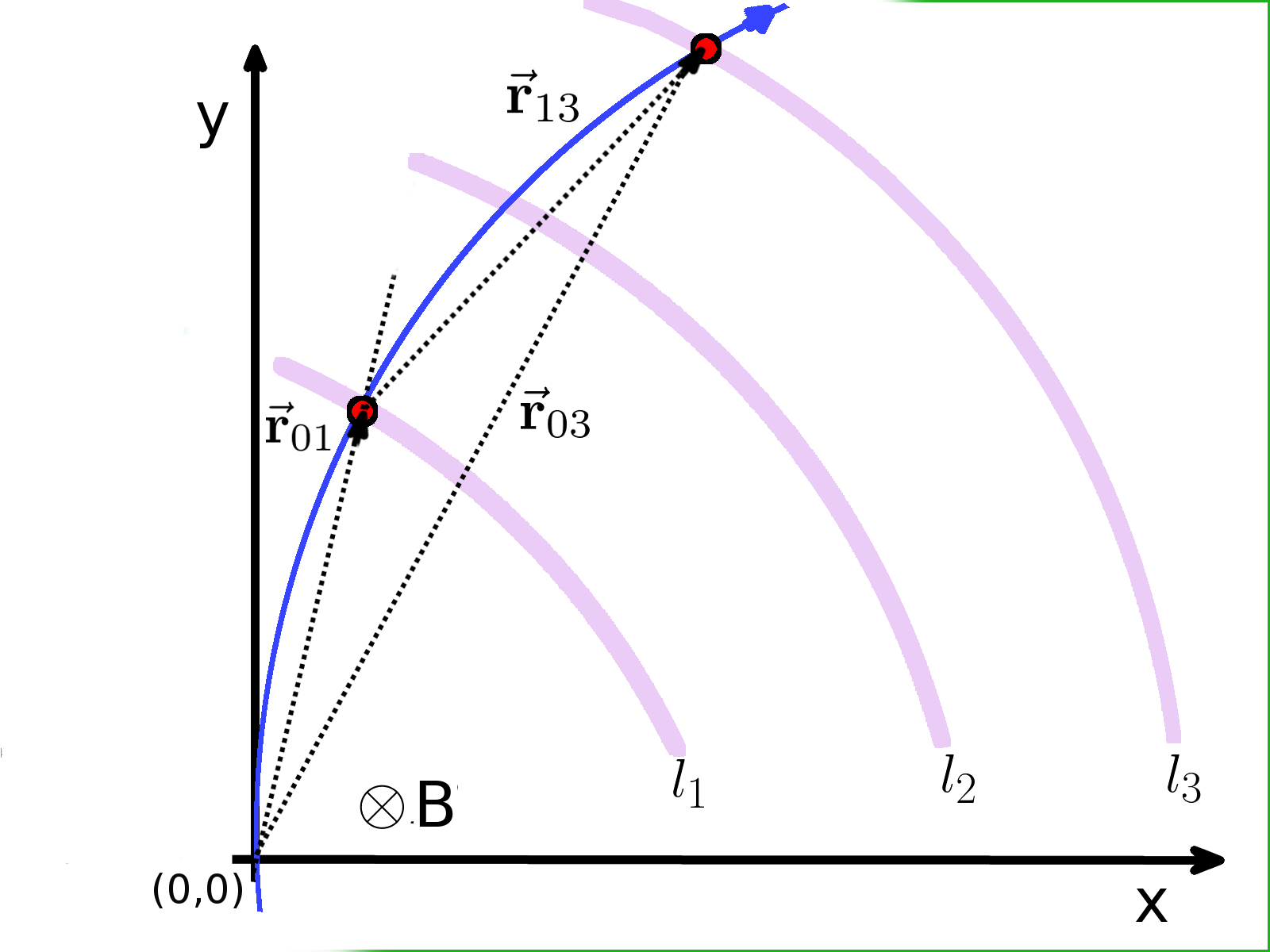}}\\
  \subfloat[]{\includegraphics[width=0.35\linewidth, trim={5cm 0 0 0cm}, clip]{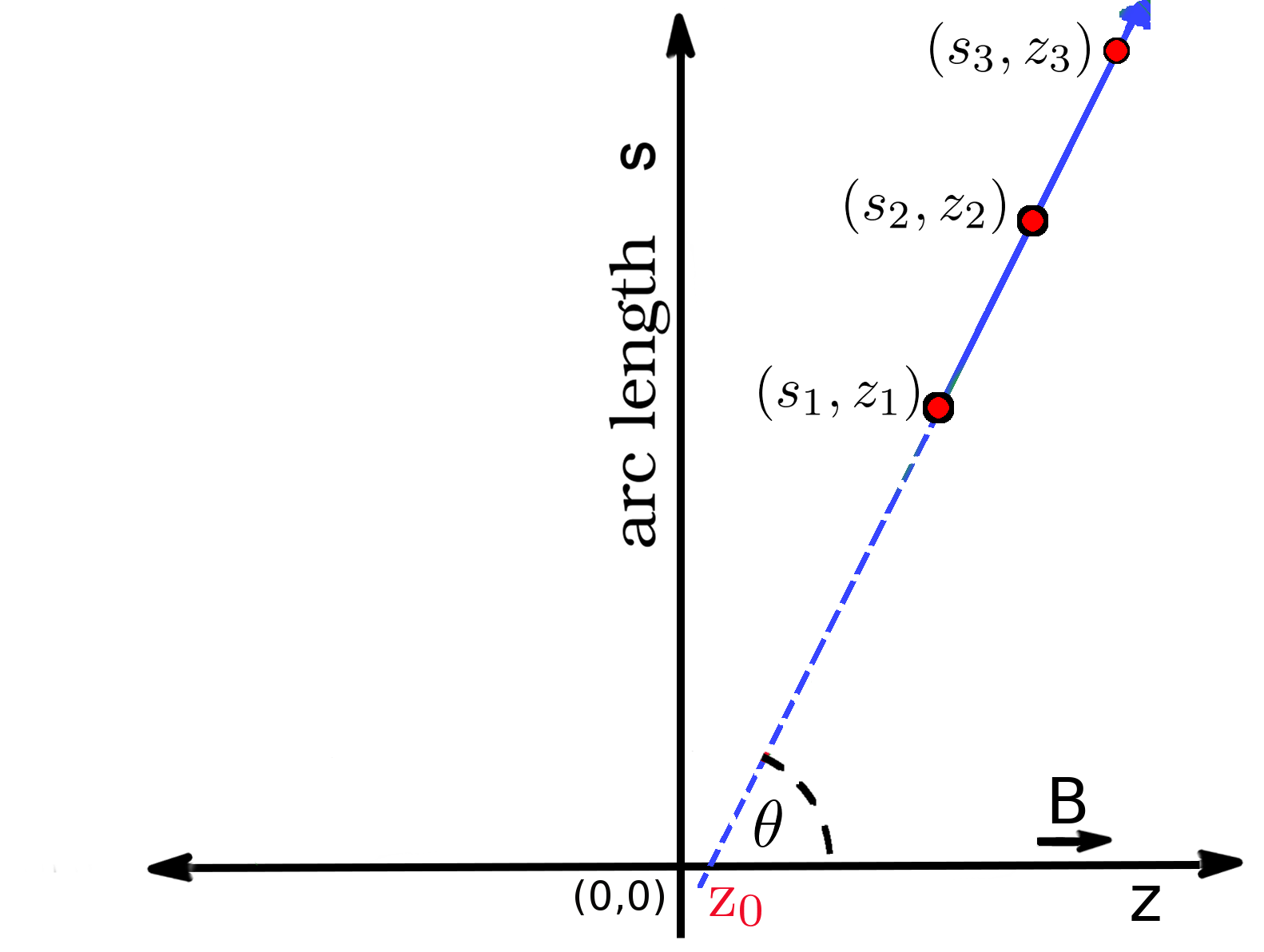}}
  \caption{Reconstruction of a charged particle track in a uniform magnetic field $B$ with the TTT, in the transverse $x$-$y$ plane: (a) without beamline constraint and (b) with beamline constraint, and in the longitudinal $s$-$z$ plane: (c), with $s$ being the arc-length.
  }
  \label{fig:TTTR}
\end{figure*}
In this section, we first describe the reconstruction algorithm for the TTT barrel. In \autoref{sec:TTTREC}, the required changes for the TTT endcaps are described.

\subsection{Reconstruction algorithm for TTT barrel} \label{sec:reconstruction_barrel}

The track reconstruction consists of three parts: the fast track candidate finding, the curvature consistency cut for applying the beamline constraint, and the calculation of the track parameters for applying the final cuts, which determine the track reconstruction efficiency and purity.

The processing time of all three parts is critical for a track trigger. 
Hence, the fast and simple pre-selection cuts are applied before the more
complex (and therefore more time-consuming) calculation of the track parameters.

\subsubsection{Pre-selection Cuts} \label{sec:preselection_cuts}
A triplet is formed by combining three hits, one each from the three layers, $l_1,\, l_2,\, l_3$ of the TTT; 
see \autoref{fig:TTTR}.
Hits belonging to the same track form a straight line in the longitudinal plane, and due to the close stacking of the triplet detector layers, an almost straight line in the transverse plane. 
This approximation is valid if the track radius $R_\textrm{track}$ is much larger than the gap size, i.e.\ $R_\textrm{track} \gg d_r$.

The algorithm starts by searching valid hit combinations in the inner ($l_1$) and outer ($l_3$) layers:\footnote{Spherical coordinates are used, with the positive $z$-coordinate
defined by one of the two beam directions; the nominal interaction point defines the origin.
}
\begin{align}
 \Delta{\phi} &=  |\phi_{3} - \phi_{1}| < \Delta \phi_{13}^\mathrm{cut}, \label{eq:phi13}\\
\Delta{z} &= |z_{3} - z_{1}| < \Delta z_{13}^\mathrm{cut}. \label{eq:z13}
\end{align}
Here, $\phi_i$ are the azimuthal hit angles in the transverse plane, and $z_i$ are the longitudinal hit positions.
These cuts are applied on all hit combinations in a small detector region.
The maximum bending angle cut, $\Delta \phi_{13}^\mathrm{cut}$, takes the magnetic field into account and depends on the minimum transverse momentum to be reconstructed by the TTT.
The cut parameter $\Delta z_{13}^\mathrm{cut}$, defined in the longitudinal plane, depends on the length of the luminous region.\footnote{Alternatively, one could also cut on the polar angle difference $\Delta{\theta} = |\theta_{3} - \theta_{1}| < \Delta \theta_{13}^\mathrm{cut}$, which is computationally a bit more evolved. 
} 

Next, a hit in the middle layer is searched for in a window defined by:
\begin{align}
    \Delta \phi_{2} &= |\phi_2 - 0.5 \, (\phi_1 + \phi_3)| < \Delta \phi_{2}^\mathrm{cut} \label{eq:dPhi2}, \\
\Delta z_{2} &= |z_2 - 0.5 \, (z_1 + z_3)| \: \: <  \Delta z_{2}^\mathrm{cut} / \sin{(\theta)}^{n_z}. \label{eq:dz2}
\end{align}

The residua $\Delta \phi_2$ and $\Delta z_2$ are calculated by 
comparing the middle layer hit position with the interpolated position determined by the hits in the first and third layers.
The $\Delta \phi_2$ cut defines maximum curvature,
see also \autoref{fig:TTTConcept}\,(a), and should also be selected according to the minimum transverse momentum threshold.
The $\Delta z_2$ cut is much tighter and corresponds to a straight line cut in the longitudinal plane; see also \autoref{fig:TTTConcept}\,(b).
The cut value is mainly determined by the amount of MS at the middle tracking layer and therefore, depends on the polar angle. The optimal value for the $\sin(\theta)$ exponent is empirically determined to be $n_z \approx 1-1.5$, see \autoref{table:opt_cutsTTTBr}.

The purity of the track candidates at this level of the track selection is with  $\gtrsim 50 \%$ (see also performance study of $HH \rightarrow 4b$ events and \autoref{fig:PurVspT_ECExt}) already quite high, such that tighter cuts based on computationally more expensive track parameters can be performed. 

\subsubsection{Curvature Consistency Cut}
To further reduce fake candidates and tracks from secondary interactions with the detector material, we employ a simple method called \textit{curvature consistency} cut, which is described in the following.

First, the \emph{triplet track curvature}, $\kappa_{123}$, is calculated from the radial hit position vectors: 
\begin{align}
\kappa_{123} := \frac{1}{\textrm{R}_{123}} 
&=\frac{2\,(\vec{\mathbf{r}}_{23} \times \vec{\mathbf{r}}_{12})_z}{|\vec{\mathbf{r}}_{23}|\,|\vec{\mathbf{r}}_{12}|\,|\vec{\mathbf{r}}_{13}|}
\label{eq:kappa_123}
\end{align}
where $\vec{\mathbf{r}}_{ij}$'s are the position vectors in the transverse plane pointing from hit $i \text{ to } j$, see \autoref{fig:TTTR}(a).
Similarly,  a \emph{beamline-constrained curvature}, $\kappa_{013}$, is defined by requiring that the curve intersects the beamline and the hits in the layers $l_1$ and $l_3$, see 
\autoref{fig:TTTR}(b): 
\begin{align}
\kappa_{013} := \frac{1}{\mathrm{R}_{013}} 
&= \frac{2\,(\vec{\mathbf{r}}_{13} \times \vec{\mathbf{r}}_{01})_z}{|\vec{\mathbf{r}}_{01}| \, |\vec{\mathbf{r}}_{13}|\,|\vec{\mathbf{r}}_{03}|} \label{eq:kappa_013}
\end{align}
The sign of $\kappa$ denotes the direction in which a particle is bent:
$\kappa > 0 \text{ and } \kappa < 0$ correspond to the clockwise and anti-clockwise motion of the particle along the circle, respectively.

The curvature consistency cut requires that both curvatures agree within uncertainties, i.e.\ $\kappa_{123} \approx \kappa_{013}$ and it is implemented by defining the pull:  
\begin{align}
\frac{\Delta\kappa}{\sigma_{\kappa}} &=\frac{\kappa_{123} - \kappa_{013}}{\sigma_{\kappa}}  ,\label{eq:pull}
\end{align}
where $\sigma_{\kappa}$ is the uncertainty on the curvature given by:
\begin{align}
\sigma_{\kappa}^2 = \sigma_{\kappa_{123}}^2 + \sigma_{\kappa_{013}}^2 \approx  \sigma_{\kappa_{123}}^2 \ .
\end{align}

Note that both curvatures have largely different uncertainties, i.e.\ $\sigma_{\kappa_{123}} \gg \sigma_{\kappa_{013}}$. 
The triplet curvature uncertainty is very large due to the short lever arm spanned by the triplet, which is about twice the gap size $|\vec{\mathbf{r}}_{13}| \approx 2 d_r$.
In contrast, the uncertainty of the beamline-constrained curvature, $\kappa_{013}$, is about an order of magnitude smaller due to the very large lever arm of about $\SI{1}{\meter}$. 

The triplet curvature uncertainty, $\sigma_{\kappa_{123}}$ depends on the amount of MS in the material of the middle tracking layer, and the hit uncertainties of all triplet hits:
\begin{align}
\sigma_{\kappa}^2 &= {\sigma_{\kappa}^\text{hit}}^2 +  {\sigma_{\kappa}^\text{MS}}^2 \label{eq:sigma_kappa2} \ ,
\end{align}
which are given by:\footnote{
Note that equations~\ref{eq:sigma_hit} and~\ref{eq:sigma_MS} assume that the bending angle in the magnetic field is not very large for the triplet.}

\begin{align}
{\sigma_{\kappa}^\text{hit}} & \approx \frac {\sqrt{6} \, \sigma_\perp} { d_r^2 }, \label{eq:sigma_hit} \\
{\sigma_{\kappa}^\text{MS}} &
\approx
\frac{ \Theta_\text{MS}}{d_r \cdot  \sin{\theta }} .
\label{eq:sigma_MS} 
\end{align}
Here, $\sigma_\perp$ is the spatial hit uncertainty in all three layers perpendicular to the track direction, and $\Theta_\text{MS}$ is the MS uncertainty at the middle triplet layer. 
The latter depends on the particle momentum $p = q \cdot B/(\kappa \cdot \sin{\theta})$ (see \autoref{sec_track_parameter}), the velocity $\beta \approx 1$ and
the radial material thickness in units of the radiation length, $\xi^\text{BR}/X_0$, as described
by the modified
Highland Formula\,\cite{HighLandFormula}:
\begin{align}
\Theta_\text{MS}^\text{BR} &\approx 
\frac{13.6 \,\si{\MeV \per c}}{\beta p} \sqrt{\frac{\xi^\text{BR}}{X_{0}\,\sin{\theta}}} , \label{eq:HighlandsFormula}
\end{align}
which includes the polar angle dependence of the effective thickness.

The curvature consistency condition is now defined as:
\begin{align}
|\Delta\kappa| < n \cdot \sigma_{\kappa}, \label{eq:curvature_consistency}
\end{align}
where $n$ is an acceptance cut,  typically chosen in the range of $3-5$ standard deviations.

The MS uncertainty computation requires knowledge of the transverse momentum and polar angle, whose calculation, together with other important track parameters, are explained in the following.

\subsubsection{Track Parameter Calculation and Final Cuts} \label{sec_track_parameter}
The best transverse momentum estimate is obtained from the beamline constraint.
The transverse momentum is therefore derived from the curvature, $\kappa_{013}$:

\begin{align}
\pT\,[\si{\GeV\per c}] &= \frac{0.3\,B}{\kappa_{013}}\,[\si{\tesla \cdot \m}]. \label{eq:pT_013}
\end{align}

The calculation of the polar angle requires the arc length of the triplet track, $s_{13}$:
\begin{align}
s_{13}:= s_3 - s_1= \frac{2}{\kappa_{013}} \; \asin{\left(\frac{r_{13} \; \kappa_{013}}{2}\right)} 
\label{eq:track_para_first}
\end{align}
For small bending angles: $r_{13} \; \kappa_{013} \ll 1$, the arc length can be approximated by $s_{13} \approx  r_{13}$.

The polar angle is then given by the difference of the $z$ positions of hits in $l_1$ and $l_3$:
\begin{align}
\theta &= \acot{\left(\frac{z_3 - z_1}{s_{13}}\right)}
\end{align}
By extrapolating the $z$ position of the first triplet hit to the beamline, the $z_0$ parameter is calculated, see \autoref{fig:TTTR}(c):
\begin{align}
z_0 &= z_1 - {s_1} \; \cot {\theta}
\end{align}

Finally, the initial azimuth angle at track origin, $\phi_0$, is calculated from the hit position in the first TTT layer:
\begin{align}
\phi_0 &= \phi_1  +  \asin{\left(\frac{r_{01} \; \kappa_{013}}{2}\right)}
\label{eq:track_para_last}
\end{align}
Note that the distance of the closest approach to the beamline (dca) is zero by construction. 
 
The final track selection cuts which determine the acceptance of the TTT are the minimum transverse momentum cut and the cut on the luminous region:
\begin{align}
\pT >& \; \pT^\textrm{min} \\
|z_0| <& \; z_0^\textrm{cut}.
\end{align}

 The implementation of the TTT algorithm in fast processing hardware is discussed in \autoref{sec:fast_process}.

\subsection{Reconstruction algorithm for TTT endcaps}
\label{sec:TTTREC}
The reconstruction algorithm for the TTT endcap is similar to that of the TTT barrel, but some pre-selection cuts must be adapted to the different geometry.

The endcap discs are equidistantly spaced along $z$.
Since all hits within the same endcap disc have the same $z$-position, the $z$-position cuts of the barrel (\autoref{eq:z13} and \autoref{eq:dz2}) are replaced by cuts on the pseudorapidity:
\begin{align}
    \Delta \eta &= |\eta_3 - \eta_1| < \Delta \eta_{13}^\mathrm{cut} \\
    \Delta \eta_2 &= |\eta_2 - 0.5(\eta_1 + \eta_3)| < \Delta \eta_2^\mathrm{cut},
\end{align}
where $\eta_i = - \ln{\tan{(\theta_i/2)}}$ with $\theta_i = \arctan{(r_i/z_i)}$.

Due to the vertical orientation of the discs, the MS uncertainty has a different polar angle dependence than in the barrel, given by:
\begin{align}
\Theta_\text{MS}^\text{EC} &\approx
    \frac{13.6 \,\si{\MeV \per c}}{\beta p} \sqrt{\frac{\xi^\text{EC}}{X_{0}\,\cos{\theta}}}  \label{eq:ECHighlandsFormula} ,
\end{align}
where $\xi^\text{EC}/X_0$ is the thickness of the endcap in units of the radiation length. 
Another difference comes from the fact that the radial size of an endcap triplet decreases with increasing pseudorapidity, thus reducing the momentum resolution at large pseudorapidities.

\section{Implementation of the Algorithm}
\label{sec:TTTAlg_Implementation}

In this section, we describe a possible hardware implementation of the algorithm and the TTT emulation used for the performance studies in \autoref {sec:trk_perf}  and \autoref{sec:trig_perf}.

\subsection{Hardware Implementation}  \label{sec:fast_process}

\begin{figure}[!htb]
  \centering
  \includegraphics[width=0.98\linewidth]{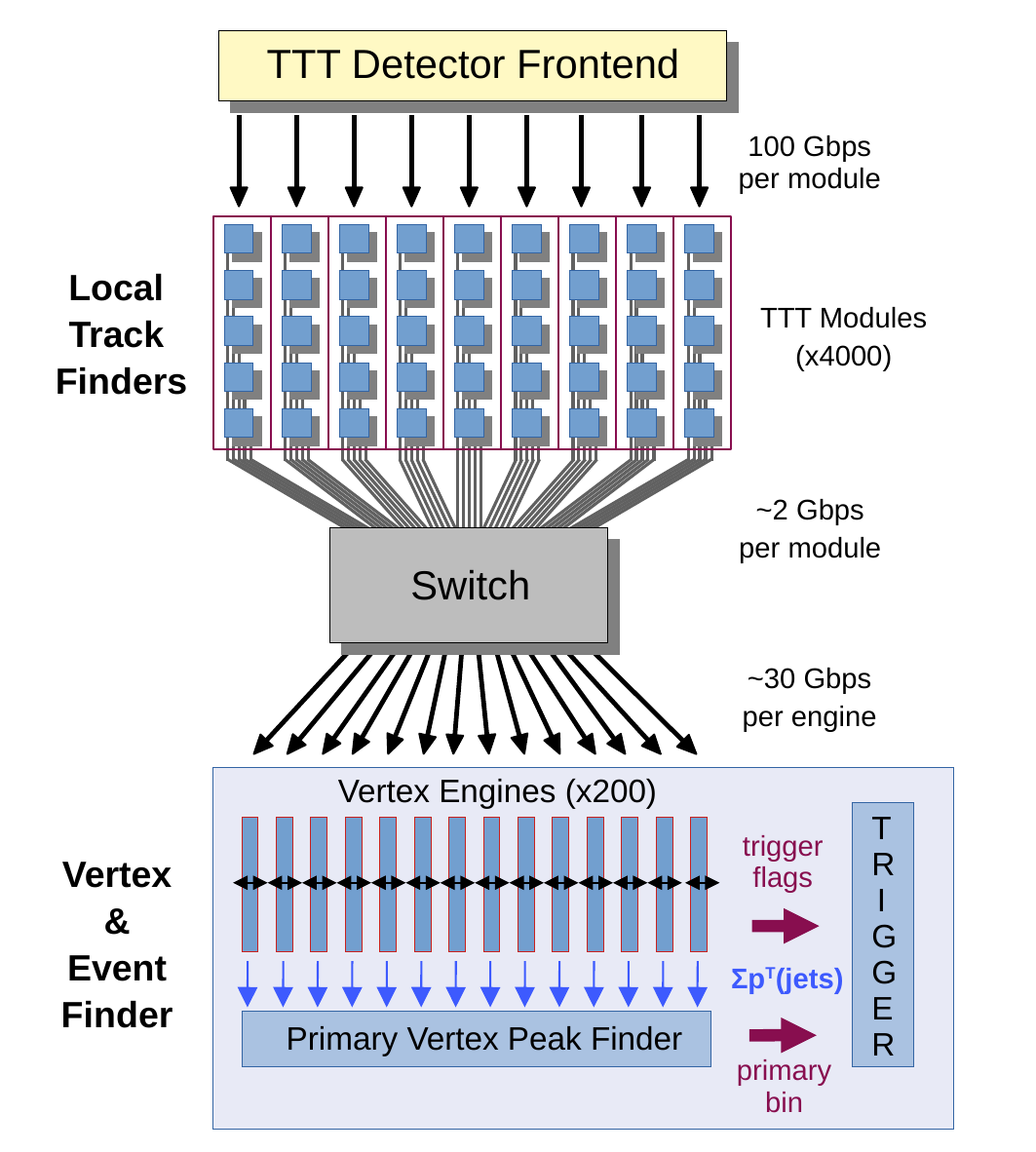}
  \caption{Sketch of the TTT Hardware Implementation.
  }
  \label{fig:HW_implementation}
\end{figure}

The proposed TTT hardware (HW) design follows a modular concept shown in \autoref{fig:HW_implementation}, which implements a high degree of parallelisation.
Hit cluster positions are sent from the Frontend via high-speed links to a farm of \emph{Local Track Finders}, where tracks are reconstructed and track parameters are calculated. 
For performance reasons, the \emph{Local Track Finder} must be realised as an application-specific integrated circuit (ASIC) and the latency is estimated to be \SI{500}{ns}, see \autoref{sec:app_hw}.

Next, the track parameter output is sent via a switch to the \emph{Vertex \& Event Finder}, which consists of about 200 \emph{Vertex Engines}. 
The \emph{Switch} can be regarded as a $z_0$-selector which distributes the tracks to the \emph{Vertex Engine} according to their $z_0$-track parameter.
Every \emph{Vertex Engine} processes tracks originating from a small region of the interaction region and reconstructs track objects (e.g.\ track-jets), which serve two tasks:
Firstly, they are used as input for an event-specific trigger decision. 
In the case of the $HH \rightarrow 4b$ decay studied here, the trigger decision would be based on the presence of at least four track-jets, see \autoref{sec:reference}.
Secondly, they are used to calculate an estimator for the \emph{Primary Vertex Peak Finder}. The estimator is here chosen to be the transverse momentum sum of all filtered track-jets; see also \autoref{sec:PrimaryVertex}.
Finally, the trigger decision is based on the information of the \emph{Primary Vertex Peak Finder} and the trigger decision of the \emph{Vertex Engines}. 
The algorithm is highly parallelizable.
The \emph{Vertex \& Event Finder} is 
assumed to be implemented either in FPGAs or using a High-Performance Computer.

A detailed description of the proposed HW Implementation, including an evaluation of the trigger latency, is given in \autoref{sec:app_hw}.

\subsection{Emulation of the TTT-algorithm}
\label{sec:TTTAlg_Emulation}
For the performance studies presented in the following sections, the track reconstruction algorithm described in \autoref{sec:TTTR} was implemented in C\texttt{++} on a CPU.

To speed up the triplet hit selection described in section\,\ref{sec:TTTR}, at first, the hits are sorted based on the $\phi$ sector they belong to; see \autoref{sec:simulation}.
In a given $\phi$ sector, the algorithm then loops over the hits in the outer layer and checks for valid hit combinations, first in the inner and then in the middle layer.\footnote{
The chosen geometry of the castellated design in the barrel ensures that no two tracks with $\pT \geq \SI{2}{\GeV/c}$ share hits in two adjacent $\phi$ detector modules.
Since the TTT endcap is based on a rather simplified design without castellation, hits from the neighbouring $\phi$ sectors are also considered for track reconstruction.}
The hit selection step is repeated for all $\phi$ sectors.

The pre-selection cuts at this stage are loose.
The last and most important cut to reject fake tracks is the curvature consistency cut.
This cut is applied last since it involves the computationally intensive calculation of trigonometric functions and the calculation of the track quality. 
\section{Optimisation of Track Selection Cuts}
\label{sec:RecOpt}
In this section, the optimisation study for the track selection cuts is presented.
The study aims to optimise the track purity while maintaining a high track reconstruction efficiency.

Tracks are reconstructed from hit clusters measured in the TTT layers using the reconstruction algorithm described in \autoref{sec:TTTR}.
Reconstructed tracks are categorised as \emph{matched} or \emph{fake} based on the following criteria:
If a track comprises only hits arising from the same simulated charged particle, it is classified as correctly reconstructed (matched); otherwise, the track is classified as wrongly reconstructed (fake).
Note that this categorization is very strict since in-flight particle decays in the TTT layers, and particles interacting with the TTT layers (e.g.\ hadronic showers) cannot be matched by definition.

For the following studies, all reconstructed tracks are required to fulfil the TTT acceptance cuts: a minimum transverse momentum $\pT^\text{min}=\SI{2}{\GeV/c}$, a pseudorapidity of less than 2.5 defined by the acceptance of the TTT layers, and a longitudinal vertex position of less than $z_0^\text{cut} = \SI{100}{\mm}$.

\subsection{Optimisation of Pre-selection Cuts}
\begin{figure}[!htb]
  \centering
    \subfloat[$|\Delta\phi_2|$ vs $\sin{\theta}$]{\includegraphics[width=0.89\linewidth, trim={0.01cm 0 0.7cm 0.5cm}, clip]{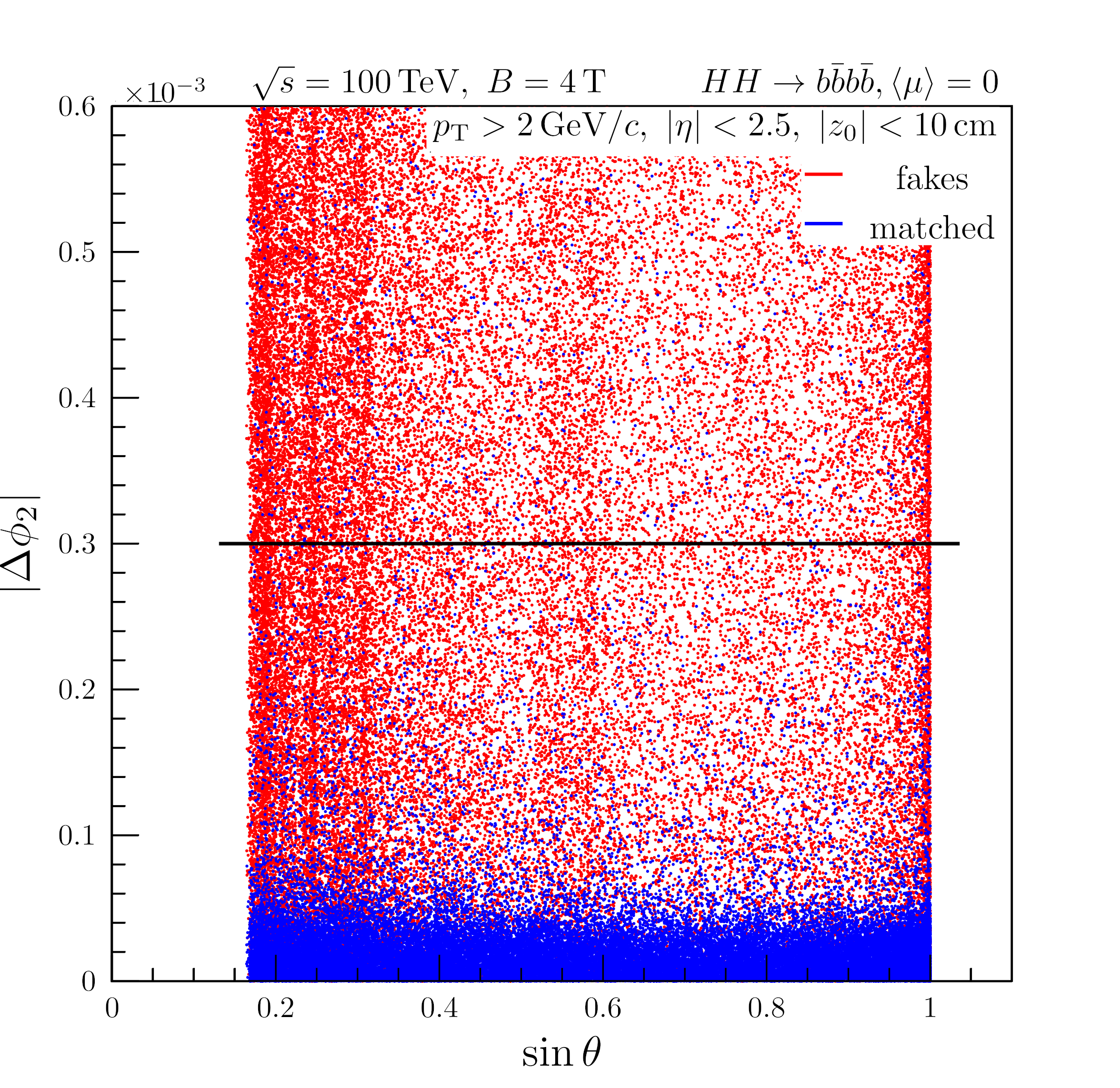}}\\
    \subfloat[$\Delta z_2$ vs $\sin{\theta}$]{\includegraphics[width=0.89\linewidth, trim={0.01cm 0 0.7cm 0cm}, clip]{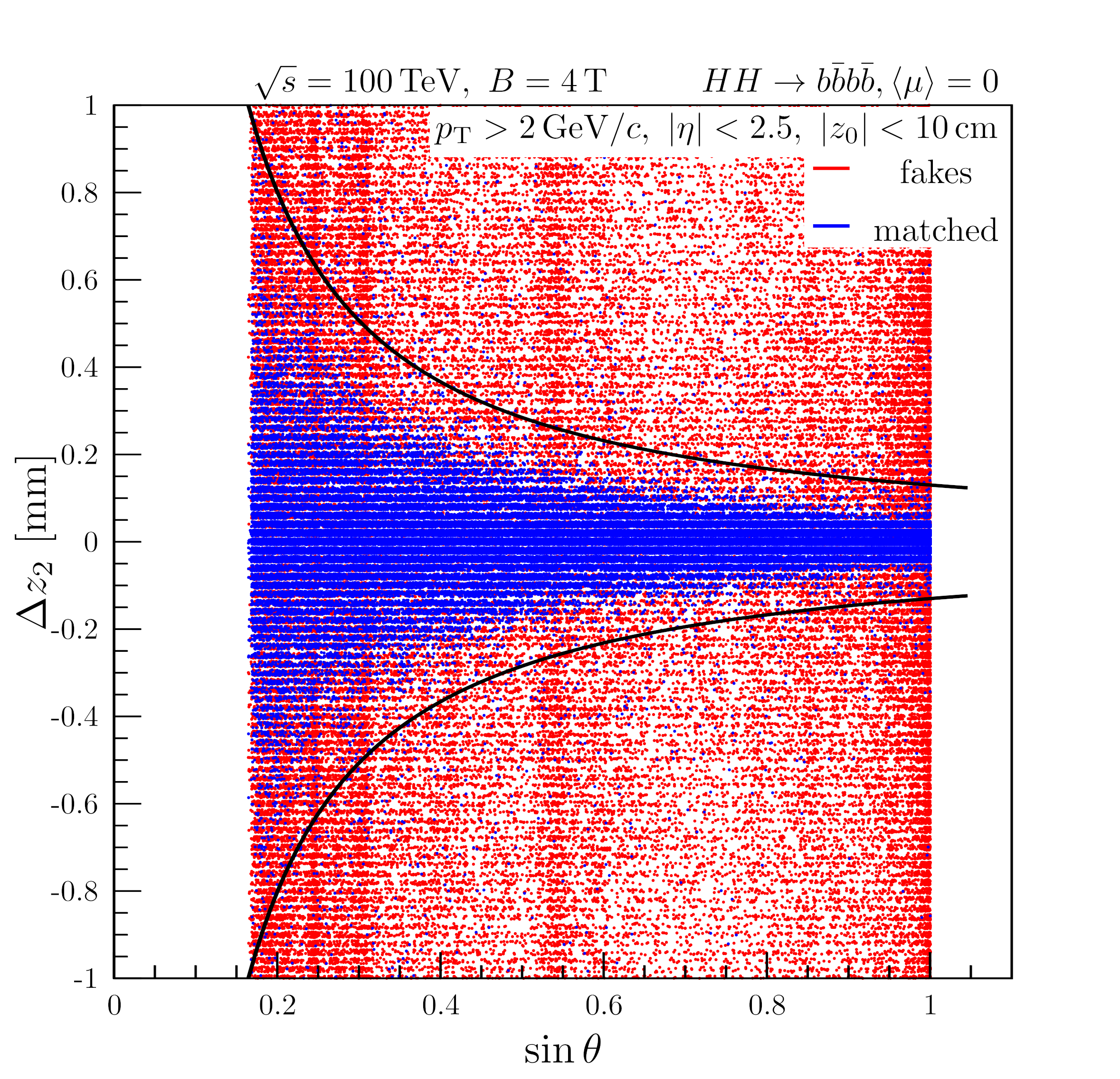}}
  \caption{ $\Delta\phi_2$ and $\Delta z_2$ as a function of  the reconstructed $\sin \theta$ for wrongly reconstructed triplets (fakes) in \emph{red} and correctly reconstructed triplets (matched) in \emph{blue} for the TTT \emph{extended design} with $d_r=\SI{30}{\mm}$ and for $HH \rightarrow 4b$ events.
  The \emph{black solid line} shows the effect of the middle layer pre-selection cuts described by \autoref{eq:dPhi2} and \autoref{eq:dz2}. 
  Track candidates are selected with $\Delta\phi < \SI{0.021}{rad}$, $\Delta z_{13} < \SI{380}{\mm}$, $\pT > \SI{2}{\GeV\per c}$, $|\eta|<2.5$, and $|z_{0}|<\SI{10}{cm}$.
  }
  \label{fig:Redundancy}
\end{figure}

At the pre-selection level, the  $\Delta \phi_2^\text{cut} \text{ and } \Delta z_2^\text{cut}$ cuts, see \autoref{eq:dPhi2} and~\ref{eq:dz2}, ensure that all three hits essentially line up in the transverse and longitudinal plane.
For the TTT \emph{extended design}, the $\Delta\phi_2 \text{ and } \Delta z_2$ values of matched and fake track candidates are shown in \autoref{fig:Redundancy} as function of $\sin \theta$, before applying the $\Delta \phi_2^\text{cut} \text{ and } \Delta z_2^\text{cut}$ cuts. The $\Delta \phi_2$ values of the matched candidates show almost no dependence on the polar angle, whereas the $\Delta z_2$ values scatter with $\approx 1/\sin{\theta}$
around the median, as expected for a detector barrel layer.
The black lines show the selected cut values, which remove $>  \SI{90}{\percent}$ of fakes while keeping essentially all matched track candidates.
At the pre-selection level, all cuts are chosen conservatively such that they neither compromise the tracking efficiency nor the acceptance.

The optimal pre-selection cuts depend on the gap size of the TTT. For the TTT barrel, they are determined for radial gap sizes in the range $d_r=\SI{20}{}-  \SI{40}{\milli \metre}$, and listed in 
\autoref{table:opt_cutsTTTBr}.
Similarly, the pre-selection cuts have been optimised for the endcap. The results are shown for the TTT endcap for gap sizes in the range $d_z=\SI{53}{}-  \SI{106}{\milli \metre}$ in 
\autoref{table:opt_cutsTTTEC}.

\begin{table}[!htb]
\centering
\caption{Optimised pre-selection cuts for the TTT barrel for different radial gap sizes. 
The cut values are defined in \autoref{sec:reconstruction_barrel}.}
\label{table:opt_cutsTTTBr}
\begin{tabular}{lcccccc}
\toprule
\multirow{2}{*}{\begin{tabular}[c]{@{}c@{}}Optimized\\  cuts\end{tabular}} &
 \multicolumn{5}{c}{TTT gap size $d_r\,[\si{\mm}]$}  \\ \cmidrule{2-6}
  
                                         & 20 & 25 & 30 & 35 & 40   \\ \midrule
$\Delta \phi_{13}^\mathrm{cut}$\,[rad]          & 0.014        & 0.018        & 0.021        & 0.025        & 0.028              \\ 
$\Delta z_{13}^\mathrm{cut}$\,[mm]              & 250          & 320          & 380          & 430          & 480           \\ 
$\Delta \phi_{2}^\mathrm{cut}$\,[rad] & \multicolumn{5}{c}{\num{3e-4}}                                          \\ 
$\Delta z_2^\mathrm{cut}$\,[mm]       & 0.1          & 0.12         & 0.13         & 0.15         & 0.16              \\ 
$n_z$                                    & 0.9         & 1.1         & 1.13        & 1.18        & 1.3     \\ \bottomrule             
\end{tabular}
\end{table}
\begin{table}[!htb]
\centering
\caption{Optimised selection cuts for the TTT endcap for different $z$ gap sizes.
The cut values are defined in \autoref{sec:reconstruction_barrel} and \autoref{sec:TTTREC}.}
\label{table:opt_cutsTTTEC}
\begin{tabular}{lcccccc}
\toprule
\multirow{2}{*}{\begin{tabular}[c]{@{}c@{}}Optimized\\  cuts\end{tabular}} &
 \multicolumn{5}{c}{TTT gap size $d_z\,[\si{\mm}]$}  \\ \cmidrule{2-6}
  
                                         & 53 & 67 & 80 & 93 & 106   \\ \midrule
$\Delta\phi_{13}^\mathrm{cut}$\,[rad]          & 0.014        & 0.018        & 0.021        & 0.025        & 0.028              \\ 
$\Delta \eta_{13}^\mathrm{cut}$\,[rad]              & 0.005          & 0.006          & 0.007          & 0.008          & 0.009           \\ 
$\Delta \phi_{2}^\mathrm{cut}$\,[rad] & \multicolumn{5}{c}{\multirow{2}{*}{\num{5e-4}}} \\ 
$\Delta \eta_2^\mathrm{cut}$\,[rad] & \multicolumn{5}{c}{}      \\ \bottomrule
\end{tabular}
\end{table}

\subsection{Curvature Consistency Cut}

The most important cut to reject fake track candidates is the curvature consistency cut (\autoref{eq:curvature_consistency}), which relies on the correct estimation of the curvature uncertainty $\sigma_\kappa$.
In the case of dominating MS uncertainties, $\sigma_\kappa$ depends on the effective material thickness; in the case of dominating spatial hit uncertainties, $\sigma_\kappa$ depends on the pixel size and gap between the TTT layers.

For the TTT barrel, the hit and the MS uncertainties defined in \autoref{eq:sigma_hit} and \ref{eq:sigma_MS} can be rewritten in a simplified form: 
\begin{align}
    \sigma_{\kappa}^\text{hit} \cdot d_r^2 &\approx K_\text{hit} \\
    \sigma_{\kappa}^\text{MS} \cdot |d_r| &\approx \frac{K_\text{MS} \cdot \kappa_{013}}{\sqrt{\sin{\theta}}}.
\end{align}
Here, two new parameters are introduced:
$K_\text{hit}$ depends on the pixel size, whereas $K_\text{MS}$ depends on the magnetic field strength and the relative radiation length of the middle TTT barrel layer.
Note that the parameters $K_\text{hit}$ and $K_\text{MS}$ are generalised. 
They are independent of the gap size and polar angle and need to be determined only once. 

For a pixel size of \SI{40}{\micro \meter} and
a radiation length of \SI{1.5}{\%}\,$X_0$ one obtains:
\begin{align}
K_\text{hit} & \approx  \SI{0.028}{mm},  \\
K_\text{MS} & \approx \SI{1.38}{mm} .
\end{align}

The same generalised curvature consistency parameters can also be used for the TTT endcap if the different orientation of the detector (discs instead of the barrel) and the radius-dependent pixel size are taken into account: 
\begin{align}
    \sigma_{\kappa}^\text{hit} \cdot d_z^2 &\approx \frac{K_\text{hit}}{\tan^2{\theta}} \, \frac{r_2}{[\SI{40}{mm}]} \\
    \sigma_{\kappa}^\text{MS} \cdot |d_z| &\approx \frac{K_\text{MS} \cdot \kappa_{013}}{\sqrt{\cos{\theta}} |\tan{\theta}|},
\end{align}

\begin{figure}[!htb]
  \centering
  \includegraphics[width=0.98\linewidth, trim={0.2cm 0 0.1cm 0.2cm}, clip]{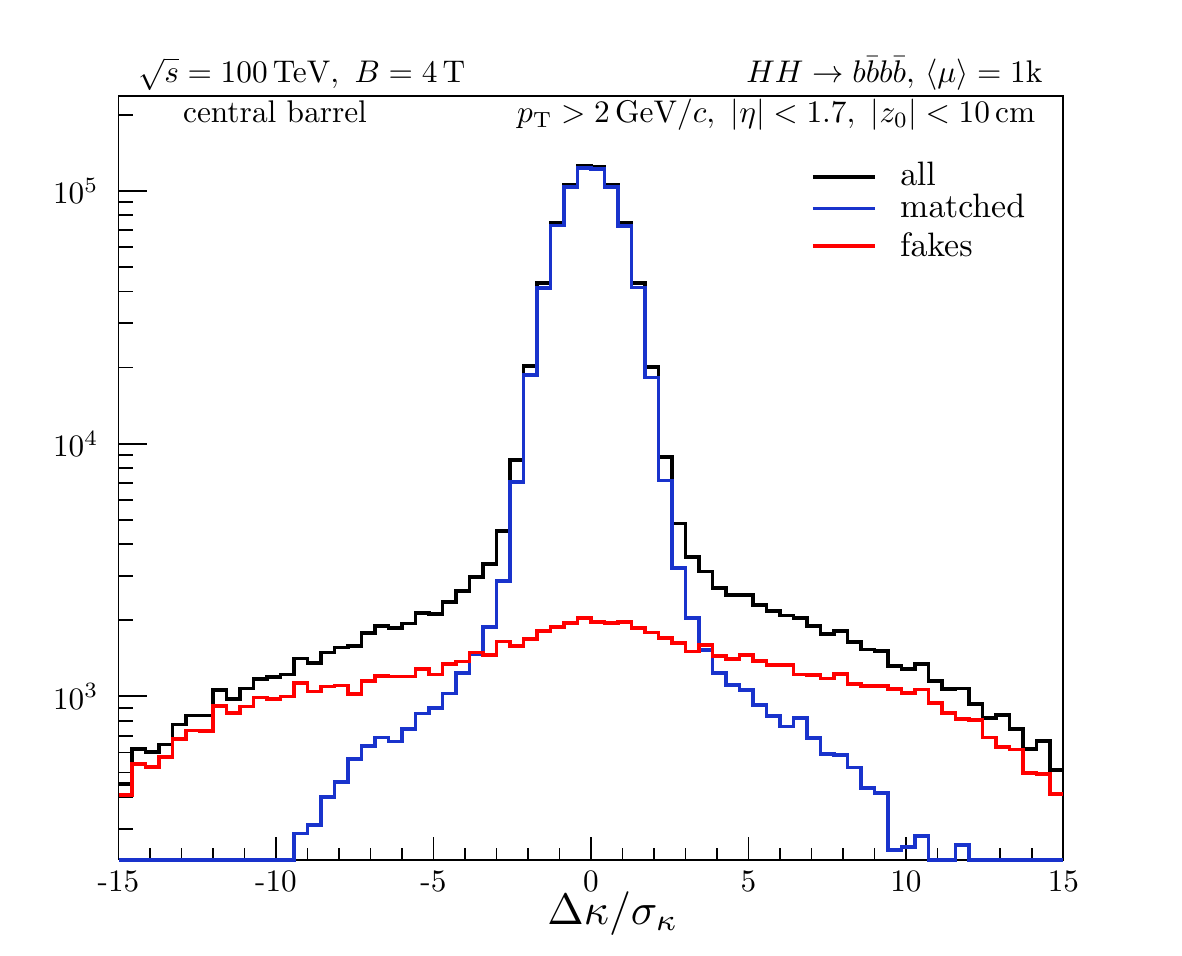}
  \caption{$\kappa$ pull distribution for all reconstructed tracks (\emph{black}), matched tracks (\emph{blue}) and fake tracks (\emph{red}) in $HH \rightarrow 4b$ events with $\langle\mu\rangle = 1000$.
  Only tracks in the central barrel region ($|\eta| < 1.7$) are shown. All cuts except the curvature consistency cut are applied.
  }
  \label{fig:pull}
\end{figure}

To check the correctness of the curvature uncertainty, 
the $\kappa$ pull, defined in \autoref{eq:pull}, is shown in \autoref{fig:pull} for central tracks.
A clear peak at $\Delta\kappa/\sigma_{\kappa} = 0$ with an $RMS=2.75$ is visible.
The core of the peak, which has a fitted sigma of 1.025, is from matched tracks, which originate from the beamline. 

The matched tracks also show tails on both sides, which originate from a small contamination of secondary decays. These tracks have a non-zero impact parameter $d_0$, which is not accounted for in the beamline constraint and, thus, dilutes the $\kappa$ pull distribution.

At this selection level, the contribution of fake tracks is already quite low and has a rather flat distribution.
Fake tracks consist of two contributions; one is purely combinatorial background, and the other is from secondary particles produced in particle interaction with the TTT tracking layers. 
These secondary particles 
are in \autoref{fig:pull} responsible for the peak in the fake distribution around $\Delta\kappa/\sigma_{\kappa} = 0$.

\begin{figure}[!htb]
  \centering
  \includegraphics[width=0.99\linewidth, trim={0.15cm 0 0.1cm 0cm}, clip]{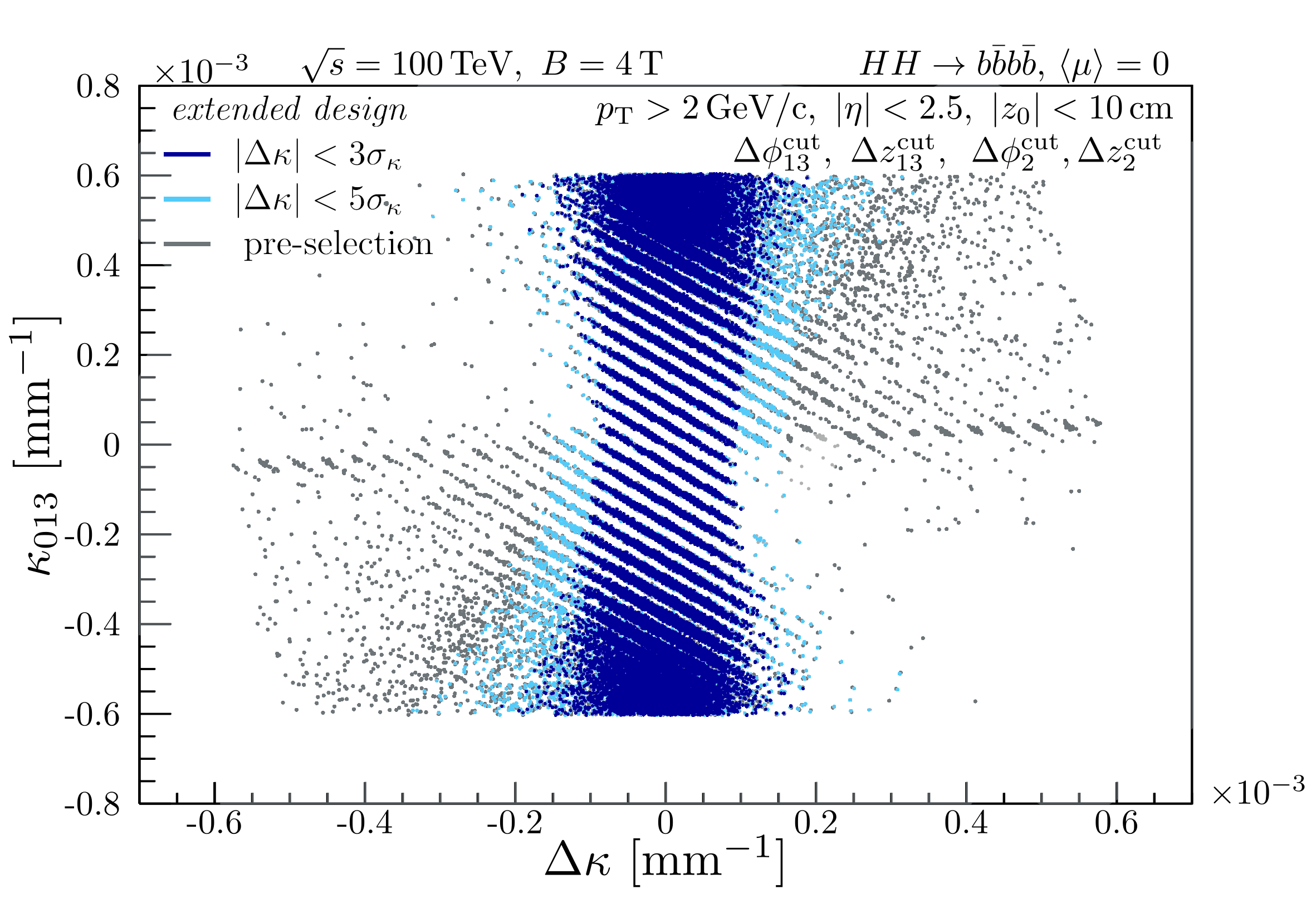}
  \caption{Reconstructed curvature, $\kappa_{013}$, as a function of the difference between the curvatures determined using two independent methods for tracks reconstructed with $\pT > \SI{2}{\GeV\per c}$, $|\eta|<2.5$, and $|z_{0}|<\SI{10}{cm}$ from the $HH \rightarrow 4b$ events.
  The effect of the momentum consistency cut for three different $\kappa$ cuts, namely, a $3\sigma_{\kappa}$ cut in \emph{dark blue}, a $5\sigma_{\kappa}$ cut in \emph{light blue}, and only the pre-selection cuts without $\kappa$ cut in \emph{grey}, is  presented for the TTT \emph{extended design} with $d_r=\SI{30}{\mm}$.}
  \label{fig:KappaCut}
\end{figure}

For the \emph{extended design},
\autoref{fig:KappaCut} shows the correlation of the beamline-constrained curvature $\kappa_{013}$ versus the curvature difference, $\Delta\kappa=\kappa_{123}-\kappa_{013}$.
Track candidates are shown with different colours depending on the level of consistency: 
\begin{itemize}
\item $|\Delta\kappa| < 3 \, \sigma_{\kappa}$ in \emph{dark blue} (tight selection),
\item $|\Delta\kappa| < 5 \, \sigma_{\kappa}$ in \emph{light blue} (loose selection),
\item only pre-selection  in \emph{grey}.
\end{itemize}
The distribution shows an ``hourglass" region populated by good track candidates, which also fulfil 
curvature consistency cuts. 
This ``hourglass" shape comes from the combined hit-position and MS uncertainties. 
Entries outside the ``hourglass" region which do not fulfil the final
curvature consistency cut are mainly due to the secondary electrons and positrons arising from late photon conversions in the FCC-hh tracking detector.
The curvature consistency cut is not only important for separating correct and wrong hit combinations but also quite effective in rejecting a significant fraction of secondary particles.
Secondary particles created in the material at a certain distance from the beamline have a strong positive correlation between $\kappa_{013}$ and $\Delta \kappa$ and are visible in \autoref{fig:KappaCut} as tilted bands going from the lower left to the upper right (see also figure~7.10 in \cite{Tkar_thesis}).
In addition, many parallel fine stripes are visible in the plot, which results from discretisation effects due to the finite pixel size and the small gap between TTT layers.


\section{Tracking Performance}
\label{sec:trk_perf}

In this section, the tracking performance is presented for the TTT \emph{extended} and  \emph{endcap designs} using a MC sample consisting of $\sim 3500~ HH \rightarrow 4b$ signal events; see \autoref{sec:MC_sample} for details.

\begin{figure}[!htb]
  \centering
  \includegraphics[width=1.0\linewidth, trim={0.0cm 1.0cm 1.6cm 0.7cm}, clip]{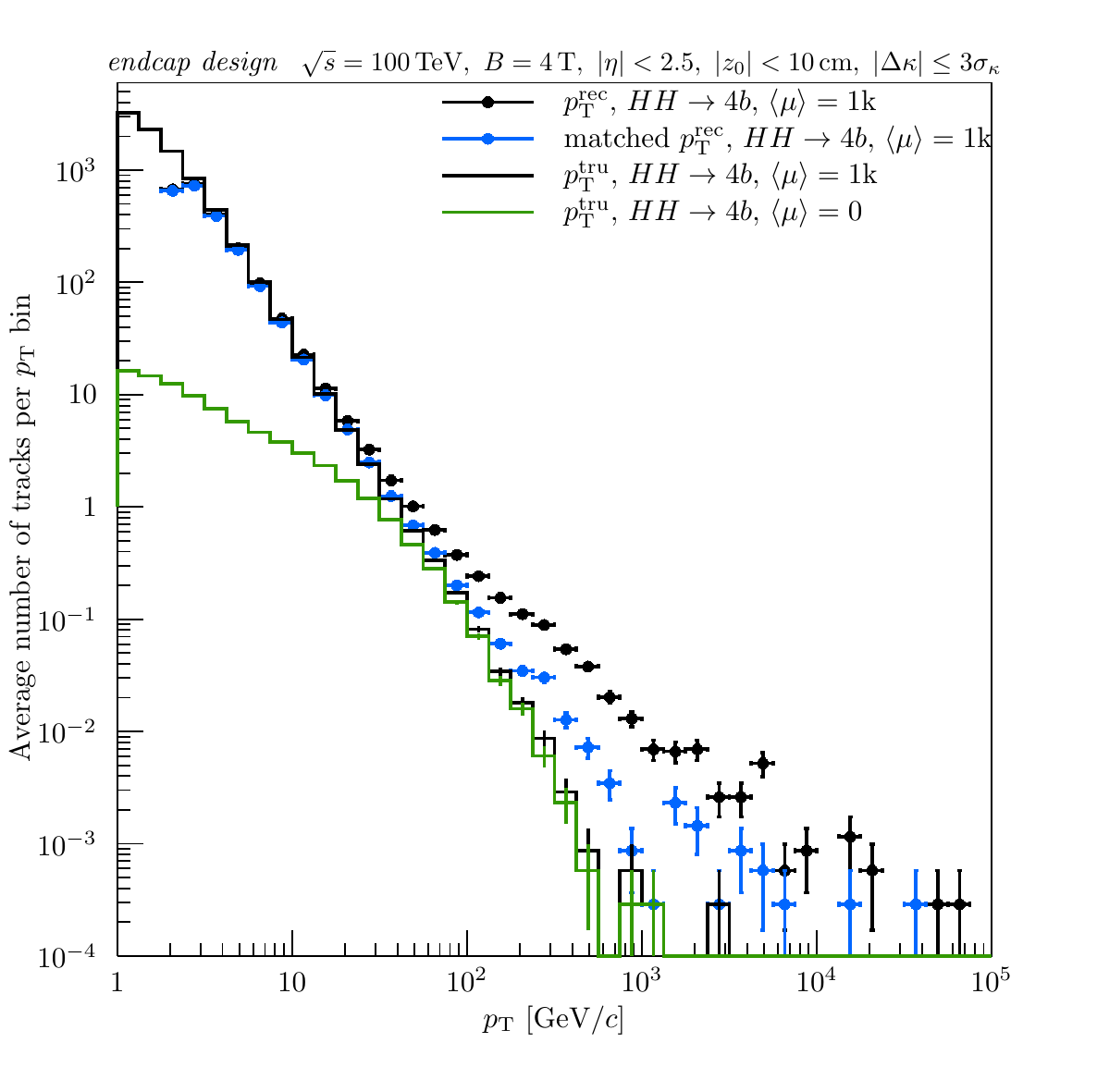}
  \caption{ 
  Transverse momentum distribution of charged particles in 
  $HH\rightarrow4b$ events normalised to the number of events. 
  The truth transverse momentum ($\pT^\text{tru}$) distribution is shown without pileup (\emph{green line}) and with an average pileup of $\langle \mu \rangle =1000$ (\emph{black line}).  
  The transverse momentum distribution of  $HH\rightarrow4b$ with $\langle \mu \rangle =1000$ events as reconstructed by the TTT ($\pT^\text{rec}$) is shown for all (i.e.\ both matched and fake) reconstructed tracks (\emph{black points}) and for matched tracks (\emph{blue points}) only.
  The vertical error bars display the statistical uncertainties.
  Note that tracks below \SI{2}{\GeV/c} are not reconstructed by the TTT.
  }
  \label{fig:pT_dist}
\end{figure}

In \autoref{fig:pT_dist}, the simulated (true) transverse momentum ($\pT^\text{tru}$) distributions of all charged particles originating from the luminous region  ($|z_0|< \SI{10}{\cm}$) are shown for $HH \rightarrow 4b$ events with (\emph{black}) and without (\emph{green}) pileup.
At low and medium transverse momentum, $\pT^\text{tru} \lesssim \SI{30}{\GeV\per c}$, 
the distribution is steeply falling and dominated by tracks from minimum bias collisions.
This is the region where the track resolution is dominated by MS.

The $\pT$ distribution as reconstructed by the TTT (\emph{black points}) is also shown in \autoref{fig:pT_dist}.
The true transverse momentum distribution is very well reproduced,
except for very high transverse momenta.
The excess of reconstructed tracks observed for $\pT \gtrsim \SI{10}{\GeV/c}$ results from a combination of two effects.
Firstly, the huge amount of low-$\pT$ tracks from minimum bias events create fake tracks; this combinatorial background is approximately flat in $1/\pT$ and becomes visible in the figure as the difference between the \emph{black} and \emph{blue points} at high  $\pT$.
Secondly, the curvature uncertainty from hit position uncertainties increases with $\pT$ and dilutes the momentum resolution. 
Due to the steeply falling $\pT$ distribution, migration effects distort the distribution and become visible in the figure at very high $\pT$, visible as the difference between the \emph{blue dots} and the \emph{black line}.

The tracking performance of the TTT is further evaluated based on the following figure of merits:
\begin{itemize}[leftmargin=*]
    \item The \emph{track reconstruction efficiency} measures how efficiently the TTT reconstructs charged particles.
    It is defined as the fraction of correctly reconstructed tracks (matched tracks) out of all simulated tracks that fall within the detector acceptance of $|\eta|\leq2.5$ and originate near the beamline, i.e.\ $|d_{0}| \leq \SI{0.5}{\cm}$ and $|z_{0}| \leq \SI{10}{\cm}$ with $\pTtru \geq \SI{2}{\GeV/c}$ (\emph{truth} tracks).
    \item The \emph{track purity}
    is defined as the fraction of the correctly reconstructed tracks  (matched tracks) from all the tracks reconstructed by the TTT. 
    The track purity is anti-correlated to the reconstruction efficiency.
    \item {The \emph{track parameter resolution} is a quality measure for the track parameter determination and 
    derived from the difference of the TTT reconstructed and the simulated track parameter values.}
    Of special importance are the transverse momentum ($\pT$) resolution and the $z_0$ resolution, which are key for the pileup suppression.
\end{itemize}

\subsection{Track reconstruction efficiency}

The track reconstruction efficiency depends on the selection cuts, the particle type and the track parameters 
and, therefore, on the sample used for the study.
Charged particles produced in the chosen $HH \rightarrow 4b$ events are mainly pions, kaons and other stable mesons that leave hits in the tracker.
B-mesons that decay semi-leptonically also give rise to electrons and muons of high momentum.
Charged particles produced in minimum bias events are mostly pions of low momentum.

Thus, the overall tracking efficiency for the $HH \rightarrow 4b$ sample presented here is largely determined by the pion efficiency.

\begin{figure}[!htb]
\centering
\subfloat[efficiency vs $\pTtru$]{\includegraphics[width=0.85\columnwidth,trim={0.2cm 0.2cm 1.1cm 1.1cm}, clip]{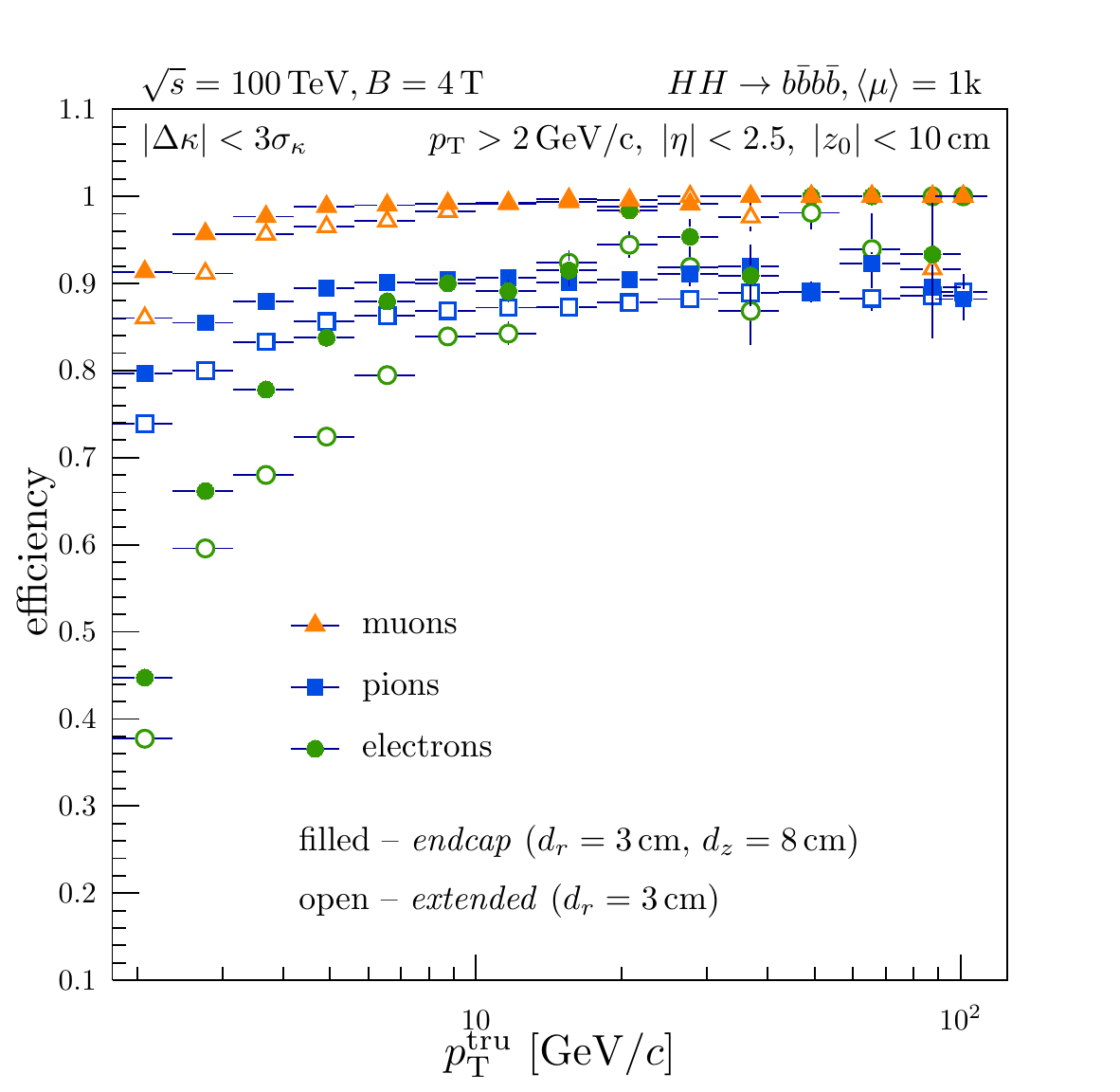}}\\
\subfloat[efficiency vs $\eta^\mathrm{tru}$]{\includegraphics[width=0.85\columnwidth,trim={0.2cm 0.2cm 1.1cm 1.1cm}, clip]{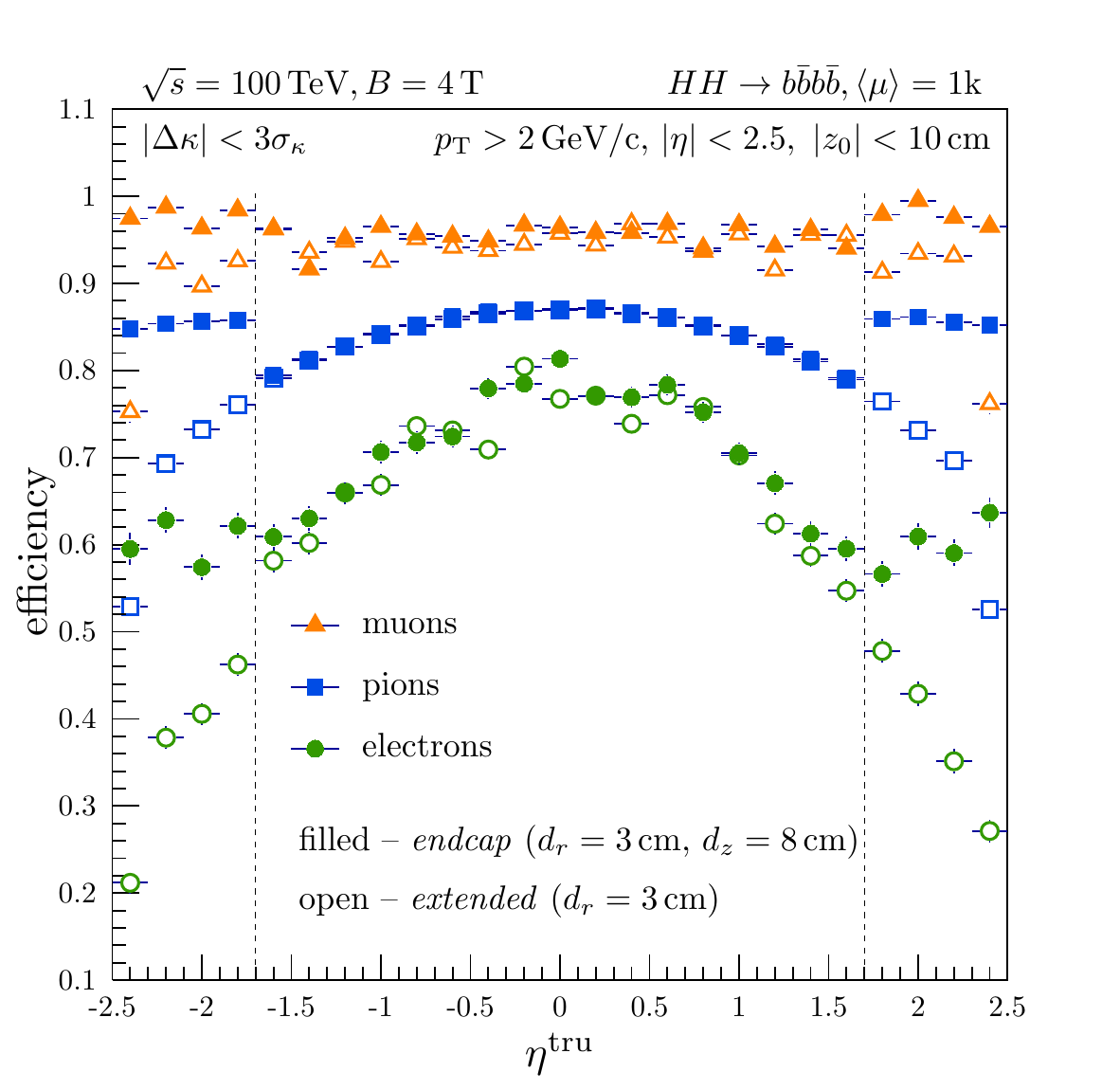}}
\caption{Tracking efficiency of muons in \emph{triangles}, pions in \emph{squares} and electrons in \emph{circles} for $HH \rightarrow 4b, ~\langle\mu\rangle = 1000$ events as a function of the true $\pT$\,(a) and $\eta$\ (b) for the \emph{endcap} and the \emph{extended designs} indicated with filled and open markers respectively.
Dashed vertical lines in (b) indicate the transition between the barrel layers and the endcap discs in the \emph{endcap design}.
The vertical error bars display the statistical uncertainties.
}
\label{fig:EffVsEtapT}
\end{figure}

In \autoref{fig:EffVsEtapT}\,(a) and (b), the TTT track reconstruction efficiency is shown for the \emph{endcap} and \emph{extended designs} as a function of the true transverse momentum ($\pTtru$) and pseudorapidity ($\eta^\mathrm{tru}$), separately for muons, pions and electrons from the sample. 

\label{sec:trk_eff}

Muons show a very high tracking efficiency of $\geq \SI{97}{\percent}$ over a wide $\pT$ range. 
For $\pTtru > \SI{10}{\GeV\per c}$, the muon tracking efficiency even rises to $\sim \SI{100}{\percent}$.
A significant efficiency loss is only observed for low-momentum muons, where ionisation loss and MS in the material in front of the TTT are significant.
For the same reason, the efficiency is lower in the large $|\eta|$-regions, in particular for the \emph{extended design}.

Charged pions are reconstructed with an efficiency of around $85 -\SI{90}{\percent}$. Inelastic nuclear interactions are the major source of inefficiency affecting all transverse momenta. Again, this efficiency loss increases at large $|\eta|$-values and is particularly prominent for the \emph{extended design}. 
In contrast, the endcaps in the \emph{endcap design} show the same pion reconstruction efficiency as in the central region.

Electrons have the lowest tracking efficiency ranging from $50 - \SI{90}{\percent}$ depending on the $\pT$.
For electrons, the efficiency loss is mainly due to bremsstrahlung before reaching the TTT layers.
The efficiency loss is large, in particular,  for low-momentum electrons, which can lose so much energy that they either do not reach the TTT layers or, if detected, the track reconstructed parameters do not fulfil the track acceptance cuts ($\pT > \SI{2}{\GeV \per c}$, vertex region) or the curvature consistency cut.
Two conclusions can be drawn from the track efficiency study. 
Firstly, the main source of track reconstruction inefficiency is the material in front of the TTT. 
Secondly, the tracking efficiency is higher for the \emph{endcap design} than for the \emph{extended design}
for all particle types.
Again, this is due to the material in front of the TTT.

\subsection{Track purity} \label{sec:track_purity}

\begin{figure}[!htb]
\centering
\subfloat[\emph{endcap}]{\includegraphics[width=0.85\linewidth,trim={0.2cm 0.2cm 1.1cm 1.1cm}, clip]{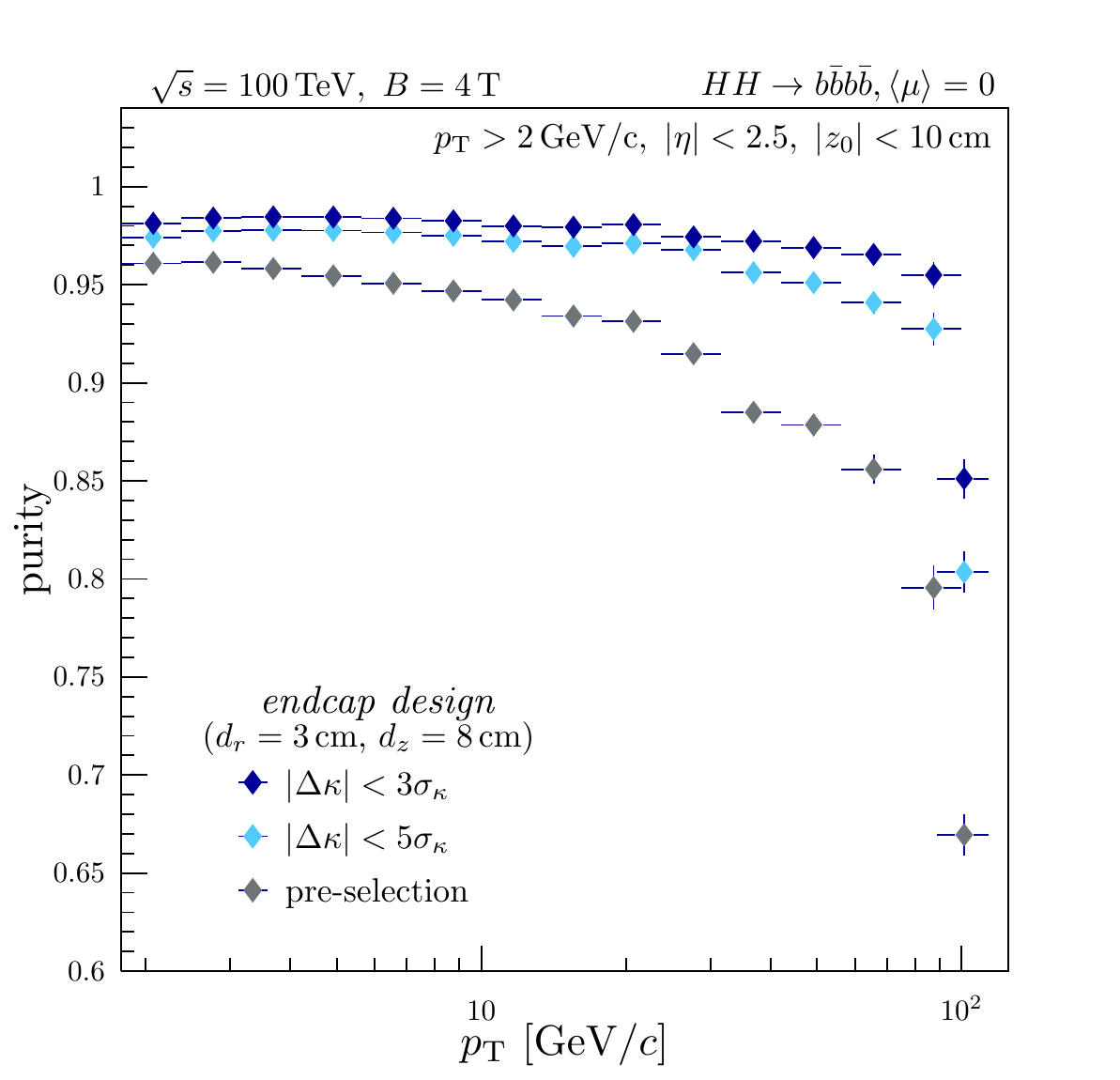}}\\
\subfloat[\emph{extended}]{\includegraphics[width=0.85\linewidth,trim={0.2cm 0.2cm 1.1cm 1.1cm}, clip]{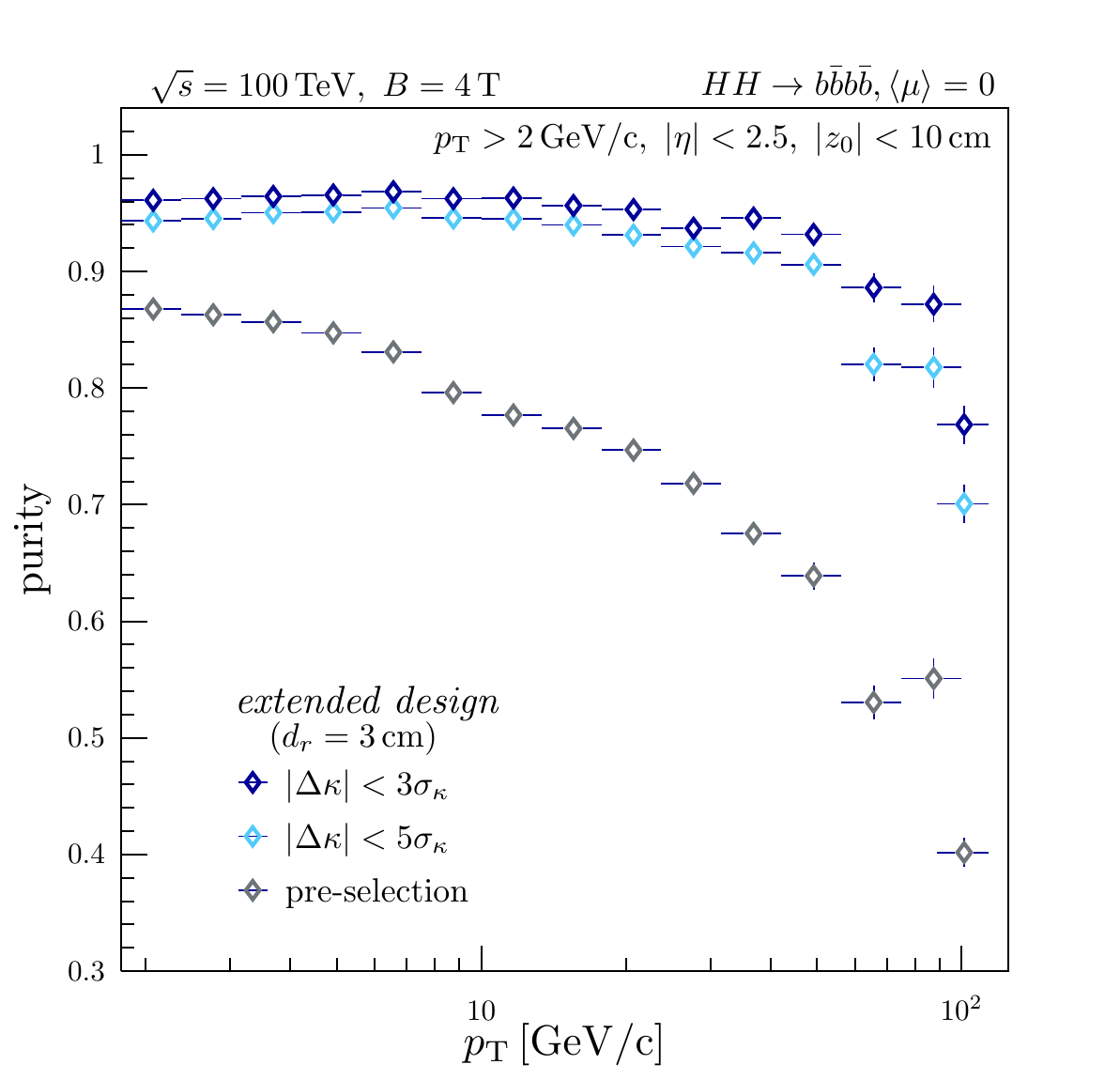}}
\caption{Track purity as a function of reconstructed $\pT$ for three different cuts, namely, a $3\sigma_{\kappa}$ cut in \emph{dark blue}, a $5\sigma_{\kappa}$ cut in \emph{light blue}, and the pre-selection cut in \emph{grey}, for the \emph{endcap} (a) and \emph{extended} (b) designs using $HH \rightarrow 4b$, $~\langle \mu \rangle = 0$ events. The tracks are reconstructed for $\pT> \SI{2}{\GeV \per c}$, $|\eta|<2.5$, and $z_{0}<\SI{10}{\cm}$.
The vertical error bars display the statistical uncertainties.
}
\label{fig:PurVspT_ECExt}
\end{figure}

The impact of the curvature consistency cut on the track purity is shown in \autoref{fig:PurVspT_ECExt} for both the \emph{endcap} and \emph{extended design}. 
The track purity is shown for $HH \rightarrow 4b$ signal events ($\langle \mu \rangle=0$) as a function of the reconstructed transverse momentum of the track 
 for three cases: no curvature consistency cut,  a loose $5\sigma_\kappa$ cut and a tight $3\sigma_\kappa$ cut.
By applying the curvature consistency cut
(i.e.\ beamline constraint), the track purity significantly improves.
The track purity is higher for the \emph{endcap design} and well above $\SI{95}{\percent}$ for the tight cut up to $\pT \lesssim \SI{100}{GeV/c}$. 
For the loose curvature consistency cut, the track purities are slightly worse.
The \emph{extended design} shows a significantly worse track purity, falling below \SI{90}{\percent} for $\pT > \SI{60}{\GeV/c}$ even for the tight curvature consistency cut.
The main reason is also here the increased material in front of the extended barrel, 
in which many additional secondary and tertiary particles are produced, 
which in turn produces many additional hits in the TTT layers. 

Since the main purpose of the TTT is the discrimination of high-momentum tracks in signal events from low-momentum tracks in minimum-bias events (pileup), the tight $3\sigma_\kappa$ curvature consistency cut is used
for all subsequent studies.

\begin{figure}[!htb]
\centering
\subfloat[purity vs $\pT$]{\includegraphics[width=0.85\linewidth,trim={0.2cm 0.2cm 1.1cm 1.1cm}, clip]{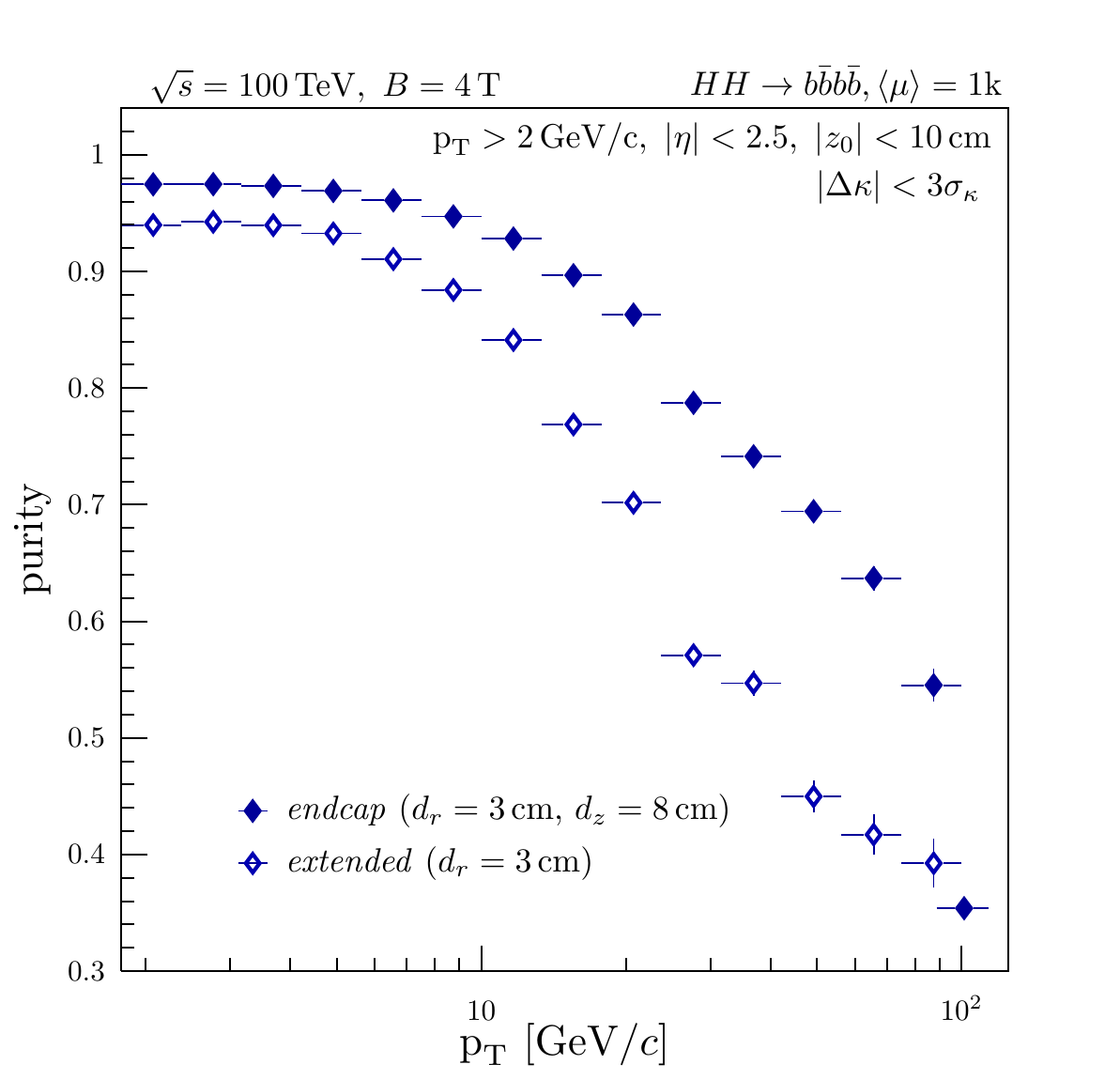}}\\
\subfloat[purity vs $\eta$]{\includegraphics[width=0.85\linewidth,trim={0.2cm 0.2cm 1.1cm 1.1cm}, clip]{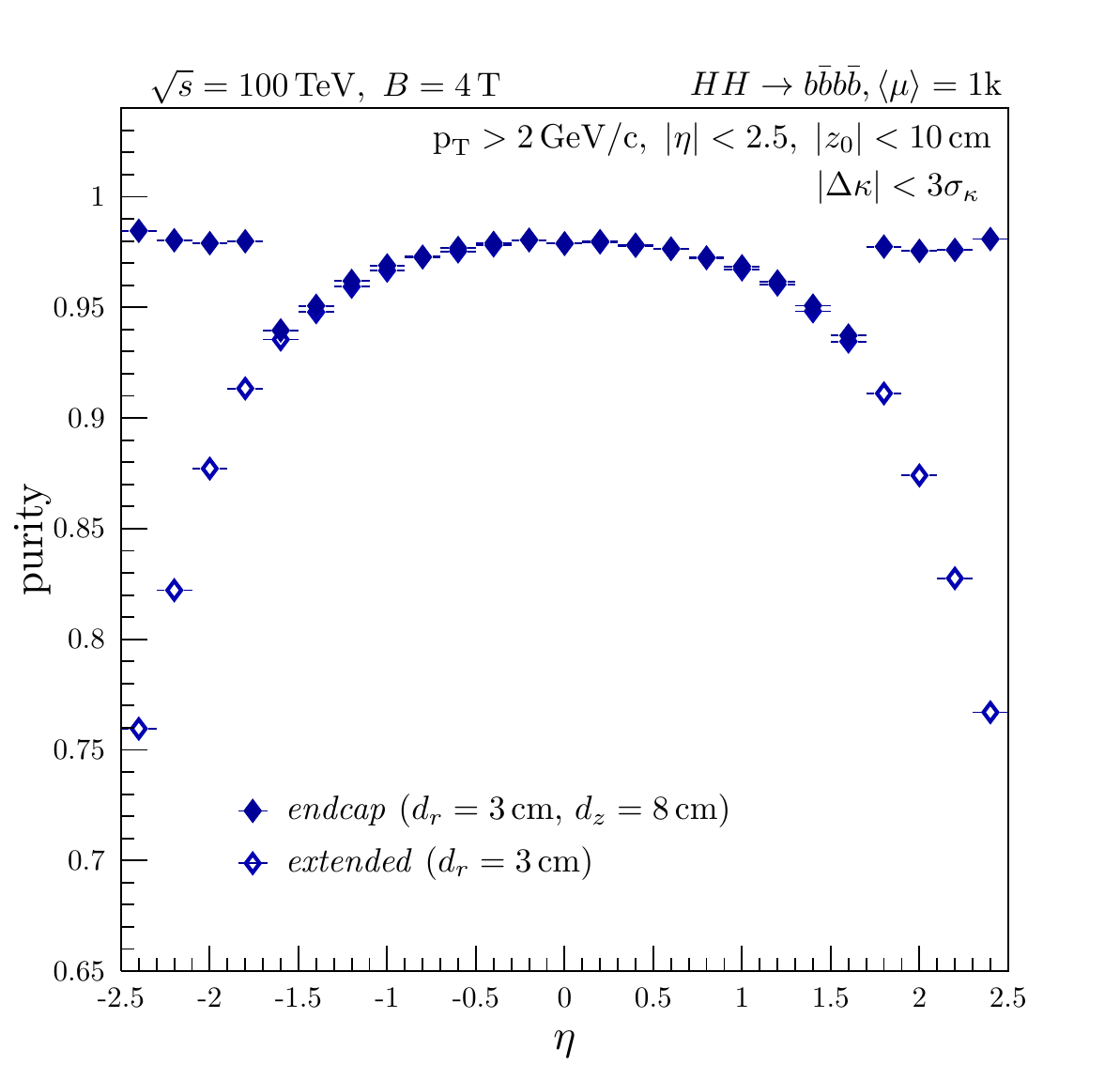}}
\caption{Track purity as a function of the reconstructed $\pT$\,(a) and $\eta$\,(b) for the \emph{endcap} (\emph{filled markers}) and the \emph{extended} (\emph{open markers}) designs considering all reconstructed tracks (including minimum-bias events) in $HH \rightarrow 4b, ~\langle \mu \rangle = 1000$.
The vertical error bars display the statistical uncertainties. 
}
\label{fig:PurVsEtapT}
\end{figure}

The impact of the 1000 pileup events on the track purity in $HH \rightarrow 4b$ events is shown in \autoref{fig:PurVsEtapT}
as a function of the reconstructed $\pT$ and $\eta$.
For the \emph{endcap design}, a track purity of $>\SI{95}{\percent}$ is obtained for $\pT < \SI{10}{\GeV\per c}$;
for higher $\pT$-values the track purity decreases, reaching about \SI{50}{\percent} at $\pT \approx \SI{100}{\GeV\per c}$. 
This decrease is caused by the huge amount
of low-momentum particles that interact with the detector material and create secondary particles, which in turn leave hits in the TTT tracking layers and lead to fake tracks.   
Since the rate of true high momentum tracks is very low, the track purity is significantly compromised by fake tracks in this region of phase space. 
Integrated over all transverse momenta, a track purity of $>\SI{92}{\percent}$ is reached in the entire pseudorapidity range, see \autoref{fig:PurVsEtapT}~(b).

For the \emph{extended design}, the track purity is significantly lower for large $|\eta|$-values. 
This results from the loose selection cuts required to retain a high tracking efficiency in this region, which is dominated by MS and also sees many secondary particles from the material in front of the TTT tracking layers.

Note that no effort was made here to improve the track purity by resolving hit ambiguities in the track reconstruction.
Algorithms for resolving hit ambiguities are computationally intensive and
it is beyond the scope of this study to answer the question of whether hit ambiguities can be resolved at a first-level trigger with the required speed.
  
Nonetheless, it will be shown that the track purities obtained by the TTT without resolving hit ambiguities are sufficient for the intended application (\autoref{sec:trig_perf}).

\subsection{Track parameter resolution}
The track parameters $p_T$ and $z_0$ are not only very important for pileup suppression but also determine the selectivity of the TTT.
\label{sec:Reso}
\begin{figure}[!htb]
\centering
\subfloat[$\sigma_{\delta_{\pT}/\pTtru}$ vs $\pTtru$]{\includegraphics[height=0.8\linewidth,trim={0.2cm 0.2cm 0.4cm 1.1cm}, clip]{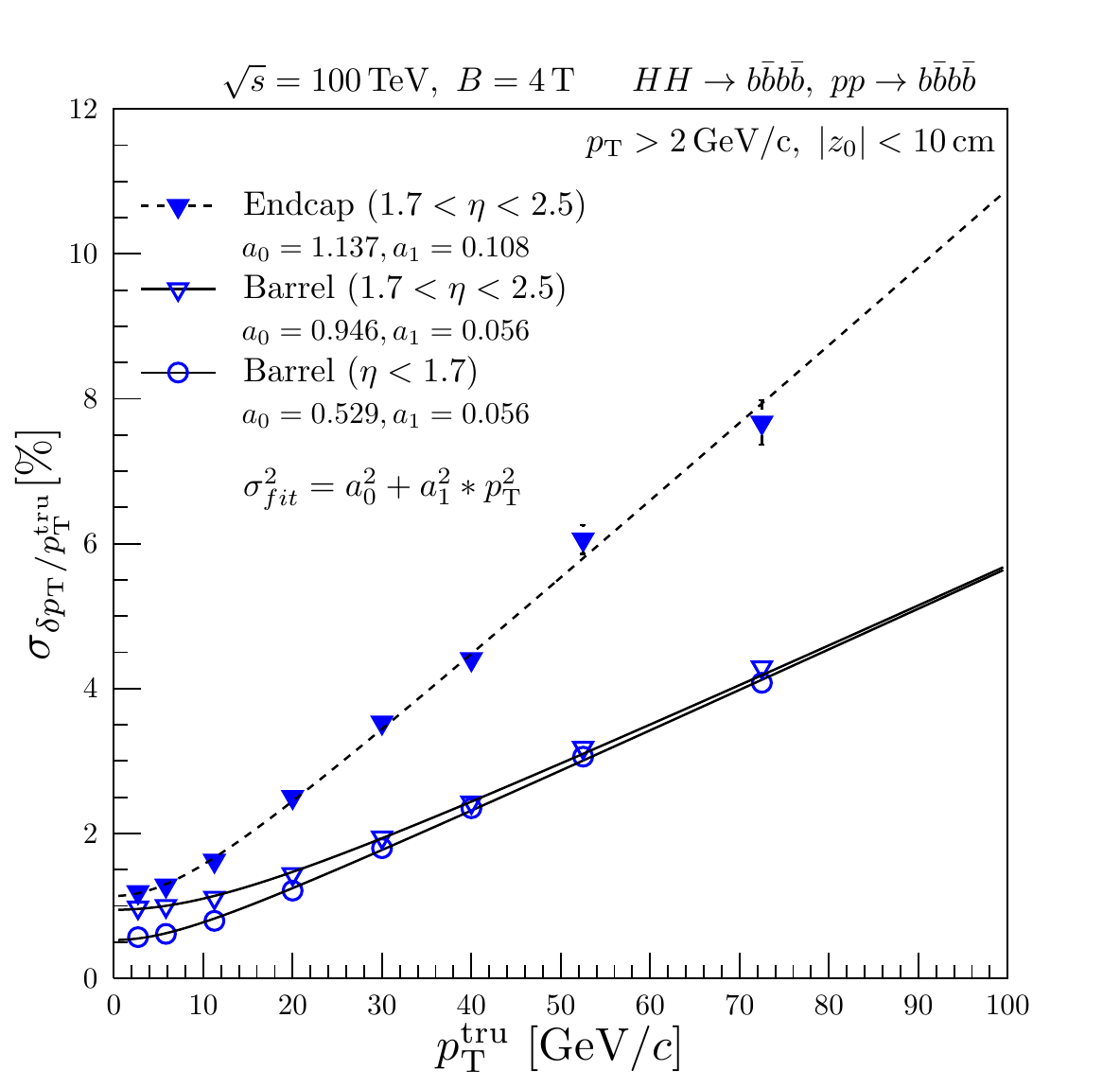}}\\
\subfloat[$\sigma_{\delta_{z0}}$ vs $\pTtru$]{\includegraphics[height=0.8\linewidth,trim={0.2cm 0.2cm 0.4cm 1.1cm}, clip]{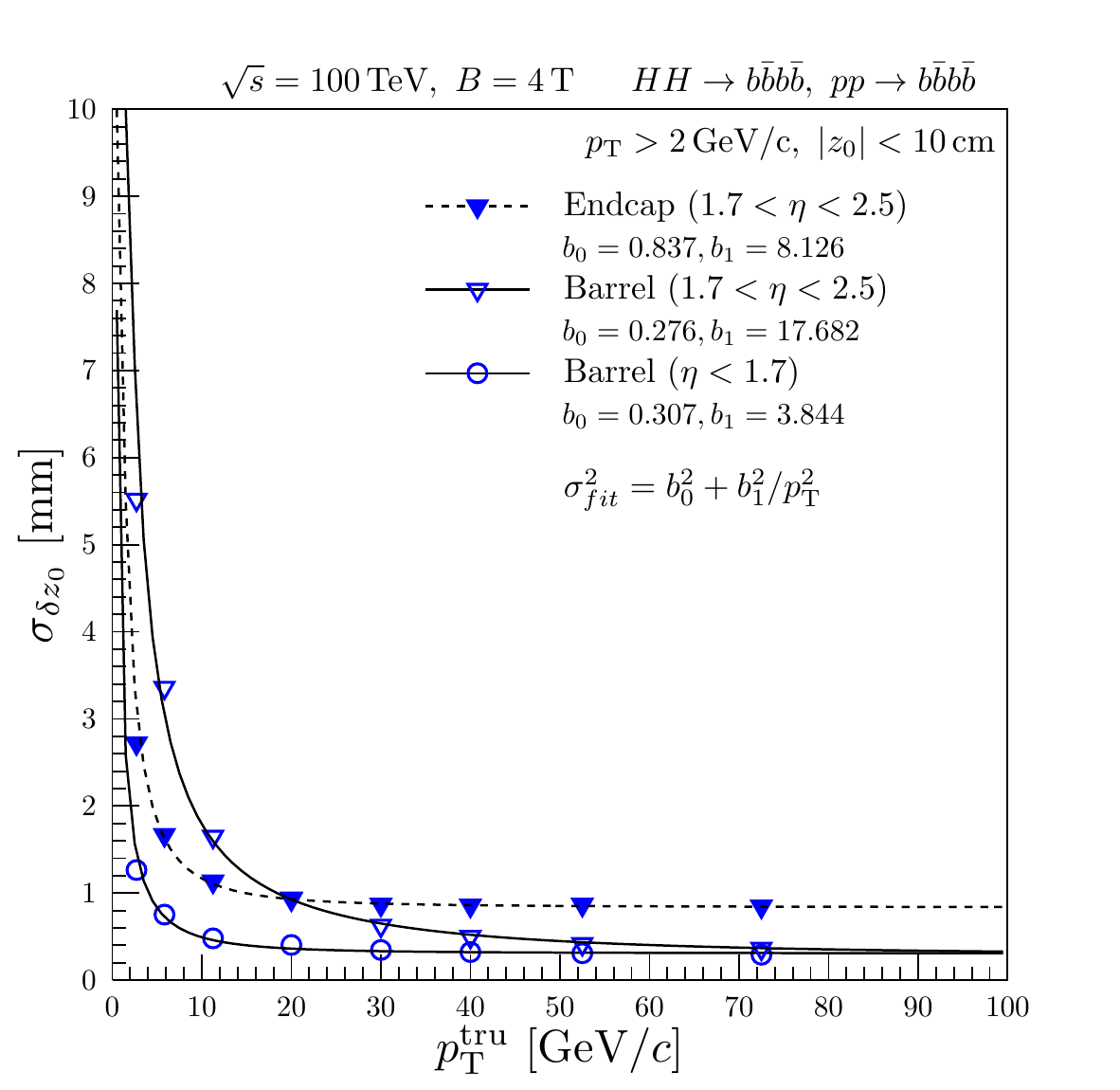}}
\caption{Relative $\pT$\,(a) and $z_0$\,(b) resolution as a function of the true $\pT$ using $HH \rightarrow 4b, \text{ and } pp \rightarrow 4b$ events for two separate pseudorapidity regions of the \emph{endcap} and \emph{extended designs}.
$\sigma_{\delta_{\pT}/\pTtru}$ and $\sigma_{\delta_{z_{0}}}$ are determined using a Gaussian fit for each of the $\pTtru$ bins.
\emph{Circles} are used for the central barrel region, $|\eta|<1.7$, whereas \emph{filled} and \emph{open triangles} are used for the endcap and barrel layers in $1.7<|\eta|<2.5$, respectively.
The lines show simple parameterisations of the $\pT$ dependence of relative momentum and $z_0$ resolution,  which are determined by fits to the data using \autoref{ref:para_pt_resolution} and \autoref{ref:para_z0_resolution}, respectively.
}
\label{fig:ResoVspT}
\end{figure}

\autoref{fig:ResoVspT} shows (a) the resolution of the relative transverse momentum,   $\sigma_{\delta_{\pT}/\pTtru}$, and (b) the $z_0$-resolution, $\sigma_{\delta_{z_{0}}}$, of the TTT tracks as a function of the true transverse momentum $\pTtru$.
They are shown for different regions of the detector, namely, the central barrel with $|\eta| < 1.7$, the forward barrel (\emph{extended design}) and endcap discs with $1.7 < |\eta| < 2.5$.

The best relative momentum resolution is obtained in the central barrel region, with $\sigma_{\delta_{\pT}/\pTtru} \approx \SI{0.5}{\percent}$ at low $\pTtru$, and then slowly degrading to  \SI{6}{\percent} for $\pTtru \approx \SI{100}{GeV/c}$.
Similar behaviour is observed for the extended barrel region ($1.7 < |\eta| < 2.5$).
However, at large $|\eta|$, the relative momentum resolution is about a factor $2$ worse for low-momentum tracks due to an increased MS contribution. 
At high transverse momenta, the relative momentum resolution is the same for all pseudorapidities.
For the \emph{endcap design}, the relative momentum resolution in the large $|\eta|$ region is significantly worse than for the extended barrel. 
The reason is the reduced lever arm to measure the track curvature in the bending plane. 
The lever arm, given by the TTT gap size projected to the bending plane, is constant for the extended barrel layers.
For the endcap discs, the lever arm depends on the pseudorapidity of the track and decreases at larger $|\eta|$.
The momentum resolution of endcap tracks is, therefore, significantly compromised at high pseudorapidity. 

The relative momentum resolution of all data points can be parameterised by the following functional form:
\begin{align}
\sigma_{{\delta_{\pT}}/\pTtru}^2 &= a_0^2 + a_1^2 \cdot (\pTtru)^2,
\label{ref:para_pt_resolution}
\end{align}
where the first term describes the contribution from MS  and the second term from the spatial hit uncertainty. 
The fit results are also shown in \autoref{fig:ResoVspT}.

The TTT achieves a $z_0$ resolution of $\lesssim \SI{1}{\mm}$  
 in the central barrel region for $\pTtru > \SI{4}{GeV/c}$ and
 in the high $|\eta|$ region for $\pTtru > \SI{20}{GeV/c}$. 
The $z_0$ resolution  can be parameterised by the following form:
\begin{align}
\sigma_{\delta_{z_{0}}}^2 &= b_0^2 + b_1^2/ (\pTtru)^2,   \label{ref:para_z0_resolution}
\end{align}
where, again, the first term describes the contribution from MS  and the second term from the spatial hit uncertainty. 
Contrary to the relative momentum resolution, the $z_0$ resolution is best for high momentum tracks where MS effects are negligible.
In the high $|\eta|$ region and for large $\pTtru$, the $z_0$ resolution is better reconstructed in the extended barrel than in the endcap disc. 
The reason is, again, the larger lever arm of the extended barrel, which helps in the track extrapolation to the beamline. For low momenta, however, the endcap disc is better in determining the $z_0$ parameter, mainly due to the smaller amount of material in front of the TTT.

The \emph{endcap} and \emph{extended designs} have both advantages and disadvantages.
On the one hand, high $\pT$ tracks, for which MS effects are insignificant, benefit from the larger lever arm provided by the extended barrel layers.
On the other hand, the \emph{endcap design} has a higher track purity compared to the \emph{extended design}; see \autoref{sec:track_purity}.

To summarise, 
an overall relative momentum resolution of better than \SI{6}{}--\SI{10}{\percent} and sub-mm $z_0$ resolution is obtained for a wide $\pT$ range.
As it will be shown in \autoref{sec:trig_perf}, this resolution is fully sufficient to trigger on physics processes of interest at the electroweak scale.


\section{Track Trigger  performance for  \texorpdfstring{$HH \rightarrow 4b$}{HH to 4b}}
\label{sec:trig_perf}

The trigger performance crucially depends on the capability to correctly identify physics objects (jets, isolated tracks/leptons, missing transverse energy, etc.), the ability to precisely reconstruct their energy (momentum) and to determine their origin (hard scattering process or minimum bias).
In low-pileup environments, a good trigger performance can be achieved with a calorimeter by triggering on one or several of the highest-energy objects, which usually come from the hard interaction. Here, a calo-trigger is particularly well suited for multi-jet final states.
However, in extremely high pileup environments such as the FCC-hh, this concept no longer works because energy deposits from several pileup collisions sum up and lead to the reconstruction of jet energies that are far larger than the actual jet energies of the hard interaction.
Therefore, trigger decisions must be preceded by an efficient pileup suppression.

Trackers have the great advantage of providing precise vertex information, making them essential for pileup suppression.
Track and vertex information, however, are usually not available at the first trigger level of hadron-collider experiments due to the high complexity of track finding and fitting.
Here, the TTT can play a key role by providing track parameters, particularly vertex information, at the first trigger level.
For a luminous region of $\pm \SI{10}{\cm}$, a vertex resolution of the order of \SI{1}{mm} is already sufficient to drastically reduce pileup.

In this section, we study the trigger performance of the TTT using the process $HH \rightarrow 4b$ as a showcase.
The reference selection, which serves as a basis for the evaluation, and the used MC samples are described in \autoref{sec:reference}.
The first step of the trigger algorithm is the identification of the hard interaction PV, which is described in \autoref{sec:PrimaryVertex}. 
In the second step, all tracks originating from the hard interaction PV are used to form the final trigger decision (\autoref{sec:TrigEffRate}).
The TTT performance is compared to a calo-trigger in \autoref{sec:TrigCalo},
followed by a discussion of a \SI{4}{MHz} trigger scenario (\autoref{sec:4MHz}).
Finally, results from a signal significance study are presented that underline the benefits of the TTT (\autoref{sec:signalsignificance}).

\subsection{Reference Selection} \label{sec:reference}
For all the following studies, we define a reference sample consisting of $HH \rightarrow 4b$ events as signal and $pp\rightarrow 4b$ events as background.\footnote{Only the dominant background process is considered. The reference samples are generated without a pileup of minimum bias events and without any trigger selection.}
The corresponding cross sections assuming SM couplings are $\sigma(pp \rightarrow HH \rightarrow 4b)=\SI{0.411}{pb}$ (NNLO, gluon gluon fusion only) \cite{ggFKfactor} and $\sigma(pp \rightarrow 4b)= \SI{2.328e4}{pb}$ (NLO) \cite{Bishara:2016kjn}, respectively.
The reference sample is required to satisfy the (offline) physics analysis cuts described in chapter~3 of \cite{Tkar_thesis}.
The main reference selection criteria are:
\begin{itemize}
\item at least four jets with a minimum transverse energy of ($55$, $40$, $35$, $20$) \si{\GeV} in the pseudorapidity range  $|\eta| \leq 2.5$,
\item at least four b-tagged jets with a minimum transverse energy of \SI{20}{\GeV} in the pseudorapidity range  $|\eta| \leq 2.5$,
\item two Higgs candidates, reconstructed from any $HH\rightarrow (b\overline{b}) (b\overline{b})$ jet pairing, fulfilling the invariant mass cut $|M_{H_{1(2)}^\textrm{cand.}} - 125| < \SI{30}{\GeV} $. 
\end{itemize}
About \SI{9}{\percent} of the generated $HH \rightarrow 4b$ sample fulfil these selection criteria for a b-tagging efficiency per jet of \SI{80}{\percent}. 
After applying the above selection cuts, the signal-to-background ratio is 1:2376.
Assuming SM couplings,
the associated projected signal significance, defined as $\text{S}/\sqrt{\text{B}}$ with S (B) being the number of signal (background) events,
is derived from simple event counting to be $21.3$  for an integrated luminosity of \SI{30}{\atto\barn^{-1}} and full detector readout~\cite{Tkar_thesis}.

\subsection{Identification of the Primary Hard Interaction Vertex}
\label{sec:PrimaryVertex}
\begin{figure}[!htb]
\centering
\includegraphics[width=0.8\columnwidth]{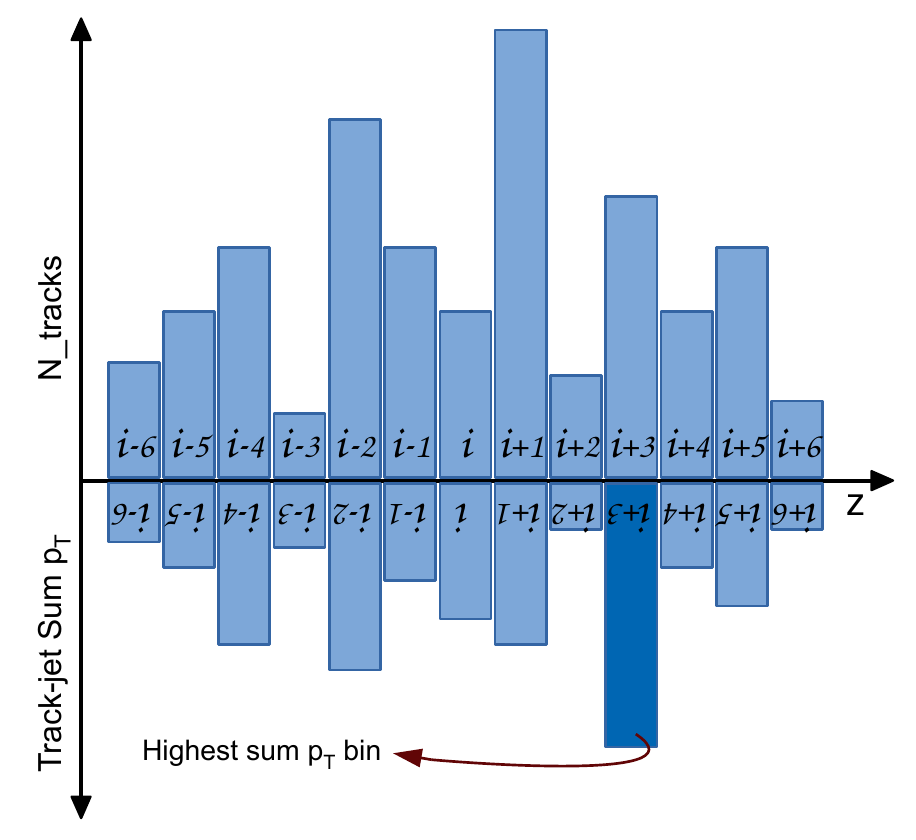}
\caption{Illustration of the identification of the hard interaction PV. 
For reasons of clarity, the bins are not shown overlapping; see text for more details.}
\label{fig:PB}
\end{figure}

For identifying the hard interaction PV, the luminous region of $-\SI{10}{cm} \leq z_0 \leq \SI{10}{cm}$ is subdivided into equidistant overlapping $z$-bins, where the overlap fraction is $2/3$ of the bin size ($1/3$ left and $1/3$ right). 
The overlap regions are implemented to mitigate migration effects at bin boundaries. 
TTT tracks are then filled and grouped according to the $z_0$ track parameter.\footnote{
Due to the bin overlap, a TTT track is filled into three adjacent bins, except for the first and last bins of the histogram.}
This grouping is illustrated by the histogram in the top part of \autoref{fig:PB}.
TTT-tracks belonging to the same group are then
clustered into track-jets (\emph{TTT-jets}) using the anti-$k_{t}$ jet  algorithm\,\cite{FASTJET} with a radius parameter $\Delta R_\textrm{jet} = 0.4$  and minimum jet transverse momentum of $p_\mathrm{jet}^\mathrm{min} = \SI{5}{\GeV \per c}$.
To suppress the impact of high $\pT$ fake tracks, which are dominant for $\pT > \SI{100}{\GeV \per c}$, see \autoref{fig:pT_dist},
a ceiling cut of $\pT = \SI{100}{\GeV \per c}$ is applied before feeding tracks into the jet  algorithm.

\begin{figure*}[!htb]
\centering
\subfloat[$HH \rightarrow 4b$]{\includegraphics[width=0.42\linewidth,trim={0.0cm 0.0cm 0.0cm 0.0cm}, clip]{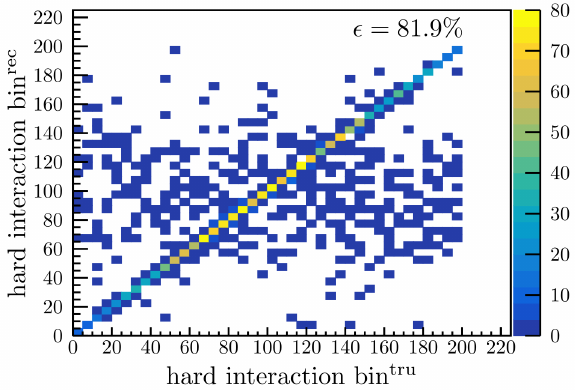}}
\quad \quad \quad 
\subfloat[$pp \rightarrow 4b$]{\includegraphics[width=0.42\linewidth,trim={0.0cm 0.0cm 0.0cm 0.0cm}, clip]{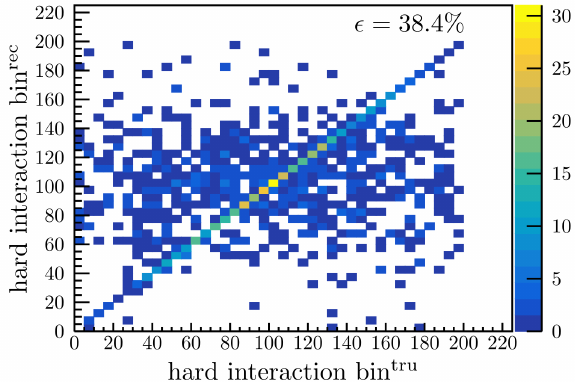}}
\caption{Correlation between true and reconstructed hard-interaction PV bin using the \emph{endcap design}
for (a) $HH \rightarrow 4b$ and (b) $pp \rightarrow 4b$ samples with an average pileup of 1000 and $|\eta|<2.5$.
The bin size of the vertex histogram is \SI{3}{mm}.
}
\label{fig:PBEff}
\end{figure*}

Finally, a histogram is filled with the $z_0$ position of all TTT jets, weighted with the transverse momentum of the TTT jet; see the lower part of \autoref{fig:PB}. 
The bin with the highest weight (summed $\pT$) defines the hard interaction PV.\footnote{In \cite{Tkar_thesis}, an alternative implementation for vertex identification was also studied.}

Using the \emph{endcap design} geometry and choosing a bin size of  \SI{3}{\milli \meter} for the vertex histogram,\footnote{
Note that for a vertex histogram bin width of \SI{3}{mm} every bin sees a pileup subtraction of about \SI{98.5}{\percent}, by definition.}
 the correlation between the reconstructed and true hard interaction bin 
is shown in \autoref{fig:PBEff}
for~(a) the $HH\rightarrow 4b$ signal and~(b) the $pp \rightarrow 4b$ background reference samples, which have both been simulated and reconstructed including minimum bias events with $\langle\mu\rangle=1000$.
A clear correlation is seen, in particular for the signal sample.
The hard interaction PV is successfully reconstructed if
the distance to the true bin position is $\le 1$.
The reconstruction efficiency of the hard interaction PV is about \SI{82}{\percent} for the signal sample, in comparison to $\sim \SI{38}{\percent}$ for the QCD background.
This difference is explained by the, on average, much larger track $\pT$  in the signal sample, which allows for better discrimination against pileup.

In the following, only TTT-jets reconstructed in the hard interaction PV bin are considered for the trigger decision.

\subsection{Trigger Rate and Trigger Efficiency}
\label{sec:TrigEffRate}
\begin{figure}[!htb]
\centering
\includegraphics[width=0.9\linewidth,trim={0.3cm 0.2cm 0.10cm 1.1cm}, clip]{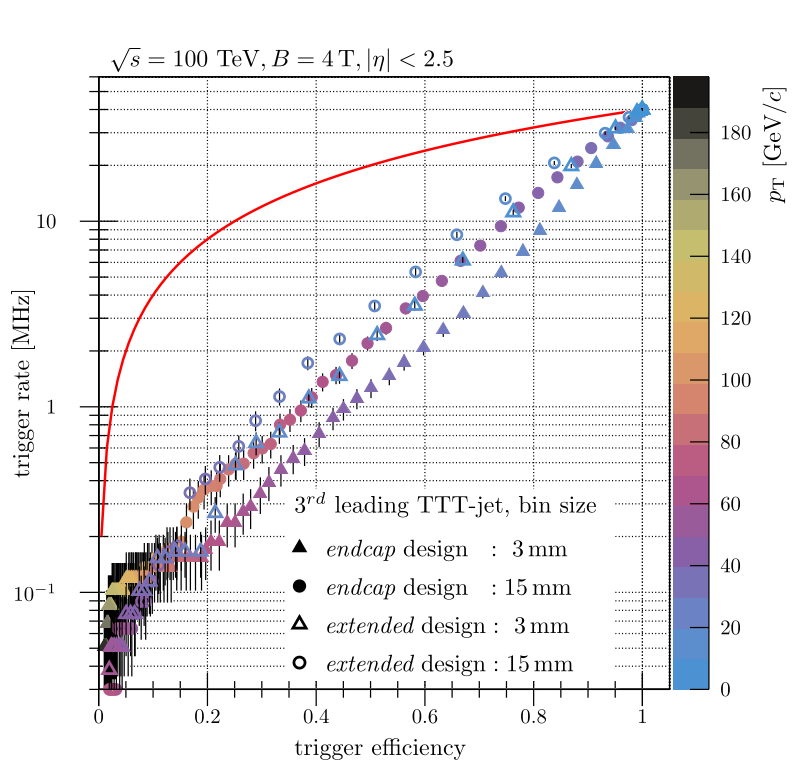}
\caption{Correlation of the trigger rate and trigger efficiency for $HH \rightarrow 4b$ events as function of
the transverse  momentum threshold on the
$3^\mathrm{rd}$ leading TTT-jet.
Results are shown for:
\emph{extended design} (open symbols), \emph{endcap design} (full symbols);
vertex histogram bin size \SI{3}{\mm} (triangles), \SI{15}{\mm} (circles).
The colour bar indicates the value of the trigger threshold. 
The average pileup is $(\langle \mu \rangle = 1000)$. 
The solid red line indicates the correlation for random pre-scaling of the trigger. 
}
\label{fig:TTT_TrigEff}
\end{figure}
The trigger performance is described by two parameters: the trigger efficiency, which should be as high as possible, and the trigger rate, which should be as low as possible.
The trigger efficiency is here determined from the $HH \rightarrow 4b$ signal reference sample, including minimum bias events with $\langle\mu\rangle = 1000$; the trigger rate is determined from a large minimum bias sample, also with $\langle\mu\rangle = 1000$.
To evaluate the trigger performance, the correlation between the trigger efficiency and the trigger rate is studied.

The  $HH \rightarrow 4b$  events have a multi-jet final state.
The discrimination power of the five highest-$\pT$ TTT-jets has been studied in \cite{Tkar_thesis}.
The highest discrimination power
 was found for the transverse momentum of the $3^\mathrm{rd}$ leading TTT-jet, $\pT^\text{jet-3}$, which is chosen as basis for the trigger decision.
The trigger efficiency is then given by the fraction of $HH \rightarrow 4b$ events passing the $\pT^\text{jet-3}$-cut while the trigger rate is given by the fraction of accepted pileup events, multiplied with the bunch crossing frequency, which is assumed to be \SI{40}{\MHz}.

\autoref{fig:TTT_TrigEff} shows for $HH \rightarrow 4b$ events the correlation between the trigger rate and the trigger efficiency as a function of the trigger threshold, which is varied between $0-$\SI{200}{\GeV\per c}.
The trigger performance is shown for the \emph{extended} and  \emph{endcap design} and for vertex histogram bin sizes of  $\SI{3}{\mm}$ and $\SI{15}{\mm}$.
In addition, a line of constant $\text{S}/\text{B}$ (selectivity) is shown. This curve describes the effect of a randomly pre-scaled trigger~\cite{Schultz-Coulon:1999vpi}.

The results show an approximately exponential increase in the trigger rate as a function of the trigger efficiency. 
The highest selectivity is obtained for the \emph{endcap design} and a vertex histogram bin size of \SI{3}{mm}. 
For a large region of trigger thresholds, the \SI{15}{mm} bin size implementation has a factor $2$ higher trigger rate than the \SI{3}{mm} implementation.
This comes from the fact that the \SI{15}{mm} implementation has  $5$ times more pileup per bin, by construction.
This result underlines the importance of pileup suppression for the trigger.

A similar performance difference is seen between the \emph{extended} and the \emph{endcap designs}; the latter provides an almost factor of $2$ better reduction of the trigger rate. 
The reason is the significantly better $z_0$ resolution of low-momentum tracks for $|\eta|>1.7$ in the \emph{endcap design}, which increases both the track purity and the efficiency of identifying the hard interaction PV.

For the \emph{endcap design} and a vertex histogram bin size of \SI{3}{mm}, the highest $S/B$ is reached at a trigger threshold of  
$\pT^\text{jet-3}>\SI{60}{GeV/c}$
corresponding to  
about \SI{20}{\percent} trigger efficiency and a trigger rate of $\approx 1/200$  times the bunch crossing rate. At this threshold, the $S/B$ is $40$ times higher compared to random pre-scaling of the trigger rate (\emph{red} curve in \autoref{fig:TTT_TrigEff}).

\subsection{Comparison with Calo-Trigger}
\label{sec:TrigCalo}
\begin{figure}[!htb]
\centering
\includegraphics[width=0.9\linewidth,trim={0.3cm 0.2cm 0.2cm 1.1cm}, clip]{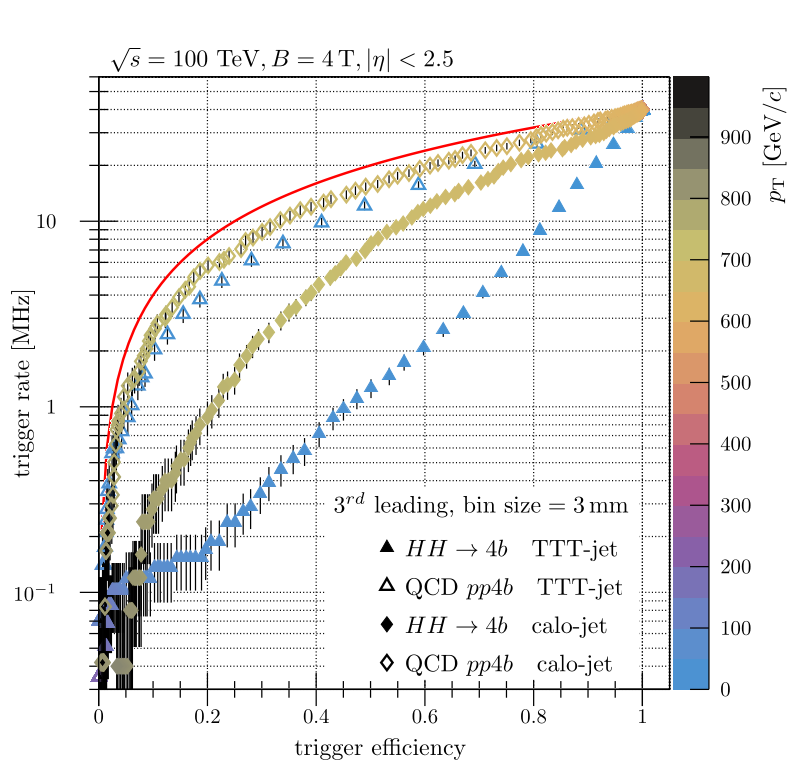}
\caption{Correlation of the trigger rate and trigger efficiency for $HH \rightarrow 4b$ events (full symbols) and $\mathrm{pp} \rightarrow 4b$ events (open symbols) as function of
the transverse  momentum threshold on the
$3^\mathrm{rd}$ leading TTT-jet.
The results are shown for TTT-jets using the \emph{extended design} (triangles) and calo-jets (diamonds). See also description of \autoref{fig:TTT_TrigEff}.}
\label{fig:TrigEff}
\end{figure}

To assess the performance of the TTT and its ability to suppress pileup, the TTT trigger performance is compared with a calorimeter-based jet trigger (\emph{calo-jets}). 
The calorimeter is emulated based on specifications from the FCC-hh reference detector~\cite{CDR_FCC}, and details of the calo-trigger emulation are described in \autoref{App:CaloEmu}.

The main differences between TTT-jets and calo-jets are:
\begin{itemize}
\item TTT-jets are based on reconstructed charged particles only, whereas calo-jets are based on energies measured in the calorimeter cells from all particles reaching the calorimeter.
\item TTT-jets are pileup suppressed using the vertex histogram technique, whereas calo-jets include all signal and pileup particles. 
\end{itemize}

For the comparison, calo-jets are reconstructed using the same jet  algorithm as for TTT-jets and sorted according to transverse energy.
\autoref{fig:TrigEff} shows the trigger rate as a function of the trigger efficiency for the $HH \rightarrow 4b$ and $pp\rightarrow4b$ processes using the $3^\mathrm{rd}$ leading jet for both the TTT and the calo-trigger.

For the calo-trigger, the highest selectivity is $S/B \approx 12$, which is reached at a threshold of 
$\pT^\text{calo-3}>\SI{750}{\GeV \per c}$,
corresponding to  
about \SI{15}{\percent} trigger efficiency and a trigger rate of about \SI{0.5}{\mega\hertz}.
This value is to be compared to the highest selectivity of $S/B \approx 40$ for the TTT.
Both the TTT  and the calo-trigger show higher selectivity for the $HH \rightarrow 4b$ process than for the $pp \rightarrow 4b$ process.
However, the TTT has the superior ability to distinguish the two processes.

Furthermore, the TTT has a significantly lower $\pT$-threshold ($\lesssim \SI{50}{\GeV/c}$) compared to the calo-trigger ($\sim \SI{750}{\GeV/c}$),
which mainly triggers events where the Higgs bosons are boosted.
Since pileup contributes about $\SI{90}{\percent}$ to the calo-jet energies, the calo-trigger is extremely sensitive to any fluctuations in the pileup rate. 
Small fluctuations in the pileup rate (e.g.\ from bunch-to-bunch variations) would cause significant changes in the calo-trigger rate and readout bandwidth, whereas the TTT shows high robustness.

\subsection{\SI{4}{MHz} Trigger Scenario}
\label{sec:4MHz}
Assuming a trigger rate of \SI{4}{MHz}
the trigger performances of the TTT and the calo-trigger are compared.
\autoref{table:TrigPerfTTT} summarises the results obtained by triggering on the $2^{\mathrm{nd}}$, $3^{\mathrm{rd}}$, and $4^{\mathrm{th}}$ leading jet.
For each setup, the corresponding trigger efficiency and trigger threshold are given.

The highest trigger efficiency (\SI{69}{\percent}) is obtained for the $3^{\mathrm{rd}}$ leading TTT-jet. 
But also, the triggers based on the $2^{\mathrm{nd}}$ or $4^{\mathrm{th}}$ leading TTT jet have a very similar performance. 
The trigger efficiency of the calo-trigger reaches at most \SI{44}{\percent} using the $2^{\mathrm{nd}}$ calo-jet.  
The trigger efficiencies using the $3^{\mathrm{rd}}$ or $4^{\mathrm{th}}$ calo-jet are smaller.  
For all studied setups, the calo-trigger thresholds have to be set above \SI{700}{GeV}, much higher than the energy scale of the objects to be triggered.
In case of fluctuating pileup rates, the thresholds would need to be further increased. 
The calo-trigger efficiencies given in \autoref{table:TrigPerfTTT} are therefore optimistic.

\begin{table}[!htb]
\centering
\caption{TTT and calo-trigger efficiencies for the $HH \rightarrow 4b$ reference sample with $\langle \mu \rangle = 1000$.
The efficiencies are calculated for a trigger rate of \SI{4}{\MHz} for different triggering jets ($2^\mathrm{nd}$, $3^\mathrm{rd}$ and $4^\mathrm{th}$ leading jets).
For each setup, the corresponding value of the trigger threshold is quoted.
}
\begin{tabular}{@{}llrrr@{}}
  \multicolumn{5}{@{}l}{Trigger performance at \SI{4}{\MHz}}  \\
\toprule
  & leading trigger jet & $2^\mathrm{nd}$ & $3^\mathrm{rd}$ & $4^\mathrm{th}$ \\ \midrule
\multirow{2}{*}{\begin{tabular}[c]{@{}l@{}}calo-trigger\end{tabular}}                    & efficiency [\si{\percent}] &44  & 37 & 33  \\
                         & threshold [\si{\GeV/c}] & 779   & 752   & 735   \\ \midrule
\multirow{2}{*}{\begin{tabular}[c]{@{}l@{}}TTT\\ (\emph{endcap} design)\end{tabular}} & efficiency [\si{\percent}] & 67 & 69  & 63 \\
                         & threshold  [\si{\GeV/c}]&  54  &  33  &  25  \\ 
\bottomrule
\end{tabular}
\label{table:TrigPerfTTT}
\end{table}

\subsection{Signal Significance Study}
\label{sec:signalsignificance}

The signal significance for an integrated luminosity of \SI{30}{a\barn^{-1}} is calculated for the events triggered by the TTT and calo-trigger assuming a trigger rate of \SI{4}{\MHz}, and for the full detector readout reference sample defined in \autoref{sec:reference}.

Since the measurement of the $HH \rightarrow 4b$ process is not background-free, the statistical significance is given by the ratio $\text{S}/\sqrt{\text{B}}$, where $\text{S}$ is the number of signal events and $\text{B}$ is the number of $pp \rightarrow 4b$ background events.
For a full assessment of the signal significance, statistical and systematic uncertainties must be considered. 
A study of systematic uncertainties is beyond the scope of this study.
Therefore, estimates of the signal significance will be based on statistical uncertainties only, following the approach in~\cite{Tkar_thesis}.

\begin{figure}[!htb]
\centering
\includegraphics[width=0.99\linewidth,trim={0.3cm 0.2cm 0.2cm 1.1cm},clip]{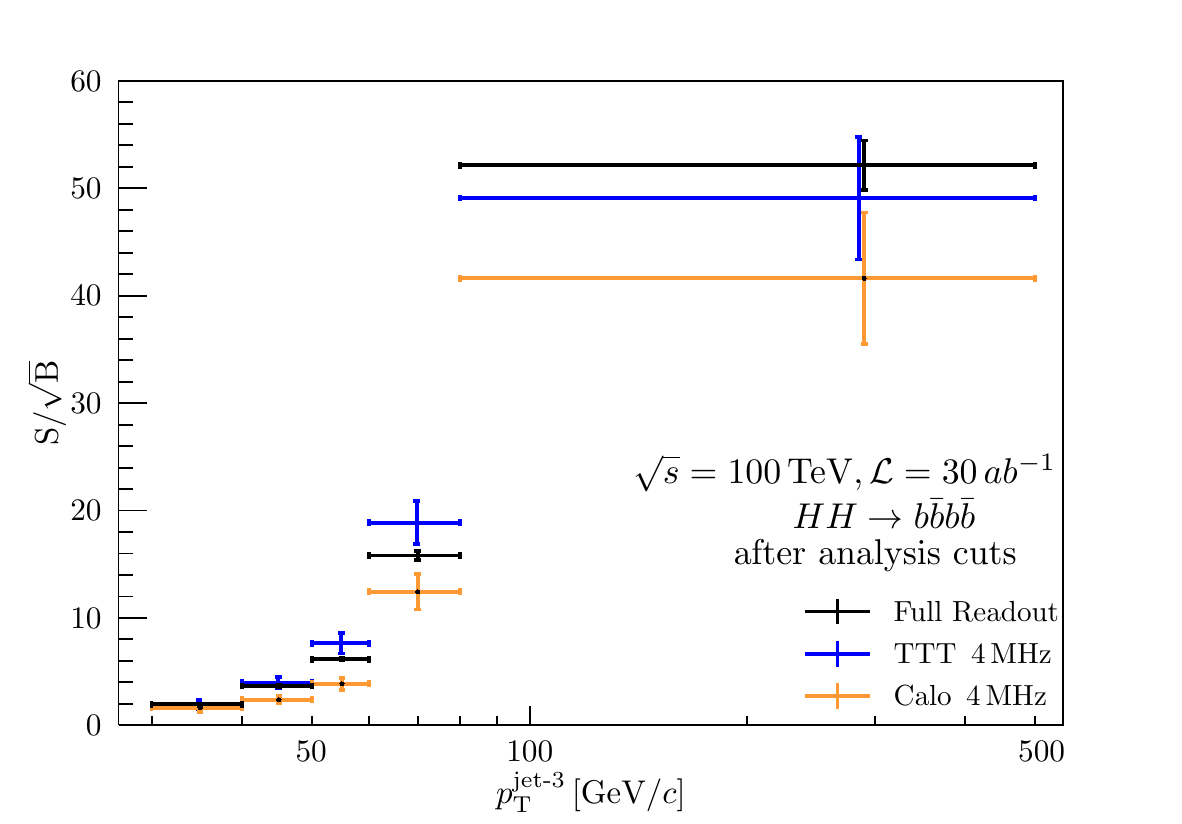}
\caption{Differential significance of the $HH \rightarrow 4b$ channel as function of the $3^\mathrm{rd}$ leading jet $\pT$.
  The sensitivities are shown for full detector readout (see \autoref{sec:reference}) assuming $100\%$ trigger efficiency (black),
  and for TTT (blue) and the calo (orange) triggered events for a trigger rate of \SI{4}{\MHz}.
The vertical error bands indicate the statistical errors.
}
\label{fig:Zi}
\end{figure}

The ratio $\text{S}/\sqrt{\text{B}}$ is largely phase space dependent. 
In such a case, the optimal signal significance (OSS) can be derived from a differential significance distribution, according to: 
\begin{align}
\text{OSS} = \sqrt{\sum_i{\frac{\text{S}_i^2}{\text{B}_i}}},  \label{eq:sig_opt}
\end{align}
with $\text{S}_i$ and $\text{B}_i$ being the number of signal and background events in small phase space elements, respectively.

The differential significance as a function of the transverse momentum of the third jet, $\frac{\Delta \, (\text{S}_i/\sqrt{\text{B}_i})}{\Delta \pT^\text{jet-3} }$, is shown in \autoref{fig:Zi} for the TTT and calo-trigger assuming a trigger rate of \SI{4}{\MHz}, and for full detector readout.
A large increase of the signal significance is observed for $\pT^\text{jet-3} \gtrsim \SI{60}{\GeV\per c}$ for all three scenarios.
For $\pT^\text{jet-3} \ge \SI{80}{\GeV\per c}$, a signal significance of $\text{OSS} \approx 50$ is reached.  
Over the full $\pT$ range, the TTT provides a significantly higher signal significance than the calo-trigger.
Note that despite the trigger efficiency of \SI{69}{\percent} at \SI{4}{\MHz} trigger rate, the TTT reaches almost the same significance as the full detector readout scenario.
The reason lies in the discrimination power of the TTT and the low trigger efficiency for the $pp \rightarrow 4b$ background, which is only about \SI{20}{\percent}, see \autoref{fig:TrigEff}. 

\begin{table}[!htb]
\centering
\caption{Projection  of the $HH\rightarrow4b$ significance for an integrated luminosity of $\mathcal{L} = \SI{30}{a\barn^{-1}}, ~ \sqrt{s}=\SI{100}{\TeV}$ and $\langle \mu \rangle = 1000$.
The  $pp \rightarrow 4b$ process is considered as background.
The projections are presented for 
full readout (i.e.\ \SI{100}{\percent} trigger rate) and triggered readout at \SI{4}{\MHz} by triggering on the $3^\mathrm{rd}$ leading jet using the TTT and using the calo-trigger. 
For the TTT, the endcap design with vertex histogram bin size of $\SI{3}{\mm}$ is used.  
}
\begin{tabular}{@{}lccc@{}}
\toprule
\multirow{2}{*}{\begin{tabular}[c]{@{}c@{}}Trigger type\\ ($3^\textrm{rd} $ jet)\end{tabular}} &
  \multicolumn{2}{c}{Trigger Efficiency [\si{\percent}]} &
  \multirow{2}{*}{\begin{tabular}[c]{@{}c@{}}Significance\\ $\sqrt{\sum_i{\frac{\text{S}_i^2}{\text{B}_i}}}$\end{tabular}} \\ \cmidrule(lr){2-3}
             & \multicolumn{1}{l}{$HH \rightarrow 4b$} & \multicolumn{1}{l}{$pp \rightarrow 4b$} &       \\ \midrule
Calo-trig. (\SI{4}{\MHz}) & 37                                     & 17                                    & $43.7 \pm 6.4$ \\
TTT (\SI{4}{\MHz})          & 69                                     & 20                                     & $53.3 \pm 6.1$ \\ \midrule
 Full readout      & 100                                      & 100                                      & $55.0 \pm 2.3$ \\ \bottomrule
\end{tabular}
\label{tab:rough_est}
\end{table}

The optimal signal significances are summarised in \autoref{tab:rough_est} for the $HH\rightarrow4b$ channel.
For full detector readout, a signal sensitivity of $55$ is expected. 
This value defines the maximum achievable signal sensitivity
and corresponds to an uncertainty of the $HH\rightarrow4b$ cross-section measurement of  $\approx \SI{2}{\percent}$.
Assuming a trigger rate of \SI{4}{\MHz}, 
almost the same signal significance, $\text{OSS} \approx 53$, is projected for the TTT, whereas for the
calo-trigger a signal significance of only  $\text{OSS} \approx 44$ is expected.
Even for a factor 10 bandwidth reduction, the TTT recovers more than $\SI{96}{\percent}$ of the statistical significance, while the calo-trigger incurs a loss of about \SI{20}{\percent} compared to the statistical significance for full detector readout.
Note that the signal significance will be lower if systematic uncertainties are included.
This applies in particular to the calo-trigger due to the mentioned instability of the calo-jet trigger thresholds.

\section{Summary} 
\label{sec:summary}

A new track trigger concept, the TTT, has been proposed for the FCC-hh to trigger interesting physics at the electroweak (EW) scale.
The TTT concept can be realised by modifying the all-silicon FCC-hh reference tracker to include three dedicated pixel tracking layers.
For $pp$ collisions at a centre of mass energy of \SI{100}{\TeV} and a luminosity corresponding to an average of 1000 minimum bias pileup events, the tracking and trigger performances of the TTT have been studied using a detailed Geant4 simulation.
The following main results have been obtained:
\begin{enumerate}[wide, labelwidth=0pt, labelindent=0pt]
    \item A full reconstruction of all tracks with a minimum transverse momentum of \SI{2}{\GeV\per c} is possible with only three highly granular tracking layers.
    The stacking of the tracking layers, together with their placement at a large distance from the interaction point, results in high track reconstruction efficiencies and purities -- even without resolving hit ambiguities between track candidates.
    The simplicity of the TTT algorithm allows for real-time reconstruction of tracks, and a possible hardware implementation has been discussed. 

  \item Thanks to the good track parameter resolution, the TTT concept allows for efficient identification of the primary interaction vertex and provides excellent suppression of minimum bias events in real-time.
    This makes the TTT concept ideal for triggering physics processes at the EW scale, in particular, signatures that do not feature prominent high $\pT$ objects that otherwise could be triggered by calorimeter or muon triggers.
    A striking example is di-Higgs production in the dominant decay channel $HH\rightarrow4b$, which
    is one of the four golden channels that will directly probe the trilinear Higgs self-coupling ($\lambda$) and hence the Higgs Potential at the FCC.
    The TTT reconstructs the primary interaction vertex in about \SI{80}{\percent} of the events correctly, while it suppresses tracks from minimum bias pileup events by more than \SI{98}{\percent}.
    
  \item 
    For a concrete trigger scenario, where \SI{10}{\percent} of the $pp$ collisions are read out,
    interesting $HH\rightarrow4b$ signal events can be triggered with an efficiency of $\sim \SI{69}{\percent}$  by using TTT track-jets.
    For the same trigger scenario, an emulated calorimeter jet trigger optimistically achieves a trigger efficiency of only $\sim \SI{36}{\percent}$.
    The TTT achieves significantly lower trigger thresholds than the calo-trigger.
    The ability of the TTT to efficiently trigger low-$\pT$ jets is of particular interest for di-Higgs production since the sensitivity to $\lambda$ is particularly large for low invariant masses of the Higgs pair\,\cite{ggFKfactor}.
    Thus, the TTT will provide enhanced sensitivity to the measurement of $\lambda$ compared to calo-triggers. 
    
\end{enumerate}

\subsection{Further Opportunities}
\label{sec:discussion}
In the presented study, the potential of the TTT has been investigated using the $HH\rightarrow 4b$ channel as a showcase.
However, other processes like $HH\rightarrow 2b 2\gamma$ and $HH\rightarrow 2b 2\tau$ could also significantly profit from the  TTT. 

Tracks reconstructed at the first trigger level might also be useful input for other or combined trigger systems.
By combining TTT and calo-trigger information, a particle flow algorithm, see for instance \cite{ATLAS:2017ghe} and references therein, could be implemented already at an early trigger level.

Furthermore, the TTT can also provide high-purity track seeds for track reconstruction at higher trigger levels, where hits from all detector layers are combined to achieve the best possible track parameter resolution. This would then also allow the reconstruction of decay vertices of $b\,(c)$-hadrons and, therefore, $b\,(c)$-tagging of jets.

\section*{Acknowledgements}
The authors thank S.\ Dittmeier for helping with the firmware implementation of the \emph{Preselector} and for reading the manuscript.
The authors are grateful to N.\ Berger and A.\ Kozlinskiy for their help in developing the simulation and reconstruction software and 
D.\ E.\ Ferreira de Lima for helping with the production of the Monte Carlo samples.

T.~Kar was financially supported by the DFG-funded Research Training Unit GRK2058 (``HighRR").

\section*{Appendix}
\appendix
\section{TTT gap size optimization}  \label{sec:geo_opt}

In the TTT concept, the track parameter resolutions $\sigma_{\delta_{\pT}}$ and $\sigma_{\delta_{z_{0}}}$ scale inversely with the gap size  as $\sim 1/d_r$.
Therefore, they can be improved by increasing the gap between the TTT layers.
Larger gap sizes, however, lead to more hit ambiguities (see \autoref{fig:TTTConcept} in \autoref{sec:TTTC}), an increased fake rate, and can thus lead to a wrong reconstruction of the PV.
The goal of this study is to find the optimal gap size that represents a compromise between the track parameter resolution and the track purity.

For this study,  $HH \rightarrow 4b$ events at $\sqrt{s} = \SI{100}{\TeV}$ and $\langle \mu \rangle = 1000$ are simulated
 for five different values of the barrel radial gap size ($d_r = 20,~ 25,~ 30,~ 35,~ \SI{40}{\mm}$)  and for five different values of the endcap gap size ($d_z = 53,~ 67,~ 80,~ 93,~ \SI{106}{\mm}$).
 The track purity of the TTT is determined for tracks in the range $\pT \in [\num{10} - \num{100}]\,\si{\GeV \per c}$.

\subsection{Optimisation of the barrel gap}

\begin{figure*}[!htb]
   \centering
   \subfloat[$\sigma_{\delta_{\pT}/\pTtru}$ vs purity]{\includegraphics[width=0.48\linewidth,trim={0.2cm 0.2cm 0.4cm 1.75cm}, clip]{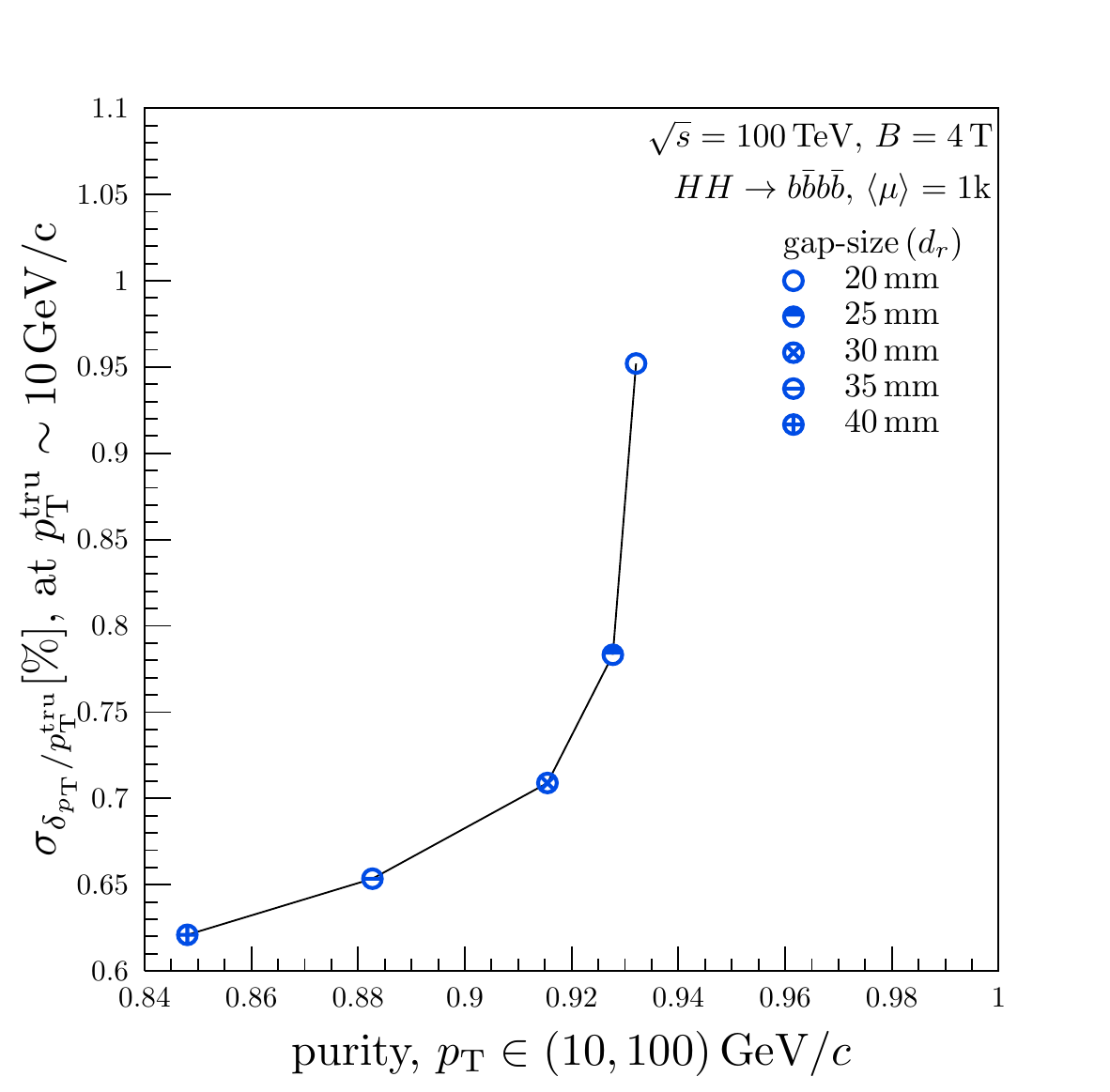}}   \subfloat[$\sigma_{\delta_{z_0}}$ vs purity]{\includegraphics[width=0.48\linewidth,trim={0.2cm 0.2cm 0.4cm 1.75cm}, clip]{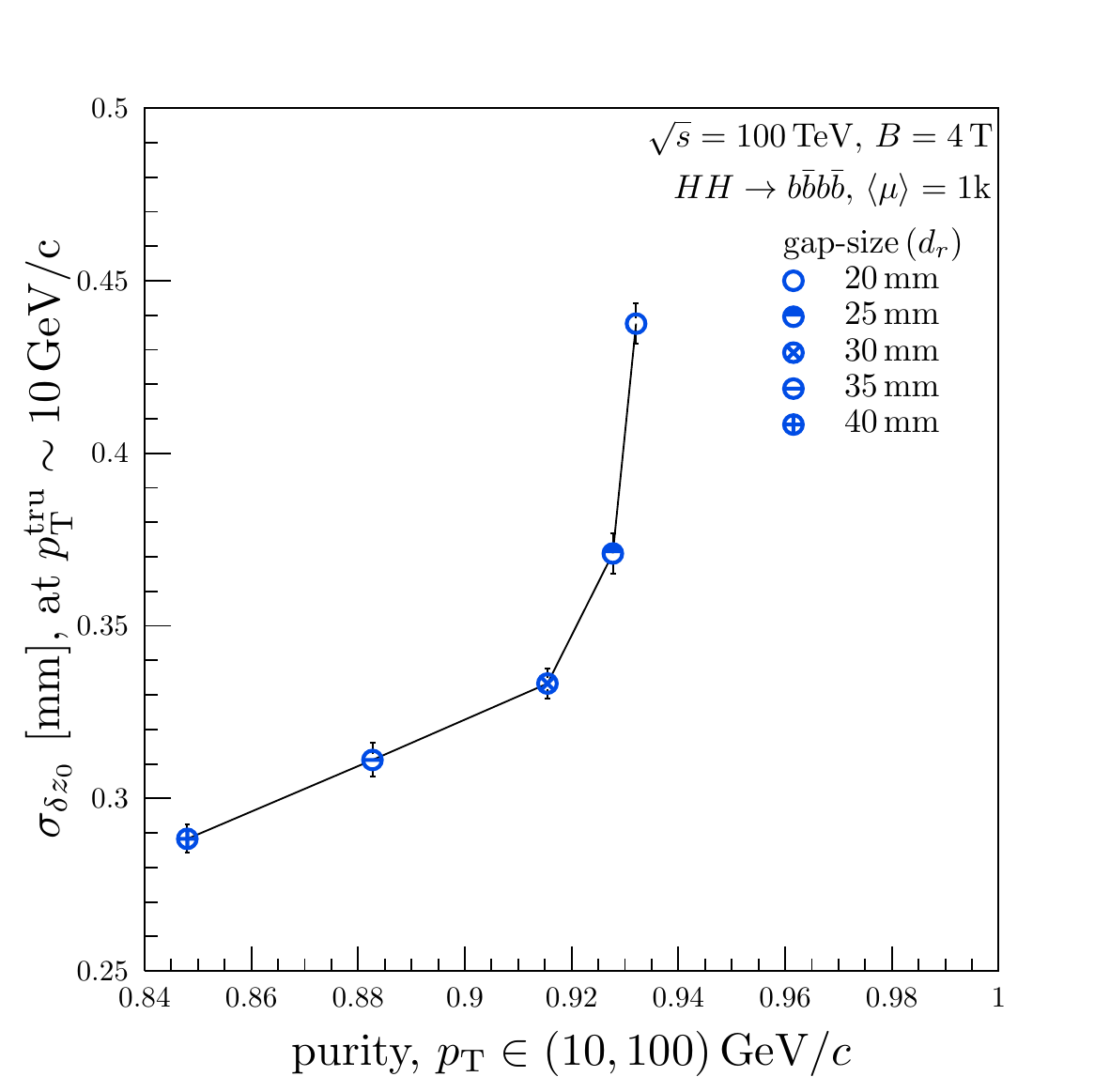}}
   \caption{Barrel gap size study: Relative momentum resolution\,(a) and $z_0$-resolution\,(b) of \SI{10}{\GeV \per c} pions as a function of track purity of $HH \rightarrow 4b$ events with a track $\pT$ in the range $10-\SI{100}{\GeV \per c}$.
   Results for five different radial gap sizes of the TTT barrel, i.e.\ $d_r = {20,25,30,35,40}\si{mm}$, using different markers,  for $\langle \mu \rangle = 1000$ are shown. Taken from\,\cite{Tkar_thesis}.}
   \label{fig:opt_dz_pt}
\end{figure*}

\autoref{fig:opt_dz_pt} shows the correlation between the relative momentum resolution and the track purity~(a) and between the $z_0$ resolution and the track purity~(b) for various gap sizes.
For both plots, only tracks in the central barrel region ($|\eta|<1.7$)
are considered.
The relative momentum and $z_0$ resolutions have been determined from single particle simulations of \SI{10}{\GeV\per c} pions.
This value is chosen since \SI{10}{\GeV\per c} particles have both a good momentum resolution as well as a good $z_0$ resolution (\autoref{fig:ResoVspT}).
Furthermore, tracks at $p_T \approx \SI{10}{\GeV\per c}$  are quite frequent (\autoref {fig:pT_dist}) in signal events so that they also have a significant weight in the PV reconstruction.

As expected, the largest gap size yields the best track parameter resolution.
However, the improvement of the track parameter resolutions by increasing $d_r$ from $ \SI{20}{\mm}$ to \SI{30}{\mm} is larger than by increasing $d_r$ from $\SI{30}{\mm}$ to \SI{40}{\mm}.
This is caused by MS in front of the TTT layers, which is limiting the achievable track parameter resolutions.

With larger gap sizes, $d_r$, the ambiguities and the hit combinatorics increase, leading to a reduction of the track purity.
This drop in the purity is very significant for $d_r > \SI{30}{\mm}$.
As a trade-off between a high track purity and good track parameter resolutions,
we, therefore, choose $d_r=\SI{30}{\mm}$ as an optimal value for the barrel gap size.

\subsection{Optimisation of the endcap gap}

\begin{figure*}[!htb]
   \centering
   \subfloat[$\sigma_{\delta_{\pT}/\pTtru}$ vs purity]{\includegraphics[width=0.48\linewidth,trim={0.2cm 0.2cm 0.4cm 1.75cm}, clip]{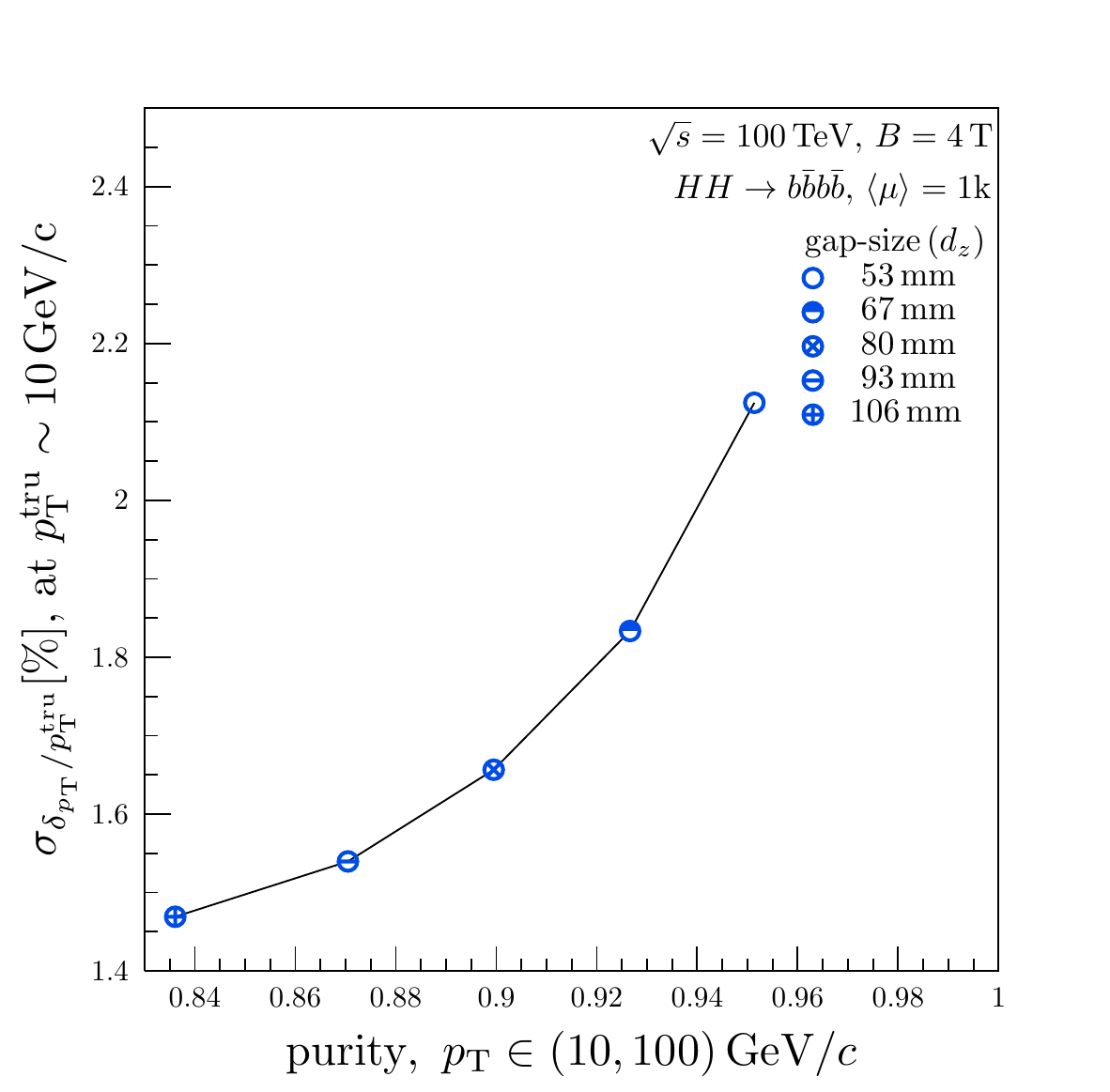}}
   \subfloat[$\sigma_{\delta_{z_0}}$ vs purity]{\includegraphics[width=0.48\linewidth,trim={0.2cm 0.2cm 0.4cm 1.75cm}, clip]{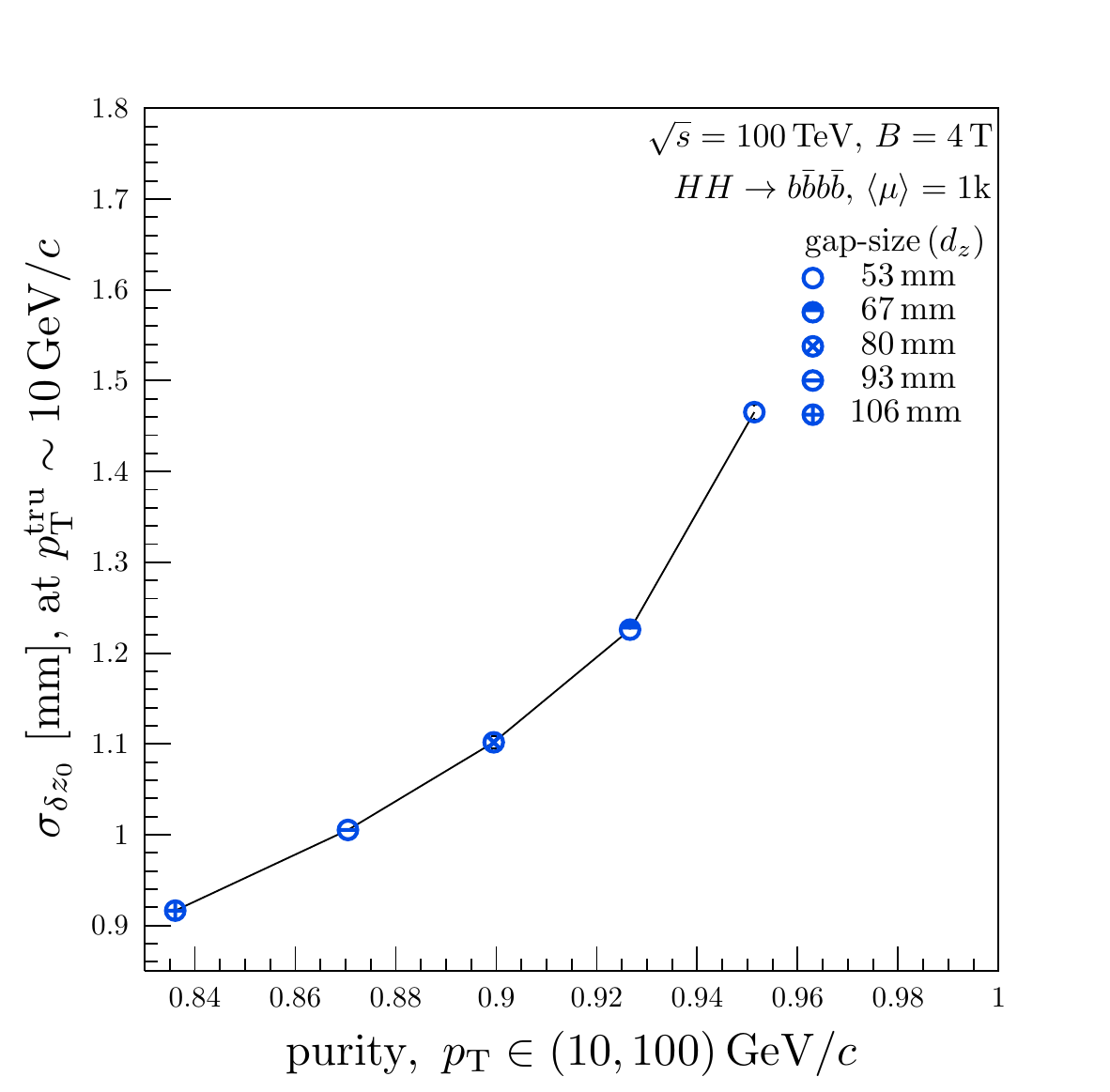}}
   \caption{Endcap gap size study: Relative momentum resolution\,(a) and $z_0$-resolution\,(b) of \SI{10}{\GeV \per c} pions as a function of track purity of $HH \rightarrow 4b$ events with a track $\pT$ in the range $10-\SI{100}{\GeV \per c}$.
   Results for five different longitudinal gap sizes of the TTT endcap disc, i.e.\ $d_z = {53,67,80,93,106}\si{mm}$, using different markers,  for $\langle \mu \rangle = 1000$ are shown.}
   \label{fig:optEC_dz_pt}
\end{figure*}

The same study is repeated for the endcap discs.
\autoref{fig:optEC_dz_pt} shows for various gap sizes
the correlation between the relative momentum resolution and the track purity~(a) and between the $z_0$ resolution and the track purity~(b).
For both figures, only tracks in the endcap region ($1.7<|\eta|<2.5$)
are considered.
The relative momentum and the $z_0$ resolution are again obtained from single particle simulations of \SI{10}{\GeV\per c} pions.

Similar to the study of the barrel gap size, the largest gap size yields the best track parameter resolution, whereas the smallest gap size provides the highest track purity.
We choose $d_z=\SI{80}{\mm}$ as an optimal value for the endcap gap size.

\section{HW Implementation of the TTT}
\label{sec:app_hw}
For the HW Implementation of the TTT, two main components are needed.
In the first stage, many local and highly parallel track finders implemented in ASICs perform the track reconstruction.
In the second stage, all tracks are sent to a \emph{Vertex \& Event Finder} for generating the trigger decision.
In the following, we sketch a possible implementation of the TTT barrel and assume that a similar scheme can also be used for processing hits from the endcaps.

The castellated design of the barrel layers enables independent reconstruction of tracks in all 58 modules, which can be implemented with a high degree of parallelisation.
To further increase parallelism, small regions for track finding are defined such that no more than 20 clusters per layer are input to the TTT track finder, which should be implemented as ASIC.
The local track finder ASIC implements both the pre-selection and final cuts.

Next, the track parameters of all reconstructed tracks are sent to the \emph{Vertex \& Event Finder} for reconstructing the hard interaction PV and making the TTT trigger decision.

\subsection{Local Track Finder ASIC} 

\begin{figure}[!htb]
  \centering
  \includegraphics[width=0.99\linewidth]{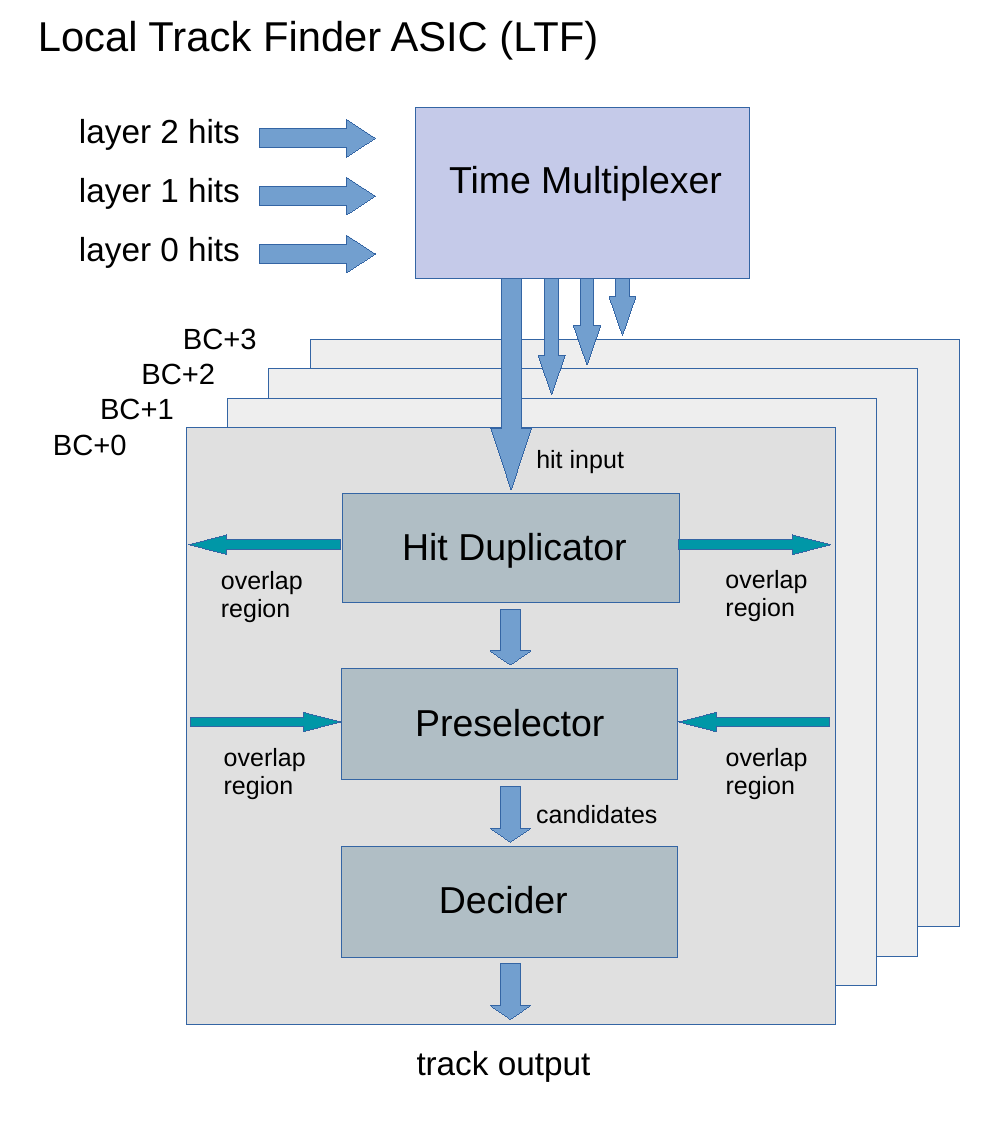}
  \caption{Block diagram of the \emph{Local Track Finder} ASIC. For description, see text.
   }
  \label{fig:LTF}
\end{figure}

It is estimated that about 14000 \emph{Local Track Finder} ASICs (LTFs) are required for the central barrel ($\pm \SI{2.4}{m}$), where each ASIC is assumed to receive hits from an instrumentation area of about $\SI{2}{\cm} \times \SI{10}{\cm} = \SI{20}{\cm^2}$  in $\phi \times z$ from all three TTT layers.
Another $2 \times 3000$ ASICs are estimated for the extended barrel or endcap region.
For the sake of simplicity, we assume that hits have already been merged into hit-clusters (e.g.\ on the sensor.
Therefore, \emph{hits} always refer to cluster positions in the following.

In the local regions, a hit position can be represented by 24 address bits, assuming a pixel size of $\SI{40}{\micro\meter} \times \SI{40}{\micro\meter}$.
The expected average input bandwidth to each ASIC is \SI{5}{Gbps} from each layer without timing information.
Assuming that hits are read out time-sorted from the sensors, the additional input bandwidth for timing information would be marginal.

A block diagram of the LTF is shown in \autoref{fig:LTF}. 
It shows all the functionalities required by the TTT algorithm discussed in \autoref{sec:TTTR}.
The LTF implements the following blocks.
\vspace{0.3cm}

The \emph{Time Multiplexer} distributes hits to several parallel processing engines, each processing hits from the same bunch crossing.
Even if we anticipate significant progress in electronics in the next decades, we believe that track reconstruction is such a complex task that it will take longer than the time between two bunch crossings (\SI{25}{ns}).
The ASIC should, therefore, implement several processing engines that perform the reconstruction time-multiplexed, see \autoref{fig:LTF}.
The number of required processing engines depends on the total latency of the track reconstruction.
We expect four time-multiplexed processing engines to suffice for the reasons discussed in the following.
\vspace{0.3cm}

\begin{figure}[!htb]
  \centering
  \includegraphics[width=0.7\linewidth]{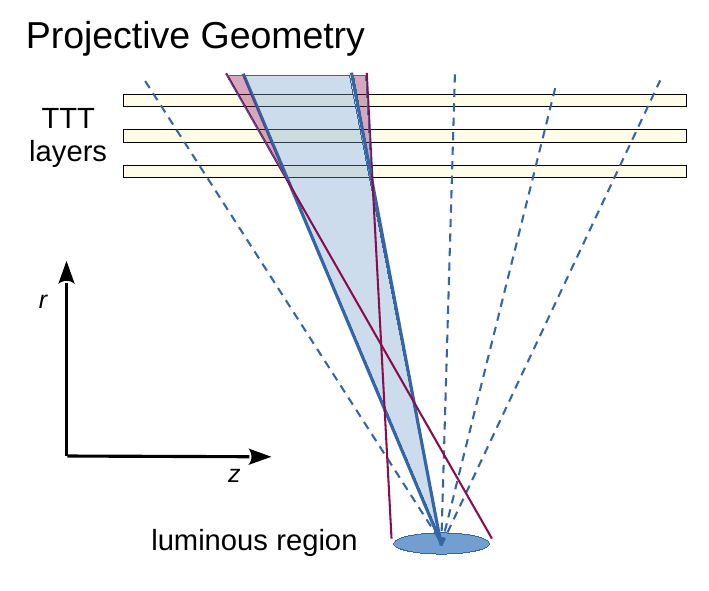}
  \caption{Sketch of a projective geometry. Sectors are defined by the dashed lines. The red areas show overlap regions caused by the finite size of the luminous region.
   }
  \label{fig:projective_geometry}
\end{figure}
The \emph{Hit Duplicator} sends hits from so-called overlap regions to up to two other ASICs to ensure full phase space coverage, see \autoref{fig:projective_geometry}.
The mapping of sensors to ASICs depends on the detector geometry and size of the sensors.
The simplest mapping is achieved if a projective detector geometry is chosen.
The size of the overlap regions is proportional to the length of the luminous region and also depends on the gap between the tracking layers; see \autoref{fig:projective_geometry}.
It is estimated that about \SI{10}{\percent} of the hits need to be sent to adjacent ASICs.
In the case of a non-projective geometry, the mapping of hits becomes more complex.\footnote{
Note that overlap regions need only to be considered in $z$-direction.
In the azimuthal direction, the castellated module design ensures that the full phase space is covered by construction.}

\vspace{0.3cm}

\begin{figure}[!htb]
  \centering
  {\includegraphics[width=0.99\linewidth]{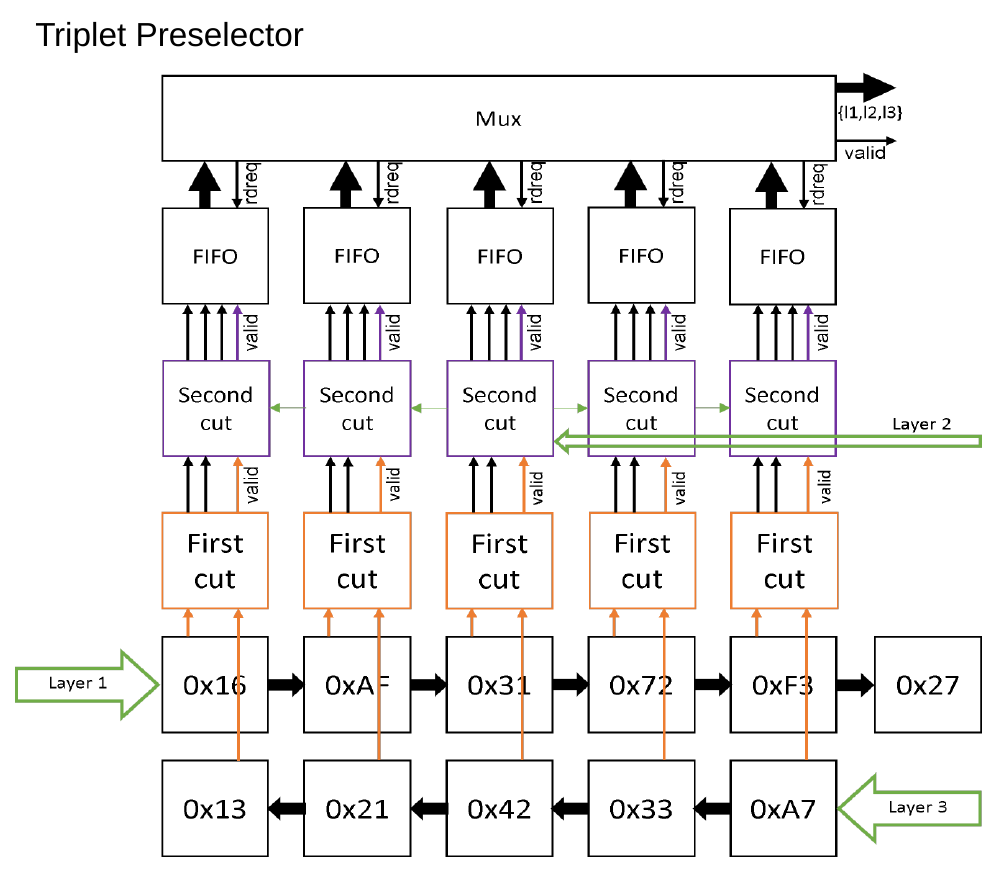} }
  \caption{Sketch of the \emph{Preselector} firmware implemented on an Arria 10 FPGA. 
    As input,  20 hits per layer are allowed at maximum. 
    For sake of visibility, the sketch shows the design for a maximum of 5 hits per layer. 
    See the text for a detailed description. 
    Taken from \cite{ref:Sheik_thesis}} 
  \label{fig:preselector}
\end{figure}

The \emph{Preselector} considers all possible hit triplet combinations and applies the pre-selection cuts.
The \emph{Preselector} is the most time-critical block since there are at maximum $20^3=8000$  possible hit combinations to be considered for a maximum input of 20 hits per layer per ASIC.
For this reason, the \emph{Preselector} algorithm was emulated in an FPGA for determining the resources and execution time \cite{ref:Sheik_thesis}.

From simulation studies, see~\autoref{sec:RecOpt},
it is known that the purity of the TTT  is rather high.
Therefore, the number of track candidates will not significantly exceed the number of input hits per layer, which is an important feature for the triplet design.
A sketch of the \emph{Preselector} implementation is shown in \autoref{fig:preselector}. 
As described in \autoref{sec:preselection_cuts}, at first, hits from the first layer are combined with hits from the third layer, corresponding to $20 \times 20 = 400$ hit comparisons in total.
The full combinatorics is implemented using two \emph{First Cut} shift registers, which are shifted in opposite directions.
Every clock cycle, up to 40 \emph{First Cut} entities apply the search window cuts defined in \autoref{eq:phi13}--\autoref{eq:z13}.
If the comparison is successful, the hits are written into \emph{Second Cut} shift registers.
There are, in total, 40 \emph{Second Cut} entities that perform the comparison between the middle layer hit position and interpolated position from the first and third layers.
The \emph{Second Cut} entities are basically a copy of the \emph{First Cut} entities and implement the cuts defined in \autoref{eq:dPhi2}--\autoref{eq:dz2}.
The validated output from the 40 \emph{Second Cut} entities are buffered in Fifos for further processing.

The \emph{Preselector} has been implemented on Arria 10 FPGA (type 10AX115N3F45l2SG).
The maximum clock frequency was found to be \SI{150}{MHz}.
The algorithm is fully pipelined, allows for up to 30 validated track candidates, and has a latency of about \SI{500}{ns}. 
\vspace{0.3cm}

The \emph{Decider} receives the TTT track candidates from the 40 Fifos on the \emph{Preselector}, calculates the track parameters, and applies the curvature consistency and acceptance cuts.
The track parameter calculations given by \autoref{eq:track_para_first}--\autoref{eq:track_para_last} are simple, can be linearised, and are efficiently implemented in Data Signal Processors (DSP).
The most complex functions are two square roots for the calculation of the hit distances between the three layers (\autoref{eq:kappa_123} and \autoref{eq:kappa_013}).
One more square root is required for calculating $\sin \theta$ in the Highland formula (\autoref{eq:HighlandsFormula}). 
No angular functions are required for the track parameter calculation if the track polar angle is parameterised as $\cot \theta$ instead of $\theta$. 
Therefore, we believe that the execution time of the \emph{Decider} is negligible compared to the \emph{Preselector}.
\hspace{0.3cm}

Considering that a state-of-the-art ASIC is about ten times faster than an FPGA, we estimate the LTF latency to be below 100ns.
That would require the implementation of four time-multiplexed processing engines per ASIC to fulfil the \SI{25}{ns} latency requirement. 
The output of the LTFs are the four track parameters: $p_T$, $\phi$, $\cot \theta$ and $z_0$. 
In 16-bit representation, an output bandwidth of about \SI{400}{\mega\byte\per\second} is expected.

\subsection{Vertex and Event Finder}
The \emph{Vertex \& Event Finder} (VEFI) receives, on average, 2200 tracks/ev from the in total \SI{\sim 20000}{} LTFs, corresponding to an input bandwidth of about \SI{700}{\giga\byte\per\second}. 
Assuming a vertex bin width of $\Delta z=\SI{1}{mm}$, the VEFI requires 200 fully parallel engines for track processing.  Each engine receives on average, about 11 tracks/ev directly from the LTFs and about 22 tracks from the neighbouring engines, which define the overlap regions. 

The most time-consuming task is the reconstruction of the (pile-up suppressed) track-jets. 
Each VEFI engine applies the trigger conditions on the track-jets and generates a trigger flag.
Furthermore, each engine calculates the total transverse momentum of all track-jets and sends this information to a \emph{Primary Vertex Peak Finder}, which identifies the primary bin in the vertex histogram, see \autoref{fig:PB}.
Finally, the TTT trigger decision is based on the trigger flags of the VEFI engines, optionally matched with the peak position
of the primary bin. 
The algorithm is highly parallel and can be implemented either on an FPGA using a pipelined architecture or on a High-Performance Computer.

\subsection{Feasibility of the Design}
The biggest challenge for the TTT concept is the high input bandwidth to the LTFs, which reduces the data rate by about $2$ orders of magnitude. Therefore, it would be very advantageous to install the farm of LTF ASICs close to the detector or even on the detector.
The design of the LTF ASICs is demanding but can already be solved with today's technology.

\section{Calorimeter Emulation}
\label{App:CaloEmu}
The calorimeter emulation used in this work is based on the reference detector design considerations for the FCC-hh presented in the conceptual design report\,\cite{CDR_FCC}.
The calorimetry starts at a radius of $R_{calo}=\SI{2}{\meter}$ and is placed inside a solenoidal magnetic field with strength \SI{4}{\tesla}.

Our simplified calorimeter emulation assumes a very fine granularity for the hadronic calorimeter of $\Delta\eta \times \Delta\phi =0.025 \times 0.025$ as specified in \cite{CDR_FCC} and extends up to a pseudorapidity of $|\eta| < 2.5$.
A 2D histogram with the above granularity is used to construct the calorimeter cells in the emulation.
Energy deposits from all charged and neutral particles, barring neutrinos of the reference samples, including an average pileup of $\langle \mu \rangle = 1000$, are accumulated in the calo-cells represented by the 2D histogram.
Note that a Geant4 simulation of the calorimeter is not done here. 
Instead, particles are taken from the event generator and smeared. 
Muons are treated as hadrons, and for the sake of simplicity, no attempt is made here to emulate the charge deposition of muons.
The bending of charged particles due to the \SI{4}{\tesla} magnetic field is taken into account in the emulation by correcting charged particles initial direction $\phi_i$ by $\phi_{i} - \arcsin{\left( R_{calo}/(2R)\right)}$, where $R\,[\si{\meter}] = \pT/(0.3\,qB)\,[\si{\GeV/c\cdot\tesla^{-1}}]$ is the radius of the charged particle.
Furthermore, a minimum transverse momentum cut of \SI{1.2}{\GeV/c} is applied to prevent charged particles that won't reach the calorimeter from making energy deposits.

Finite calorimeter energy resolution effects are taken into account, and the accumulated energy deposits in each of the $\eta - \phi$ cells are smeared by an energy resolution for hadronic showers of \SI{50}{\percent}.
The corrected energy deposits for each of $\eta-\phi$ cells are fed to the anti-$k_t$ jet  algorithm\,\cite{FASTJET} with a radius parameter of 0.4 to produce emulated calo-jets.

\bibliographystyle{elsarticle-num}
\bibliography{references}

\end{document}